\documentclass[10pt,aps,prd,twocolumn,showpacs,amsmath,amssymb,nofootinbib,eqsecnum,preprintnumbers,longbibliography]{revtex4-1}

\usepackage[english]{babel}				%% longbibliography
\usepackage[T3,T1]{fontenc}				%% inverted breve
\usepackage[mathscr]{eucal} 			%% math fonts
\usepackage{accents}					%% underaccents
\usepackage{mathtools}					%% mathclap, etc.
\usepackage{graphicx}					%% includegraphics

\usepackage[ddmmyyyy]{datetime}    		%% currenttime
\usepackage[dvipsnames]{xcolor}			%% color

\newcommand\bs[1]{\boldsymbol{#1}}

\newcommand\tb[1]{\bs{#1}}

\newcommand\dd{\mathrm{d}}
\newcommand\pp{\partial}

\newcommand\eps{\varepsilon}
\newcommand\ph{\varphi}

\newcommand\imag{i}

\newcommand\lieb[2]{\left[ #1\,{,}\,#2 \right]}

\newcommand\feq{\mathrel{\phantom{=}}}

\DeclareSymbolFont{tipa}{T3}{cmr}{m}{n}				%invbreve
\DeclareMathAccent{\invbreve}{\mathalpha}{tipa}{16}
\newcommand\ubar[1]{\underaccent{\bar}{#1}}
\newcommand\ubars[1]{\ubar{\smash{#1}}}

\newcommand\cir[1]{\accentset{\circ}{#1}}

\newcommand\mus{{\breve\mu}}
\newcommand\nus{{\breve\nu}}
\newcommand\mur{{\invbreve\mu}}
\newcommand\nur{{\invbreve\nu}}
\newcommand\ks{{\breve{k}}}
\newcommand\ls{{\breve{l}}}
\newcommand\kr{{\invbreve{k}}}
\newcommand\lr{{\invbreve{l}}}
\newcommand\Ns{{\breve{N}}}
\newcommand\Nr{{\invbreve{N}}}
\newcommand\psir{{\invbreve{\psi}}}

\renewcommand{\Re}{\operatorname{Re}}

\newcommand{\artanh}{\operatorname{artanh}}

\begin{document}

%%%%%%%%%%%%%%%%%%%%%%%%%%%%%%%%%%%%%%%%%%%%%%%%%%%%%%%%%%%%%%%%%%%%%%%%%%%%%%%%%%%%%%
%% TITLE

\title{NUT-like and near-horizon limits of Kerr--NUT--(A)dS spacetimes}

\author{Ivan Kol\'a\v{r}}
\email{Ivan.Kolar@utf.mff.cuni.cz}
\affiliation{Institute of Theoretical Physics,
Faculty of Mathematics and Physics, Charles University,
V~Hole\v{s}ovi\v{c}k\'ach 2, 180 00 Prague, Czech Republic}

\author{Pavel Krtou\v{s}}
\email{Pavel.Krtous@utf.mff.cuni.cz}
\affiliation{Institute of Theoretical Physics,
Faculty of Mathematics and Physics, Charles University,
V~Hole\v{s}ovi\v{c}k\'ach 2, 180 00 Prague, Czech Republic}

\date{January 18, 2017}

\begin{abstract}
We study a class of limits of the higher-dimensional Kerr--NUT--(A)dS spacetimes where particular roots of metric functions degenerate. Namely, we obtain the Taub--NUT--(A)dS and the extreme near-horizon geometries as two examples of our limiting procedure. The symmetries of the resulting spacetimes are enhanced which is manifested by the presence of supplementary Killing vectors and decomposition of Killing tensors into Killing vectors.
\end{abstract}

%% 04.50.-h  Higher-dimensional gravity and other theories of gravity
%% 04.50.Gh  Higher-dimensional black holes, black strings, and related objects
%% 04.70.Bw  Classical black holes

\maketitle

%%%%%%%%%%%%%%%%%%%%%%%%%%%%%%%%%%%%%%%%%%%%%%%%%%%%%%%%%%%%%%%%%%%%%%%%%%%%%%%%%%%%%%
%% INTRODUCTION

\section{Introduction}
\label{sc:Intro}
Two very important vacuum solutions of the four-dimensional Einstein equations were found in the same year 1963, the Kerr spacetime \cite{Kerr:1963} and the Taub--NUT (Newman--Unti--Tamburino) spacetime \cite{NewmanTamburinoUnti:1963}. The nonstationary part (the Taub region) was known even earlier \cite{Taub:1951}. They both are one-parametric generalizations of the Schwarzschild solution, but only the Kerr solution has a clear physical interpretation.

The Kerr spacetime describes the gravitational field of a rotating black hole with a spherical horizon topology. It is well understood and agrees with many physical observations. On the contrary, the Taub--NUT spacetime has many undesirable features and pathologies such as the existence of closed timelike curves, the semi-infinite topological singularity on the axis, no (global) asymptotic flatness, and others. For this reason it is often presented as ``a counterexample to almost anything'' \cite{Misner:1967}.

Despite many attempts (see e.g. \cite{GriffithsPodolsky:2009book}) the Taub--NUT spacetime has not been satisfactory interpreted yet. The NUT parameter is often referred to as the magnetic mass or the gravitomagnetic monopole moment, due to the similarities with the theory of the magnetic monopoles \cite{LyndenBellNouriZonos:1998}. For instance, all geodesics of the stationary part (the NUT region) lie on spatial cones as the classical orbits of charged particles under the action of a charged magnetic monopole. In this analogy, the semi-infinite singularity on the axis of the Taub--NUT solution resembles the Dirac's string \cite{DemianskiNewman:1966}, i.e. the object which connects monopoles with opposite polarity. However, contrary to the semiaxis in the Taub--NUT solution, which affects the spacetime, the Dirac's string has no effect at all. The semi-infinite singularity can thus be considered as a thin massless (semi-infinite) spining rod injecting angular momentum into the spacetime \cite{Bonnor:1969}.

These solutions were generalized by inclusion of a cosmological constant as well as a charge, and presented as a single solution admitting separability of charged Hamilton--Jacobi and Klein--Gordon equations \cite{Carter:1968b} in 1968. This exceptional property was found later to be associated with the existence of a rank-two Killing tensor. The most general family of four-dimensional type D spacetimes with an aligned non-null electromagnetic field and a cosmological constant is now known as the Plebia\'{n}ski--Demia\'{n}ski class \cite{PlebanskiDemianski:1976}, which contains seven arbitrary parameters. Although the parameters are often interpreted as mass, rotation, NUT, electric/magnetic charge parameters, and a cosmological constant, they acquire their traditional physical meaning in special subcases only. In particular, there is still an ambiguity in what the parameters responsible for rotations and NUT-like patologies are. Nevertheless, Griffiths and Podolsk\'{y} were able to introduce a new coordinate system and a set of parameters that are natural for identifying many special subcases \cite{GriffithsPodolsky:2006b,GriffithsPodolsky:2007} such as the Kerr--(A)dS, the Taub--NUT--(A)dS, the C-metric, etc.

Apart from the spacetimes available as a particular subcases, some spacetimes can be obtained by taking limits such as the near-horizon limit. The near-horizon geometry of the extreme Kerr black hole \cite{BardeenHorowitz:1999} is in many aspects similar to the ${\mathrm{AdS}_2\times\mathrm{S}_2}$ geometry arising in the near-horizon limit of the extreme Reissner--Nordstr\"{o}m black hole. For example, by taking the limit of Kerr solution the symmetry of spacetime is enhanced to ${\mathrm{SL}(2,\mathbb{R})\times \mathrm{U}(1)}$, which is accompanied by the emergence of two new Killing vectors. These results were generalized to the presence of an arbitrary cosmological constant, and it was shown that the Killing tensor in the near-horizon geometry is reducible and can be expressed in terms of the Casimir operators formed by four Killing vectors \cite{Galajinsky:2010,Rasmussen:2011,ZahraniEtal:2011,Galajinsky:2011}. Recently it was noted that such a Killing tensor can be constructed even in near-horizon limits of spacetimes that do not admit separability of the charged Klein--Gordon equation \cite{BicakHejda:2015}.

Higher-dimensional solutions of the Einstein equations became popular in connection with the string theory and AdS/CFT correspondence, however, they are important also just from the mathematical point of view. The search for the higher-dimensional black hole solutions began also in 1963, when Tangherlini introduced the generalization of the spherically symmetric solution. The spacetime describing a generally rotating black hole was discovered by Myers and Perry \cite{MyersPerry:1986} in 1986. The solution was further generalized by inclusion of a cosmological constant \cite{HawkingEtal:1999,GibbonsEtal:2004,GibbonsEtal:2005}. In 2006, Chen, L\"{u}, and Pope introduced coordinates which made the inclusion of NUT parameters very natural \cite{ChenLuPope:2006,ChenLuPope:2007}. Such geometries are known now as the Kerr--NUT--(A)dS spacetimes. Even though they were found by a rather complicated way (through the Kerr--Schild form), they are direct generalizations of the Carter's separable metric \cite{KolarKrtous:2016}. Unlike in four dimensions, neither charged nor accelerated solution have been found yet, so the Kerr--NUT--(A)dS still remain the most general higher-dimensional black hole solutions with a spherical horizon topology. In 2004, Mann and Stelea found the higher-dimensional Taub--NUT--(A)dS solution \cite{MannStelea:2004,MannStelea:2006} from the ansatz constructed as radial extensions of $\mathrm{U}(1)$ fibrations over $2$-spheres.

The Kerr--NUT--(A)dS spacetimes have many exceptional properties that are related to a high degree of symmetry encoded in the existence of a tower of Killing vectors and Killing tensors, which can be constructed from a closed conformal Killing--Yano tensor \cite{KubiznakFrolov:2007,FrolovKubiznak:2007,KrtousEtal:2007a}. In particular, the geodesic motion is completely integrable \cite{PageEtal:2007,KrtousEtal:2007b}, the Hamilton--Jacobi, Dirac, and Klein--Gordon equations are separable \cite{FrolovEtal:2007,SergyeyevKrtous:2008,CarigliaEtal:2011a,CarigliaEtal:2011b}. Several limits of the Kerr--NUT--(A)dS spacetimes were studied including the near-horizon limits \cite{LuMeiPope:2009,Chernyavsky:2014,XuYue:2015} and the limits leading to warped spaces \cite{KrtousEtal:2016}, however, all possible subcases have not been identified yet.

The purpose of this paper is to investigate a particular class of limits, which lead, for example, to the Taub--NUT--(A)dS spacetime and the extreme near-horizon geometry. The paper is organized as follows: In Sec.~\ref{sc:Kerr} we briefly summarize some properties of the general higher-dimensional Kerr--NUT--(A)dS spacetimes. Also, we investigate a choice of parameters and ranges of coordinates which could describe a black hole and introduce a useful tangent-point parametrization of a polynomial which appears in the Kerr--NUT--(A)dS metric. Section~\ref{sc:limitingprocedure} is devoted to the limiting procedure itself. We present an appropriate scaling of coordinates which leads to a finite metric. Moreover, we discuss the symmetry enhancement associated with the presence of additional Killing vectors. In Sec.~\ref{sc:NUTlikeLimits}, we provide particular examples of our limiting procedure when applied to the Euclidean sector. This results in NUT-like limits such as the Taub--NUT--(A)dS spacetime. Applications to the Lorentzian sector are examined in Sec.~\ref{sc:NHLimits}. It gives rise to the extreme near-horizon limit. Finally, we study a limit when all directions degenerate in Sec.~\ref{sc:CompleteDegeneration}. We conclude with a brief summary in Sec.~\ref{sc:Conc}. An overview of the notation and useful identities are listed in Appendix~\ref{ap:ui}. The connection forms of the Kerr--NUT--(A)dS spacetime and the limiting metric are given in Appendix~\ref{ap:cononeforms}.

%%%%%%%%%%%%%%%%%%%%%%%%%%%%%%%%%%%%%%%%%%%%%%%%%%%%%%%%%%%%%%%%%%%%%%%%%%%%%%%%%%%%%%
%% KERR--NUT--(A)DS SPACETIMES

\section{Kerr--NUT--(A)dS spacetimes}
\label{sc:Kerr}

%%%%%%%%%%%%%%%%%%%%%%%%%%%%%%%%%%%%%%%%%%%%%%%%%%%%%%%%%%%%%%%%%%%%%%%%%%%%%%%%%%%%%%
%% Metric

\subsection{Metric}
\label{scc:Metric}
The Kerr--NUT--(A)dS spacetimes in $2N$ dimensions\footnote{%
We restrict ourselves to even dimensions for simplicity. The odd-dimensional case could by analyzed in a similar manner, only additional terms related to the odd dimension would be present.}
are given by the metric
\begin{equation}\label{eq:gKerrE}
  \tb{g} = \sum_{\mu}\bigg[\frac{U_\mu}{X_\mu}\,\tb{\dd}x_\mu^{\bs{2}}
  +\frac{X_\mu}{U_\mu}\Big(\sum_{k}A_{\mu}^{(k)}\,\tb{\dd}\psi_k\Big)^{\!\bs{2}}\bigg]\;.
\end{equation}
Here, greek and latin indices take values
\begin{equation}
\begin{split}
	\mu,\nu,\,\dots &=1,\,\dots,\,N\;,
	\\
	k,l,\,\dots &=0,\,\dots,\,N-1\;.
\end{split}
\end{equation}
We do not use Einstein summation convention for these indices, but we write just ${\prod_\mu}$ or ${\sum_k}$ if products and sums run over these default ranges of indices.

In order for the metric to satisfy the Einstein equations, the functions $X_\mu$ must have the form
\begin{equation}\label{eq:Xmu}
 X_\mu = \lambda\mathcal{J}(x_\mu^2)-2b_\mu x_\mu\;,
\end{equation}
where ${\mathcal{J}(x^2)}$ is an even polynomial of degree ${2N}$ in ${x}$, which can be parametrized using its roots
\begin{equation}\label{eq:Jinroots}
    \mathcal{J}(x^2) = \prod_\mu (a_\mu^2-x^2)\;.
\end{equation}
The functions $U_\mu$ and $A_{\mu}^{(k)}$ are explicit polynomial expressions in coordinates ${x_\mu}$,
\begin{equation}
    A_\mu^{(k)}=\sum_{\mathclap{\substack{\nu_1,\,\dots,\,\nu_k\\ \nu_1<\dots<\nu_k\\ \nu_j\neq\mu}}}x_{\nu_1}^2\dots x_{\nu_k}^2\;,\quad
   U_\mu=\prod_{\substack{\nu\\ \nu\neq\mu}}(x_\nu^2-x_\mu^2)\;.
\end{equation}
All these and related quantities, together with several identities they satisfy, are collected in Appendix~\ref{ap:ui}.

Metric \eqref{eq:gKerrE} contains $N$ parameters $a_\mu$ and $N$ parameters $b_\mu$. Thanks to a scaling symmetry of the metric, one of the parameters ${a_\mu}$ can be fixed to a chosen value. For example, a common Lorentzian gauge is \eqref{eq:anlambda}. The parameter $\lambda$ is related through the Einstein equations to the cosmological constant as
\begin{equation}
	\Lambda=(2N-1)(N-1)\lambda\;.
\end{equation}

Geometry \eqref{eq:gKerrE} can represent various spaces with regard to different ranges of coordinates. Moreover, some coordinates and parameters can be considered complex provided that the metric remains real.  As will be discussed in the next subsection, in the Lorentzian regime (with Wick rotated quantities), and for a particular choice of ranges of coordinates, this metric represents a black hole rotating in ${N-1}$ independent planes of rotations, with NUT parameters, and the cosmological constant. An interpretation of a general case is, however, more complicated and not sufficiently clarified.

The coordinates $\psi_k$ are Killing coordinates in angular directions of the corresponding rotational symmetries, which are described by $N$ Killing vectors
\begin{equation}\label{eq:killingpsi}
	\tb{l}_{(k)} =\frac{\tb{\pp}}{\tb{\pp} \psi_k}\;.
\end{equation}
Spatial coordinates ${x_\mu}$ are restricted between adjacent zeros of ${X_\mu}$, which correspond to fixed points of these rotational symmetries (in the Lorentzian case they are related to horizons of the temporal Killing vector). In particular, zeros of $X_1$ defining the range of $x_1$ represent (at least part of) the north and south semiaxes of the corresponding rotational symmetry. Moreover, the intervals of coordinates ${x_\mu}$ should be mutually distinct to avoid singularities due to zeros of functions ${U_\mu}$.

It is useful to define the orthonormal frame of one-forms,
\begin{equation}\label{eq:ON1forms}
	\tb{e}^\mu = \bigg(\frac{X_\mu}{U_\mu}\bigg)^{\!\!-\frac{1}{2}}\tb{\dd}x_\mu\;,
	\quad
	\tb{\hat{e}}^\mu = \bigg(\frac{X_\mu}{U_\mu}\bigg)^{\!\!\frac{1}{2}}\sum_{k}A_{\mu}^{(k)}\,\tb{\dd}\psi_k\;,
\end{equation}
and the corresponding dual vector frame [cf. \eqref{eq:AmuU1}],
\begin{equation}\label{eq:ONvectors}
\begin{split}
	\tb{e}_\mu &= \biggl(\frac{X_\mu}{U_\mu}\biggr)^{\!\!\frac{1}{2}}\frac{\tb{\pp}}{\tb{\pp} x_\mu}\;,
	\\
	\tb{\hat{e}}_\mu &= \biggl(\frac{X_\mu}{U_\mu}\biggr)^{\!\!-\frac{1}{2}}\sum_k\frac{{(-x_\mu^2)}^{N-k-1}}{U_\mu}\frac{\tb{\pp}}{\tb{\pp} \psi_k}\;.
\end{split}
\end{equation}
The metric is then simply given by
\begin{equation}\label{eq:metricON}
	\tb{g}=\sum_{\mu}\big(\tb{e}^\mu\tb{e}^\mu+\tb{\hat{e}}^\mu\tb{\hat{e}}^\mu\big)\;.
\end{equation}

Apart from the explicit symmetries $\tb{l}_{(k)}$, the metric \eqref{eq:gKerrE} also possesses hidden symmetries encoded by $N$ rank-two Killing tensors
\begin{equation}
\tb{k}_{(k)} =\sum_\mu A^{(k)}_{\mu}\bigl(\tb{e}_\mu\tb{e}_\mu+\tb{\hat{e}}_\mu\tb{\hat{e}}_\mu\bigr)\;.
\end{equation}
Moreover, all the symmetries $\tb{l}_{(k)}$ and $\tb{k}_{(k)}$ can be generated from the principal closed conformal Killing--Yano tensor \cite{KrtousEtal:2007a}
\begin{equation}\label{eq:PCCKY}
    \tb{h}=\sum_\mu x_\mu \tb{e}_\mu\wedge\tb{\hat{e}}_\mu\;.
\end{equation}

Since a linear combination of Killing vectors forms again a Killing vector, it is not clear which Killing coordinates should be regarded as periodic. Typically, these are not $\psi_k$, but it is possible to identify different coordinates associated with vectors that vanish at zeros of $X_\mu$ resembling the rotational symmetry in four dimensions. We will not discuss this in more detail, but the first step is to introduce Killing coordinates $\phi_\mu$ related to coordinates $\psi_k$ through a linear transformation labeled by some fixed values $\cir{x}_\mu$,
\begin{equation}\label{eq:psiphirel}
    \psi_k = \sum_\nu \frac{(-\cir{x}_\nu^2)^{N{-}k{-}1}}{\cir{U}_\nu}\phi_\nu\;,
    \quad
    \phi_\mu = \sum_l \cir{A}^{(l)}_\mu\psi_l\;.
\end{equation}
The associated Killing vectors are thus\footnote{In our discussion, $\cir{x}_\mu$ are arbitrary parameters. However, the metric is regular with the periodic coordinates $\phi_\mu$ only if ${\cir{x}_\mu}$ are zeros of $X_\mu$, because then the Killing vectors $\tb{\pp}_{\phi_{\mu}}$ can vanish at the endpoints ${x_\mu=\cir{x}_\mu}$.}
\begin{equation}\label{eq:newKV}
	 \cir{\tb{r}}_{(\mu)}=\sum_k\frac{{(-\cir{x}_\mu^2)}^{N-k-1}}{\cir{U}_\mu}\tb{l}_{(k)}=\frac{\tb{\pp}}{\tb{\pp} \phi_{\mu}}\;.
\end{equation}
The circle $\cir{\phantom{x}}$ above $U_\mu$ and $A^{(k)}_\mu$ indicates that $\cir{U}_\mu$ and $\cir{A}^{(k)}_\mu$ are constructed using $\cir{x}_\mu$ instead of $x_\mu$. Equivalence of both relations in \eqref{eq:psiphirel} follows from the identities analogous to \eqref{eq:AmuU1} and \eqref{eq:AmuU2}. In these coordinates the metric \eqref{eq:gKerrE} takes the form
\begin{equation}\label{eq:gKerrEphi}
  \tb{g} = \sum_{\mu}\biggl[\frac{U_\mu}{X_\mu}\,\tb{\dd}x_\mu^{\bs{2}}
  +\frac{X_\mu}{U_\mu}\Bigl(\sum_{\nu}\frac{J_\mu(\cir{x}_\nu^2)}{\cir{U}_\nu}\,\tb{\dd}\phi_\nu\Bigr)^{\!\bs{2}}\biggr]\;,
\end{equation}
and the orthonormal frame of one-forms \eqref{eq:ON1forms} and vectors \eqref{eq:ONvectors} read
\begin{equation}\label{eq:ON1formsph}
	\tb{e}^\mu = \bigg(\frac{X_\mu}{U_\mu}\bigg)^{\!\!-\frac{1}{2}}\tb{\dd}x_\mu\;,
	\quad
	\tb{\hat{e}}^\mu = \bigg(\frac{X_\mu}{U_\mu}\bigg)^{\!\!\frac{1}{2}}\sum_{\nu}\frac{J_\mu(\cir{x}_\nu^2)}{\cir{U}_\nu}\,\tb{\dd}\phi_\nu\;,
\end{equation}
and
\begin{equation}\label{eq:ONvectorsph}
	\tb{e}_\mu = \biggl(\frac{X_\mu}{U_\mu}\biggr)^{\!\!\frac{1}{2}}\frac{\tb{\pp}}{\tb{\pp} x_\mu}\;,
	\quad
	\tb{\hat{e}}_\mu = \biggl(\frac{X_\mu}{U_\mu}\biggr)^{\!\!-\frac{1}{2}}\sum_{\nu}\frac{\cir{J}_\nu(x_\mu^2)}{U_\mu}\frac{\tb{\pp}}{\tb{\pp} \phi_\nu}\;,
\end{equation}
respectively. The polynomial $J_\mu$ (and $\cir{J}_\nu$) is defined by \eqref{eq:Jmu}. The duality can be verified by employing relations analogous to \eqref{eq:JJ} (with $\cir{x}_\mu$ instead of $a_\mu$).

By analogy with Eq. \eqref{eq:newKV}, we also introduce new rank-two Killing tensors
\begin{equation}\label{eq:newKT}
	 \cir{\tb{q}}_{(\mu)}=\sum_k\frac{{(-\cir{x}_\mu^2)}^{N-k-1}}{\cir{U}_\mu}\tb{k}_{(k)}=\sum_\nu\frac{J_\nu(\cir{x}_\mu^2)}{\cir{U}_\mu}\bigl(\tb{e}_\nu\tb{e}_\nu+\tb{\hat{e}}_\nu\tb{\hat{e}}_\nu\bigr)\;.
\end{equation}

%%%%%%%%%%%%%%%%%%%%%%%%%%%%%%%%%%%%%%%%%%%%%%%%%%%%%%%%%%%%%%%%%%%%%%%%%%%%%%%%%%%%%%
%% Black hole

\subsection{Black hole}
\label{ssc:physmeancoor}
The metric \eqref{eq:gKerrE} describes various spaces for different choices of parameters and ranges of coordinates, but a lot of them are interesting just from the mathematical point of view. In this subsection we try to highlight such choices which seem to describe physically interesting spacetimes. We begin with the requirements on the signature of the metric.

The Lorentzian signature can be obtained by the Wick rotation of one $x$ coordinate, say ${x_N}$, and by choosing the corresponding parameter $b_N$ to be imaginary,
\begin{equation}\label{eq:lorsignaturexb}
x_N=\imag r\;,
\quad
b_N=\imag m\;.
\end{equation}
Thus, we assume  $r$, $m$, as well as the remaining coordinates $x_{\bar{\mu}}$ and parameters $b_{\bar\mu}$ to be real. The parameters~$a_\mu$ can still be complex, however, the polynomial ${\mathcal{J}}$ must remain real. We also rename the Killing variables
\begin{equation}\label{eq:psicoordlorentz}
   \mathcal{T}=\psi_0\;,\quad
   \chi_{\bar{k}} = \psi_{\bar{k}+1}\;.
\end{equation}
Here, the ``barred'' indices take values
\begin{equation}
\begin{aligned}
	\bar{\mu},\,\bar{\nu},\,\dots &=1,\,\dots,\,\bar{N}\;,
\\
	\bar{k},\,\bar{l},\,\dots &=0,\,\dots,\,\bar{N}-1\;,
\end{aligned}
\end{equation}
with ${\bar{N}=N-1}$.

With these conventions, the metric \eqref{eq:gKerrE} can be rewritten as
\begin{equation}\label{eq:gKerrL}
\begin{split}
  \tb{g} &= -\frac{\Delta}{\bar{J}(-r^2)}\Big(\tb{\dd} \mathcal{T} + \sum_{\bar{k}}\bar{A}^{(\bar{k}+1)}\,\tb{\dd}\chi_{\bar{k}}\Big)^{\!\bs{2}} +\frac{\bar{J}(-r^2)}{\Delta}\,\tb{\dd}r^{\bs{2}}
  \\
  &\feq + \sum_{\bar{\mu}}\Bigg[\frac{(r^2{+}x_{\bar{\mu}}^2)\,\bar{U}_{\bar{\mu}}}{\mathcal{X}_{\bar{\mu}}}\,\tb{\dd}x_{\bar{\mu}}^{\bs{2}}
  \\
  &\feq+\frac{\mathcal{X}_{\bar{\mu}}}{(r^2{+}x_{\bar{\mu}}^2)\,\bar{U}_{\bar{\mu}}}\Big(\tb{\dd}\mathcal{T} + \sum_{\bar{k}}\big(\bar{A}^{(\bar{k}+1)}_{\bar{\mu}}-r^2\bar{A}^{(\bar{k})}_{\bar{\mu}}\big)\,\tb{\dd}\chi_{\bar{k}}\Big)^{\!\bs{2}}\Bigg]\;,
\end{split}\raisetag{14ex}
\end{equation}
The barred quantities are defined by the same relations as the ordinary ones, just involving only coordinates~${x_{\bar{\mu}}}$. We also introduced new symbols for the metric functions,
\begin{equation}\label{eq:DeltacalX}
\begin{alignedat}{2}
 \Delta &= -X_N &&=-\lambda\mathcal{J}(-r^2)-2m r\;,
\\
 \mathcal{X}_{\bar{\mu}} &=-X_{\bar{\mu}} &&= -\lambda\mathcal{J}(x_{\bar{\mu}}^2)+2b_{\bar{\mu}} x_{\bar{\mu}}\;.
\end{alignedat}
\end{equation}

The metric \eqref{eq:gKerrL} has the Lorentzian signature, provided that the parameters $a_{\bar{\mu}}$, $\lambda$, and $b_{\bar{\mu}}$ are set so that the condition
\begin{equation}\label{eq:lorcond}
\frac{\mathcal{X}_{\bar{\mu}}}{\bar{U}_{\bar{\mu}}}>0
\end{equation}
is satisfied for all values of $x_{\bar{\mu}}$. It can be achieved by restricting ${x_{\bar{\mu}}}$ to intervals between adjacent roots ${}^{\mp}x_{\bar{\mu}}$ of the polynomials ${\mathcal{X}_{\bar{\mu}}}$,
\begin{equation}\label{eq:coorranges}
{}^{-}x_{\bar{\mu}}<x_{\bar{\mu}}<{}^{+}x_{\bar{\mu}}\;,
\end{equation}
in such a way that the intervals for different ${x_{\bar{\mu}}}$ do not overlap and we thus avoid the zeros of the functions $\bar{U}_{\bar{\mu}}$. Without loss of generality, we assume that
\begin{equation}\label{eq:coorcond}
	x_1^2<x_2^2<\dots<x_{\bar{N}}^2\;.
\end{equation}
Moreover, we choose the gauge condition
\begin{equation}\label{eq:anlambda}
   a_N^2=-\frac{1}{\lambda}\;,
\end{equation}
which guarantees that the metric has well-defined limit ${\lambda\rightarrow 0}$, because ${\lambda\mathcal{J}\big|_{\lambda=0}=-\bar{\mathcal{J}}}$.

\begin{figure}
    \centering
	\includegraphics[width=\columnwidth]{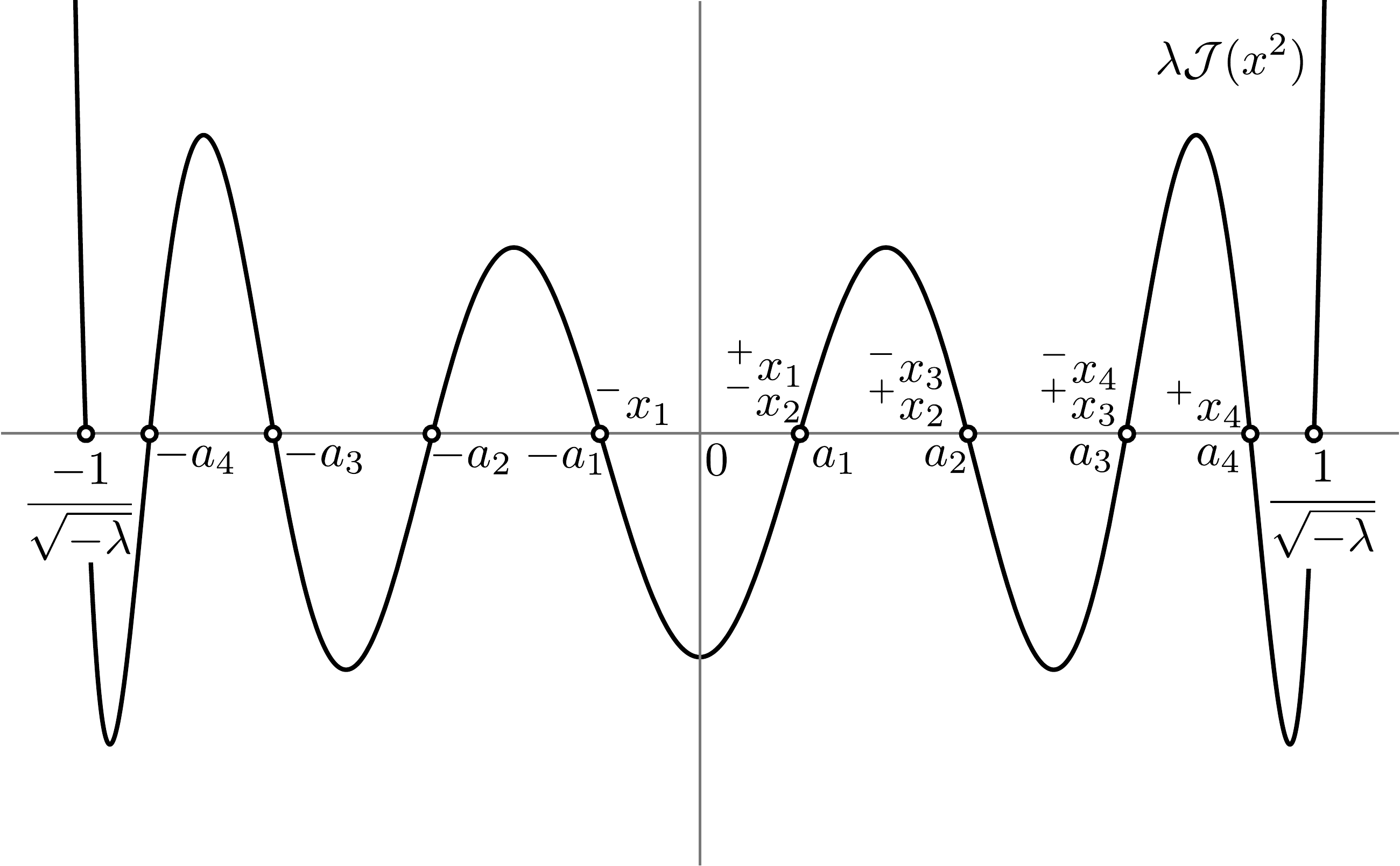}
    \caption{Graph of the polynomial $\lambda\mathcal{J}(x^2)$ for real (positive) parameters ${a_{\bar{\mu}}}$, vanishing parameters ${b_{\bar{\mu}}}$, and ${\lambda<0}$. Roots of this polynomial determine the ranges of coordinates, see \eqref{eq:xcoorrangezerob}.}
    \label{fig:rangex0}
\end{figure}

We are particularly interested in a few important examples of such a choice. First, let us consider that all parameters ${a_{\bar{\mu}}}$ are real (and positive),
\begin{equation}\label{eq:aordered}
	0<a_1<a_2<\dots<a_{\bar{N}}\;,
\end{equation}
and ${b_{\bar{\mu}}}$ vanish. Then the polynomials ${\mathcal{X}_{\bar{\mu}}=-\lambda\mathcal{J}(x_{\bar{\mu}}^2)}$ have a common set of roots $\pm a_{\bar{\mu}}$, see Fig.~\ref{fig:rangex0}.
In order to satisfy \eqref{eq:coorranges} and \eqref{eq:coorcond}, we choose the endpoints ${{}^{\pm}x_{\bar{\mu}}}$ of the coordinates $x_{\bar{\mu}}$ as follows:
\begin{equation}
\begin{split}
{}^{-}x_1 &=-a_1\;,
\\
{}^{-}x_2 &=a_1\;,
\\
{}^{-}x_3 &=a_2\;,
\\
&\vdotswithin{=}
\end{split}
\quad
\begin{split}
{}^{+}x_1 &=a_1\;,
\\
{}^{+}x_2 &=a_2\;,
\\
{}^{+}x_3 &=a_3\;,
\\
&\vdotswithin{=}
\end{split}
\end{equation}
As a result, the coordinates $x_{\bar{\mu}}$ are restricted by the intervals
\begin{equation}\label{eq:xcoorrangezerob}
	-a_1<x_1<a_1<x_2<a_2<\dots<x_{\bar{N}}<a_{\bar{N}}\;.
\end{equation}
Furthermore, we assume that the cosmological constant is sufficiently small. In particular we require that
\begin{equation}\label{eq:lambdabound}
	|\lambda|<\frac{1}{a_{\bar{N}}^2}\;.
\end{equation}
This condition implies that the roots $\pm 1/\sqrt{-\lambda}$, of the polynomial $\lambda\mathcal{J}(x^2)$ for ${\lambda<0}$ are far from the other roots and cannot influence the signature, because they are outside of the ranges of coordinates $x_{\bar{\mu}}$. On the other hand, for ${\lambda>0}$, the condition \eqref{eq:lambdabound} guarantees that the shape of the function $\lambda\mathcal{J}(x^2)$ is not modified near the origin due to the imaginary roots $\pm \imag/\sqrt{\lambda}$. In particular, it has a minimum at ${x=0}$, since
\begin{equation}
	 \big(\lambda\mathcal{J}(x^2)\big)''\big|_{x=0}=2\bar{\mathcal{A}}^{\bar{N}}\bigg(\sum_{\bar{\mu}}\frac{1}{a_{\bar{\mu}}^2}-\lambda\bigg)>0\;.
\end{equation}
For ${\lambda=0}$, real or imaginary roots corresponding to $a_N^2$ disappear.

Under the described assumptions, the metric reduces to the well-understood Myers--Perry solution with an arbitrary cosmological constant, also known as the higher-dimensional Kerr--(A)dS spacetime. This solution describes a black hole which arbitrarily rotates in $\bar{N}$ different planes, $a_{\bar{\mu}}$ are rotational parameters and $m$ is its mass. Coordinate $r$ stands for a radial-type coordinate and $x_{\bar{\mu}}$ are the latitudinal directions. Coordinates $\mathcal{T}$ and $\chi_{\bar{k}}$ correspond to timelike and spatial Killing coordinates.

Now we turn to the case of nonzero parameters~$b_{\bar{\mu}}$. These parameters are often called NUTs (or NUT charges), because they bring the NUT-like behavior in our spacetime. However, as we will discuss later, their meaning varies in several subcases.

\begin{figure}
    \centering
	\includegraphics[width=\columnwidth]{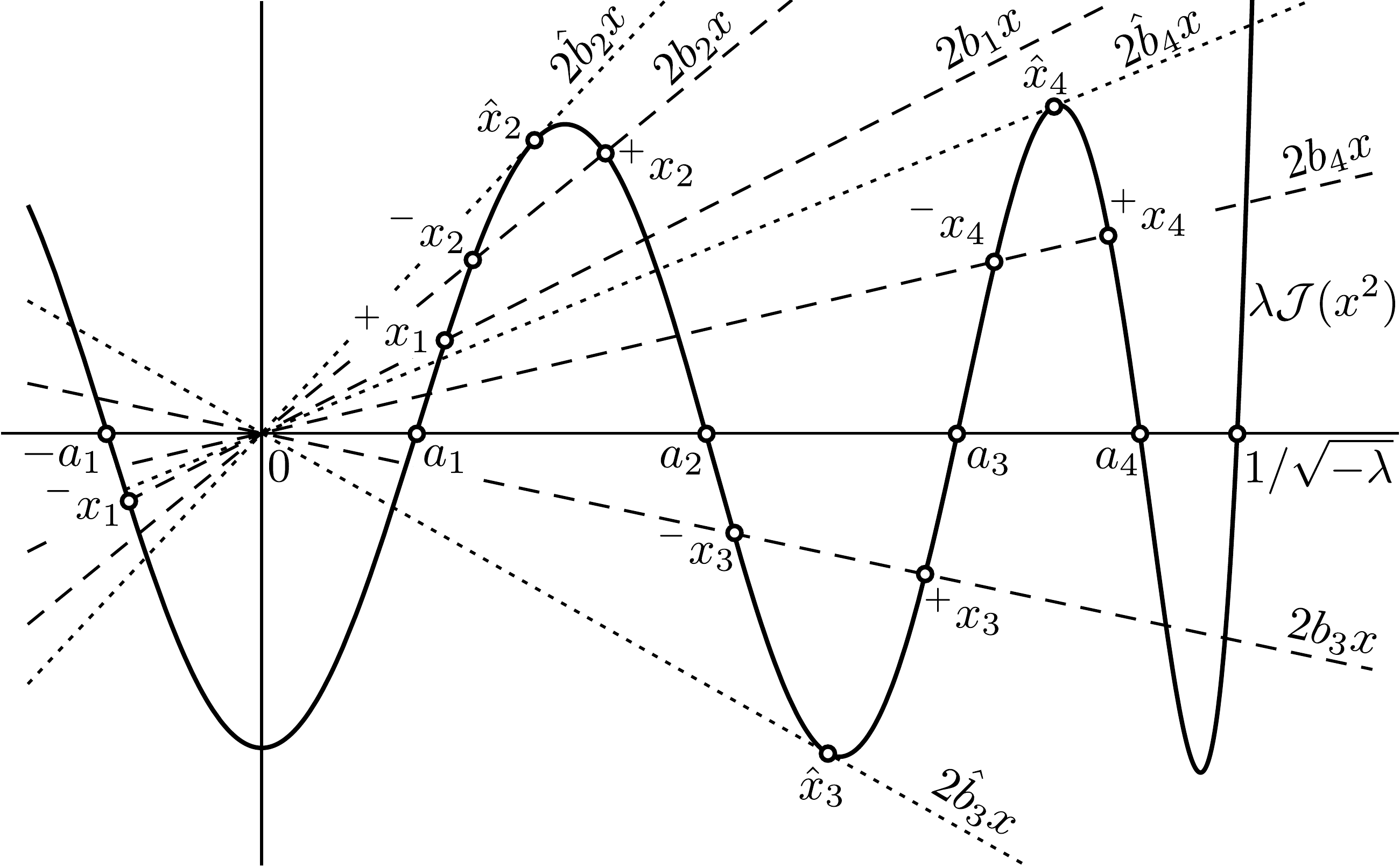}
    \caption{Graph of the polynomial $\lambda\mathcal{J}(x^2)$ for the Kerr-like choice of parameters and ranges of coordinates. Intersections of the polynomial and the lines $2b_{\bar{\mu}}x$ passing through the origin determine the ranges of coordinates, see \eqref{eq:coorranges}, \eqref{eq:xrootsineq1}.}
    \label{fig:rangex1}
\end{figure}

There are many options how to choose intervals of the coordinates to preserve the Lorentzian signature when ${b_{\bar{\mu}}\neq0}$. Our intention is not to describe all such possible choices, but rather select the ones which seem reasonable and give interesting spacetimes for the particular limiting values of parameters $a_{\bar{\mu}}$ and $b_{\bar{\mu}}$. Here, the roots of the polynomials $\mathcal{X}_{\bar{\mu}}$ are not just ${\pm a_{\bar{\mu}}}$ anymore. They do not even coincide for different indices $\bar{\mu}$, since the polynomials differ by the linear term which is proportional to the parameter $b_{\bar{\mu}}$, see \eqref{eq:DeltacalX}. Although it is impossible to find analytic expressions for the roots of polynomials $\mathcal{X}_{\bar{\mu}}$ in this case, we can learn at least something about them from the pictures of the intersections of the polynomial $\lambda\mathcal{J}(x^2)$ and the lines passing through the origin, $2b_{\bar{\mu}}x$, see Fig.~\ref{fig:rangex1}.

\begin{figure}
    \centering
	\includegraphics[width=\columnwidth]{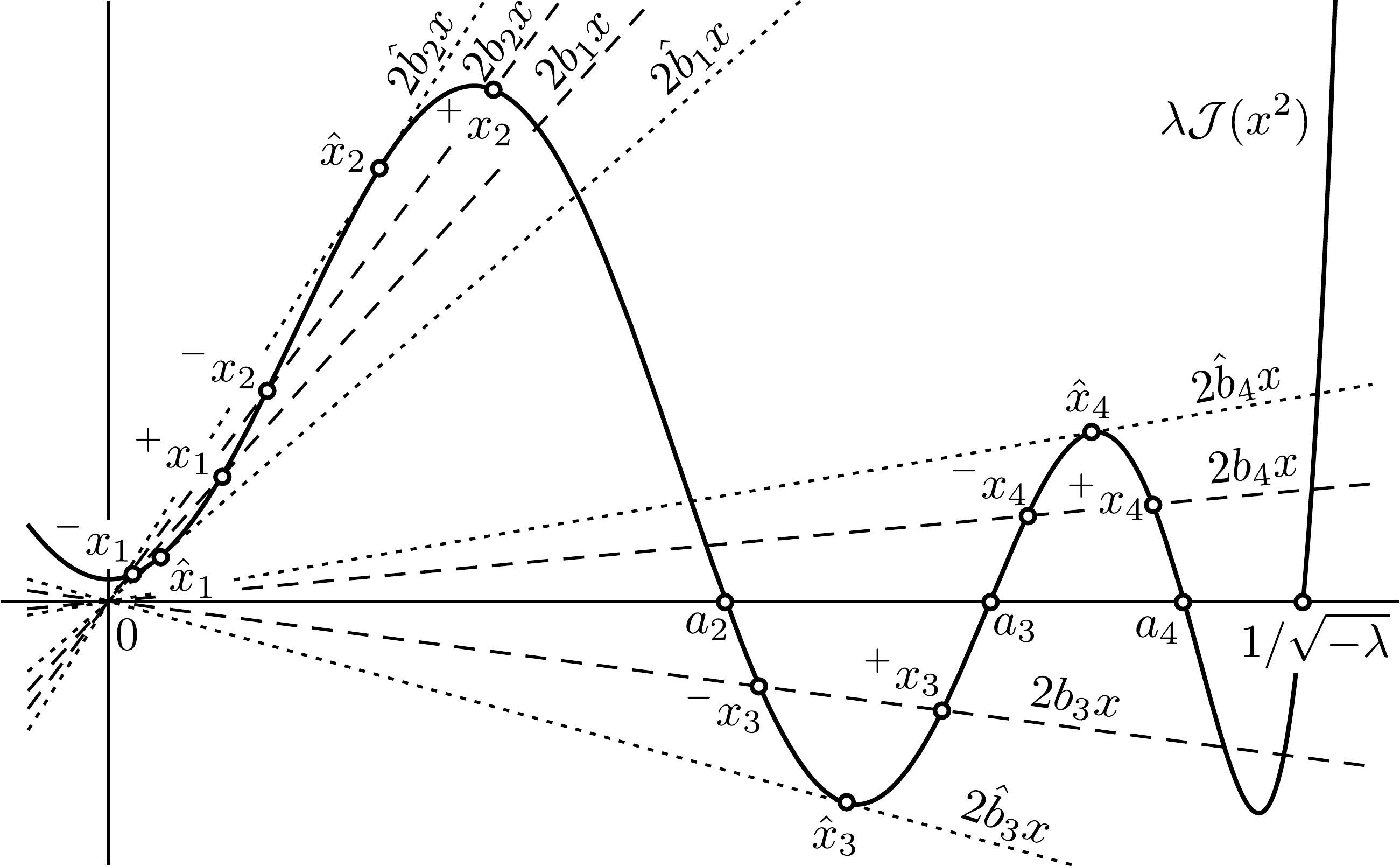}
    \caption{Graph of the polynomial $\lambda\mathcal{J}(x^2)$ for the NUT-like choice of parameters and ranges of coordinates. Intersections of the polynomial and the lines $2b_{\bar{\mu}}x$ passing through the origin determine the ranges of coordinates, see \eqref{eq:coorranges}, \eqref{eq:xrootsineq2}.}
    \label{fig:rangex2}
\end{figure}

So far we have assumed that all $a_{\bar{\mu}}$ are real, cf. \eqref{eq:aordered}, but we can extend our discussion to the case where one of the parameters $a_{\bar{\mu}}$, say $a_{1}$, becomes imaginary (and small), while the others remain real (and positive)
\begin{equation}\label{eq:a1imagsmall}
	\begin{gathered}
    a_{1}=\imag\tilde{a}_1\;,
    \\
	0<\tilde{a}_1<a_2<a_3<\dots<a_{\bar{N}}\;.
	\end{gathered}
\end{equation}
As with the previous case, this situation can also be analyzed graphically, see Fig.~\ref{fig:rangex2}.

We see from the Fig.~\ref{fig:rangex1} and Fig.~\ref{fig:rangex2} that the polynomial $\lambda\mathcal{J}(x^2)$ with \eqref{eq:aordered} and with \eqref{eq:a1imagsmall} differs qualitatively only near the origin. Far from the origin, there are just alternating ``hills'' and ``valleys'' (positive parts with maxima and negative parts with minima). Thus, we see that in order to avoid overlaps and get Lorentzian signature (see \eqref{eq:coorranges} and \eqref{eq:coorcond}) we can assume that parameters $b_{\bar{\mu}}$, except the first one, have alternating signs,
\begin{equation}\label{eq:bparsigns}
\begin{split}
	0 &<b_2 <\hat{b}_2\;,
	\\
	0 &<b_4 <\hat{b}_4\;,
	\\
	&\vdotswithin{<b_4 <}
\end{split}
\quad
\begin{split}
	\hat{b}_3 &<b_3 <0\;,
	\\
	\hat{b}_5 &<b_5 <0\;,
	\\
	&\vdotswithin{<b_5 <}
\end{split}
\end{equation}
The lines ${2b_{\bar{\mu}}x}$ then cut the polynomial $\lambda\mathcal{J}(x^2)$ through the corresponding hills and valleys. These intersections correspond to roots of ${\mathcal{X}_{\bar{\mu}}}$. We can select such intersections from among all the roots and choose them to be our endpoints ${{}^{\pm}x_{\bar{\mu}}}$,
\begin{equation}
	 {}^{-}x_{2}<\hat{x}_2<{}^{+}x_{2}<a_2<\dots<{}^{-}x_{\bar{N}}<\hat{x}_{\bar{N}}<{}^{+}x_{\bar{N}}<a_{\bar{N}}\;.
\end{equation}
We defined constants ${\hat{b}_{\bar{\mu}}}$ to be critical values of parameters ${b_{\bar{\mu}}}$ for which the roots ${{}^{\pm}x_{\bar{\mu}}}$ merge into a single double root ${\hat{x}_{\bar{\mu}}}$, i.e., they satisfy
\begin{equation}\label{eq:doublerootcal}
	\mathcal{X}_{\bar{\mu}}(\hat{x}_{\bar{\mu}})\big|_{b_{\bar{\mu}}=\hat{b}_{\bar{\mu}}}=0\;,
	\quad
	\mathcal{X}'_{\bar{\mu}}(\hat{x}_{\bar{\mu}})\big|_{b_{\bar{\mu}}=\hat{b}_{\bar{\mu}}}=0\;.
\end{equation}
Graphically, it means that ${\hat{x}_{\bar{\mu}}}$ label \emph{tangent points} at which the lines ${2\hat{b}_{\bar{\mu}} x}$ touch the polynomial ${\lambda\mathcal{J}(x^2)}$, see Fig.~\ref{fig:rangex1} and Fig.~\ref{fig:rangex2}.

In order to specify the range of the coordinate $x_1$, we have to distinguish the two situations. For $a_{\bar{\mu}}$ real (Fig.~\ref{fig:rangex1}), we choose the parameter $b_1$ so that it satisfies
\begin{equation}\label{eq:bparineq1}
    0<b_1<b_2
    %0<b_1<b_2<\hat{b}_2
\end{equation}
and we suppose that the endpoints of the coordinate $x_1$ are given by the relation
\begin{equation} \label{eq:xrootsineq1}
	-a_1<{}^{-}x_1<0<a_1<{}^{+}x_1<{}^{-}x_2\;.
\end{equation}
For $a_1$ imaginary (Fig.~\ref{fig:rangex2}), we assume
\begin{equation}\label{eq:bparineq2}
0<\hat{b}_1<b_1<b_2
%0<\hat{b}_1<b_1<b_2<\hat{b}_2
\end{equation}
and
\begin{equation} \label{eq:xrootsineq2}
	0<{}^{-}x_1<\hat{x}_1<{}^{+}x_1<{}^{-}x_2\;.
\end{equation}
We call these two alternative choices of parameters and ranges of coordinates the \emph{Kerr-like} and \emph{NUT-like} choices, respectively, for reasons which will be explained bellow. The main difference between them is that the former admits the limit of vanishing parameters ${b_{\hat{\mu}}\rightarrow 0}$ while the latter does not, because the parameter $b_1$ is bounded from below by the critical value $\hat{b}_1$.

Four dimensions are quite special, since there is just one coordinate $x$ and one parameter $b$. We drop the index~$1$ here, ${x=x_1}$, ${b=b_1}$, etc. Unlike the higher-dimensional case, the parameter $b$ is unbounded from above if ${\lambda\geq 0}$, but for ${\lambda<0}$ it must not exceed a value of a slope of a tangent line to the hill related to the root ${1/\sqrt{-\lambda}}$. Consequently, the endpoint ${}^{+}x$ is also less than the corresponding tangent point.

Returning to the arbitrary number of dimensions, it should be stressed that the choice \eqref{eq:xrootsineq2} is not always possible. The reason is that the polynomial $\lambda\mathcal{J}(x^2)$ admits tangent lines with tangent points between $0$ an $a_2$ (i.e. the lines $2\hat{b}_1 x$, $2\hat{b}_2 x$ with tangent points $\hat{x}_1$, $\hat{x}_2$) only for some particular values of parameters $a_{\bar{\mu}}$ and $\lambda$. This means that if we, for example, increase the parameter $\tilde{a}_1$ and keep the other parameters fixed, the lines $2\hat{b}_1 x$, $2\hat{b}_2 x$, and all lines in-between, approach each other until they finally cease to exist, see Fig.~\ref{fig:rangex2}. Without these lines the ranges of coordinates $x_1$, $x_2$ are not well defined.

As will be discussed in the next subsection [see \eqref{eq:doubrootcond} below], the problem of existence of tangent points $\hat{x}_1$, $\hat{x}_2$ leads to the condition
\begin{equation}\label{eq:doubrootcondlorentz}
	 \lambda\mathcal{J}(\hat{x}_{\bar{\mu}}^2)-2\lambda\hat{x}_{\bar{\mu}}^2\mathcal{J}'(\hat{x}_{\bar{\mu}}^2)=0\;,
	\;\;
	\bar{\mu}=1,2\;,
\end{equation}
which, unfortunately, cannot be solved explicitly in a general number of dimensions. In four dimensions, ${\mathcal{J}(x^2)=(\tilde{a}^2+x^2)(1/\lambda+x^2)}$ and the condition \eqref{eq:doubrootcondlorentz} reads
\begin{equation}\label{eq:doubrootcondD4}
    3\lambda \hat{x}^4+(1+\lambda \tilde{a}^2)\hat{x}^2-\tilde{a}^2=0\;.
\end{equation}
Solving this equation, we can find that the tangent point~$\hat{x}$ exists iff
\begin{equation}\label{eq:a1condD4}
	\lambda\geq 0
	\quad\text{or}\quad
	\lambda<0\;,\;\; \tilde{a}<\sqrt{\frac{7-4\sqrt{3}}{-\lambda}}\;.
\end{equation}
Therefore, under this assumption, the coordinate $x$ has well-defined endpoints ${}^{\pm}x$. In six dimensions and for vanishing cosmological constant, ${\lambda\mathcal{J}=(\tilde{a}_1^2+x^2)(a_2^2-x^2)}$ and the condition \eqref{eq:doubrootcondlorentz} gives again biquadratic equation,
\begin{equation}\label{eq:doubrootcondD6Lambdazero}
    3\hat{x}_{\bar{\mu}}^4+(\tilde{a}_1^2-a_2^2)\hat{x}_{\bar{\mu}}^2+\tilde{a}_1^2 a_2^2=0\;,
    	\;\;
	\bar{\mu}=1,2\;,
\end{equation}
Similarly, it says that the tangent points $\hat{x}_{1}$, $\hat{x}_{2}$ exist (and lie between $0$ and $a_2$) iff
\begin{equation}\label{eq:a1condD6}
	\tilde{a}_1<\sqrt{7-4\sqrt{3}}\,a_2\;.
\end{equation}

In higher dimensions (or/and with nonzero cosmological constant), the condition \eqref{eq:doubrootcondlorentz} for tangent points $\hat{x}_1$, $\hat{x}_2$, leads to higher degree polynomial equations and finding the explicit conditions of their existence is more difficult. Instead of trying to find  $\hat{x}_1$, $\hat{x}_2$ in these cases, we simply assume that such solutions exist and determine the ranges of coordinates.

With such choices of parameters and ranges of coordinates, we believe that the metric \eqref{eq:gKerrL} still represents a geometry of a rotating black hole with NUT charges and the cosmological constant. However, an exact relation of the parameters to physical quantities is still not completely clarified. In particular, it is not obvious which parameters describe rotations and which are actually responsible for effects caused by NUT charges. Nevertheless, it seems reasonable to distinguish the two regimes of the rotating black hole with NUT charges. In particular, the spacetime with real parameters $a_{\bar{\mu}}$, and small parameters~${b_{\bar{\mu}}}$,
\begin{equation}\label{eq:smallbmu}
    |b_{\bar{\mu}}|\lesssim 1\;,
\end{equation}
represents a Kerr-like (rotating) black hole with small NUT charges. On the other hand, the choice with one imaginary parameter $a_1$, and parameters ${b_{\bar{\mu}}}$ near the critical values ${\hat{b}_{\bar{\mu}}}$,
\begin{equation}\label{eq:nearcrit}
    |b_{\bar{\mu}}-\hat{b}_{\bar{\mu}}|\lesssim 1\;,
\end{equation}
corresponds to the NUT-like black hole with small rotational parameters.

The former spacetime was discussed in \cite{KrtousEtal:2016}, where the authors studied the warped metrics of two (off-shell) Kerr--NUT--(A)dS metrics which arise when the limits ${a_{\bar{\nu}}\rightarrow 0}$, ${x_{\bar{\nu}}\rightarrow 0}$ are taken in several directions $\bar{\nu}$. It was shown that such spacetimes have just one parameter less then the original Kerr--NUT--(A)dS metric, but the meaning of such parameters changes significantly. For example, if we take the limit in $\bar{N}$ directions, the resulting metric is static, though it still contains parametrized twists in the angular part of the metric (components with mixed angular directions). Thanks to the warped structure, the limiting procedure can be applied successively to untwist the angular directions. We end up with a metric containing $N$ parameters, where one contributes to the conicity and the others can be interpreted as deformations. Since these spacetimes were found under the assumption of small parameters ${b_{\bar{\mu}}}$, \eqref{eq:smallbmu}, it is not surprising that no solution with distinctive NUT-like behavior described by some NUT parameters is obtained by such a procedure.

On the contrary, the spacetime with general parameters ${b_{\bar{\mu}}}$, is not necessarily restricted just to the warped structure. In the next sections, we focus exactly on this case. Namely, we study the limits ${b_{\bar{\nu}}\rightarrow\hat{b}_{\bar{\nu}}}$, ${{}^{\pm}x_{\bar{\nu}}\rightarrow\hat{x}_{\bar{\nu}}}$ that are taken in several directions $\bar{\nu}$. Graphically speaking, it says that the corresponding lines ${2b_{\bar{\nu}}x}$ are approaching the tangents ${2\hat{b}_{\bar{\nu}}x}$ of the polynomial ${\lambda\mathcal{J}(x^2)}$. Such limits result in spacetimes which exhibit significant NUT-like behavior, for instance, the higher-dimensional generalization of the Taub--NUT--(A)dS spacetime.

The Wick rotated coordinate $x_N$ corresponds to the radial coordinate $r$, see \eqref{eq:lorsignaturexb}. Unlike the ${x_{\bar{\mu}}}$ coordinates, the coordinate $r$ does not have to be restricted between roots of the metric function, because the metric remains Lorentzian when we cross the root of the function ${\Delta}$, cf. \eqref{eq:gKerrL}, \eqref{eq:DeltacalX}. However, the coordinate $r$ changes its character. It is spatial ($\tb{\partial}_r$ is spacelike) iff ${\Delta>0}$ and temporal ($\tb{\partial}_r$ is timelike) iff ${\Delta<0}$. The surfaces ${\Delta=0}$ thus correspond to horizons, which separate the stationary and nonstationary regions of the spacetime. Here, we study the assumptions under which the spacetime admits a black hole horizon, namely the conditions on the parameters which exclude the naked singularities and other nonphysical cases for $r>0$. Nevertheless, we also discuss the analytic extension of such solutions to ${r<0}$ as it is usual in four dimensions.

\begin{figure}
    \centering
	\includegraphics[width=\columnwidth]{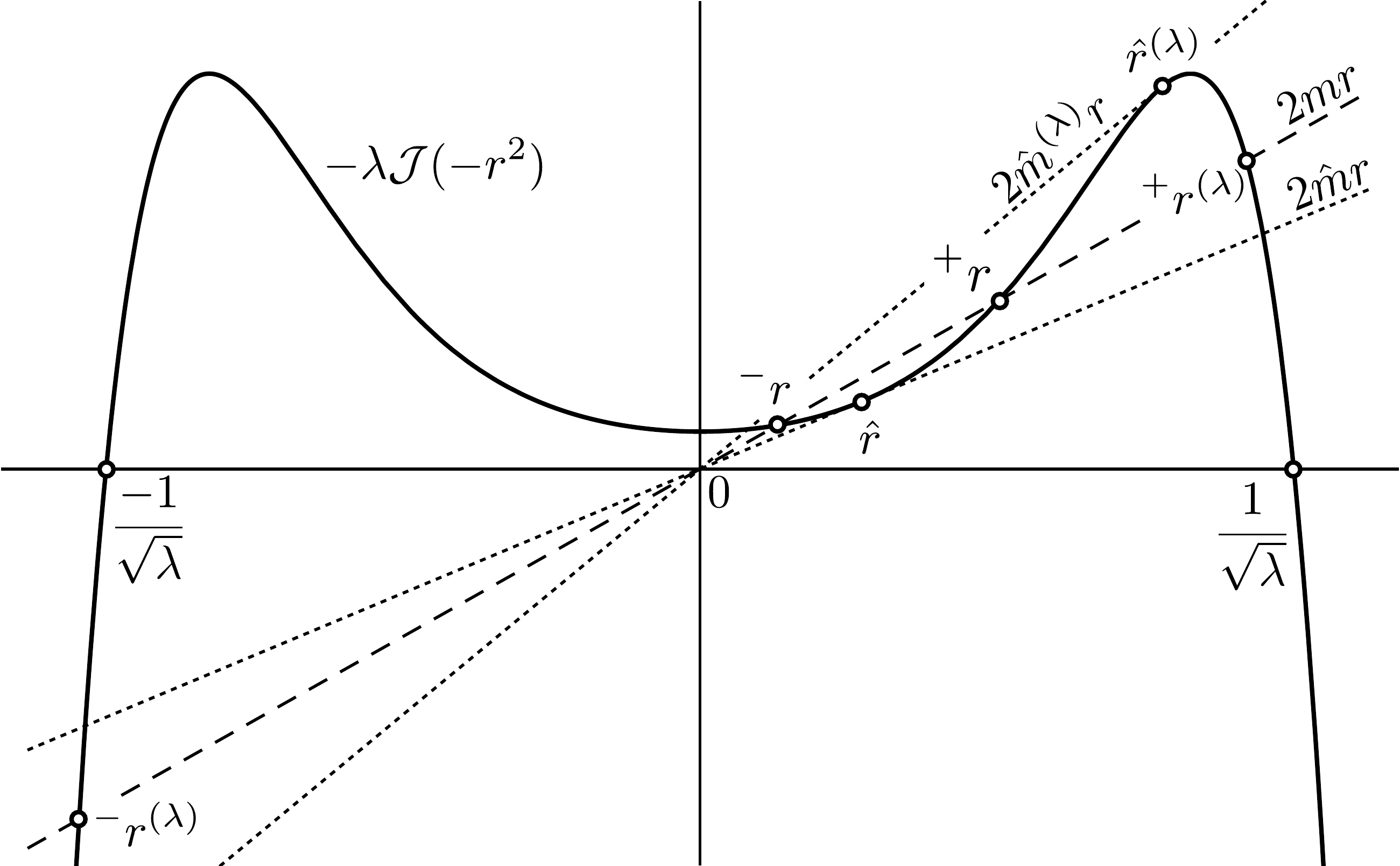}
    \caption{Graph of the polynomial $-\lambda\mathcal{J}(-r^2)$  for Kerr-like choice and an appropriate mass parameter $m$, cf. \eqref{eq:massreala}. Intersections of the polynomial and the line $2mr$ passing through the origin represent the horizons ${}^{\pm}r$, ${}^{\pm}r^{(\lambda)}$, see \eqref{eq:rangerreala}.}
    \label{fig:ranger1}
\end{figure}

First, consider the Kerr-like choice of parameters and ranges of coordinates. The graph of the metric function $\Delta$ for this situation is depicted in Fig.~\ref{fig:ranger1}.
In order to get a true black hole solution with all horizons, we choose the mass $m$ so that it satisfies
\begin{equation}\label{eq:massreala}
\begin{aligned}
	&0<\hat{m}<m\;, & \lambda &\leq 0\;,
	\\
	&0<\hat{m}<m<\hat{m}^{(\lambda)}\;, & \lambda &> 0\;,
\end{aligned}
\end{equation}
then the horizons are ordered as follows:
\begin{equation}
\begin{aligned}\label{eq:rangerreala}
	&0<{}^{-}r<\hat{r}<{}^{+}r\;, & \lambda &\leq 0\;,
	\\
	&{}^{-}r^{(\lambda)}<-\frac{1}{\sqrt{\lambda}}<0<{}^{-}r<\hat{r}<{}^{+}r & &
	\\
	&\quad\quad\quad\quad\quad<\hat{r}^{(\lambda)}<{}^{+}r^{(\lambda)}<\frac{1}{\sqrt{\lambda}}\;, & \lambda &> 0\;.
\end{aligned}
\end{equation}
Here, ${}^{-}r$, ${}^{+}r$, and ${}^{\pm}r^{(\lambda)}$ are inner, outer, and two cosmological horizons (above and below ${r=0}$) respectively. The cosmological horizons as well as the other related quantities exist only in the case of positive cosmological constant. Value $\hat{m}$ is a critical mass for which the horizons ${}^{-}r$, ${}^{+}r$ merge into a single extreme horizon~$\hat{r}$. A~similar situation also occurs with the horizons ${}^{+}r$, ${}^{+}r^{(\lambda)}$ for the critical mass $\hat{m}^{(\lambda)}$, which gives rise to an extreme horizon $\hat{r}^{(\lambda)}$. Tangent points  $\hat{r}$ and $\hat{r}^{(\lambda)}$  correspond to the double roots of the polynomial $\Delta$ for ${m=\hat{m}}$ and ${m=\hat{m}^{(\lambda)}}$,
\begin{equation}
\begin{aligned}
	\Delta(\hat{r})\big|_{m=\hat{m}} &=0 \;, &\quad	\Delta'(\hat{r})\big|_{m=\hat{m}} &=0\;,
	\\
	\Delta(\hat{r}^{(\lambda)})\big|_{m=\hat{m}^{(\lambda)}} &=0 \;, &\quad	 \Delta'(\hat{r}^{(\lambda)})\big|_{m=\hat{m}^{(\lambda)}} &=0\;.
\end{aligned}
\end{equation}

The second line of Eq. \eqref{eq:rangerreala} is not always met for the same reason as it was for \eqref{eq:xrootsineq2}. Again, the key property is the existence of the tangent lines $2\hat{m}r$, $2\hat{m}^{(\lambda)}r$. The equation for the tangent point $\hat{r}$ of the Wick rotated coordinate $r$ can be written as [see \eqref{eq:doubrootcond} below]
\begin{equation}\label{eq:doubrootcondlorentzwick}
\begin{gathered}
	\lambda\mathcal{J}(-\hat{r}^2)+2\lambda\hat{r}^2\mathcal{J}'(-\hat{r}^2) =0\;,
	\\
	 \lambda\mathcal{J}(-{\hat{r}^{(\lambda)}}^2)+2\lambda{\hat{r}^{(\lambda)}}^2\mathcal{J}'(-{\hat{r}^{(\lambda)}}^2) =0\;.
\end{gathered}
\end{equation}
In four dimensions, the solution of Eq. \eqref{eq:doubrootcondlorentzwick} can be found analytically, it exists iff
\begin{equation}\label{eq:a1condD4r}
	\lambda\leq 0
	\quad\text{or}\quad
	\lambda>0\;,\;\; a<\sqrt{\frac{7-4\sqrt{3}}{\lambda}}\;.
\end{equation}

\begin{figure}
    \centering
	\includegraphics[width=\columnwidth]{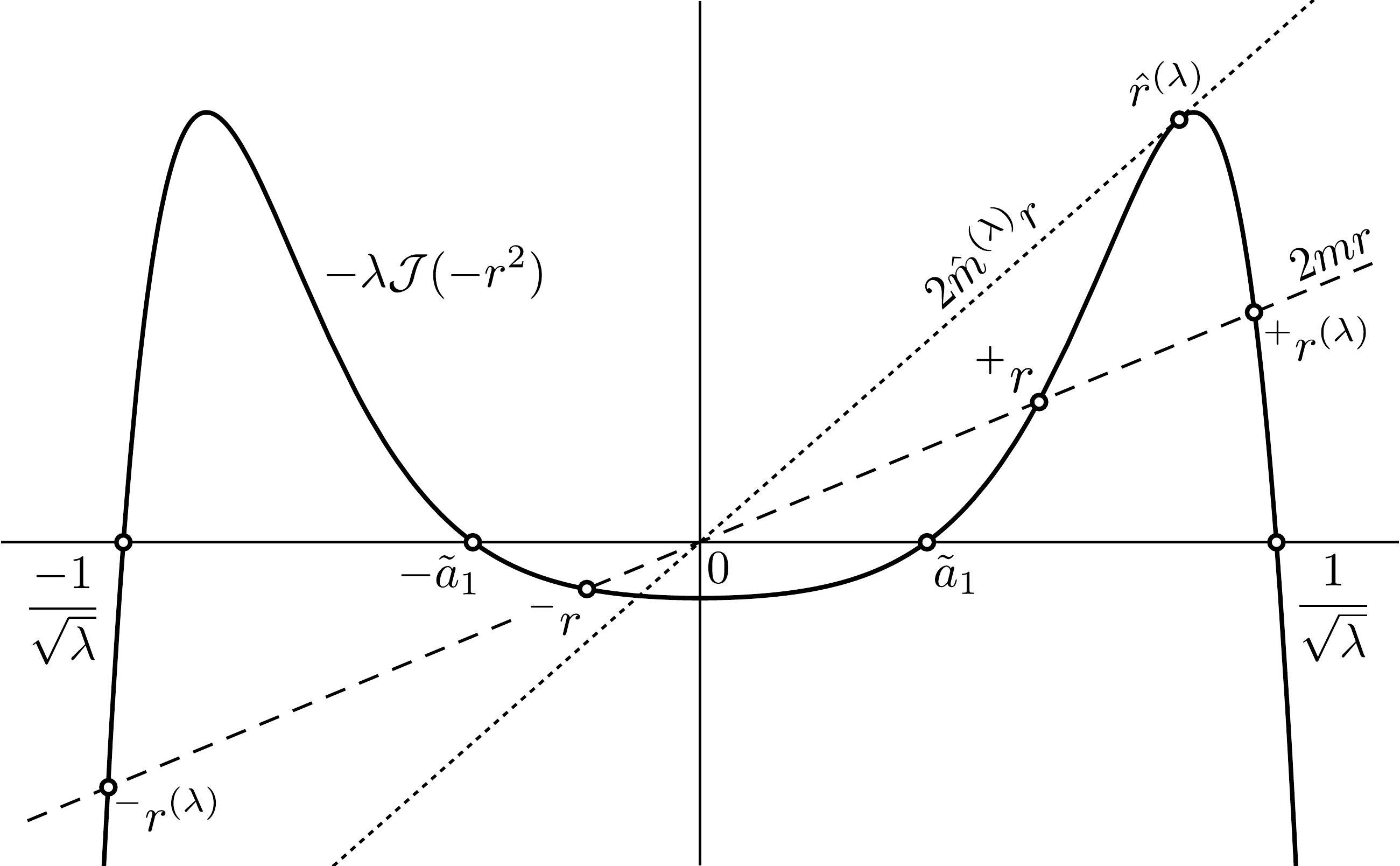}
    \caption{Graph of the polynomial $-\lambda\mathcal{J}(-r^2)$  for NUT-like choice and an appropriate mass parameter $m$, cf. \eqref{eq:massimaga}. Intersections of the polynomial and the line $2mr$ passing through the origin represent the horizons ${}^{\pm}r$, ${}^{\pm}r^{(\lambda)}$, see \eqref{eq:rangerimaga}.}
    \label{fig:ranger2}
\end{figure}

For NUT-like choice, the polynomial lies below zero near the origin, because it has two additional roots $\pm\tilde{a}_1$, see Fig.~\ref{fig:ranger2}.
Therefore, we can assume that $m$ is restricted by the relations
\begin{equation}\label{eq:massimaga}
\begin{aligned}
	&0<m\;, & \lambda &\leq 0\;,
	\\
	&0<m<\hat{m}^{(\lambda)}\;, & \lambda &> 0\;.
\end{aligned}
\end{equation}
Then, the horizons are ordered differently,
\begin{equation}\label{eq:rangerimaga}
\begin{aligned}
	&-\tilde{a}_1<{}^{-}r<0<\tilde{a}_1<{}^{+}r\;, &  \lambda &\leq 0\;,
	\\
	&{}^{-}r^{(\lambda)}<-\frac{1}{\sqrt{\lambda}}<-\tilde{a}_1<{}^{-}r<0<\tilde{a}_1<{}^{+}r & &
	\\
	 &\quad\quad\quad\quad\quad\quad\quad\quad\quad<\hat{r}^{(\lambda)}<{}^{+}r^{(\lambda)}<\frac{1}{\sqrt{\lambda}}\;, & \lambda &> 0\;.
\end{aligned}
\end{equation}
We see that the horizon ${}^{-}r$ is shifted to the negative values and the tangent line $2\hat{m}r$ disappeared. Altogether, we found that the properties of the radial coordinate (the positions of the horizons, etc.) are qualitatively similar to the four-dimensional case.

%%%%%%%%%%%%%%%%%%%%%%%%%%%%%%%%%%%%%%%%%%%%%%%%%%%%%%%%%%%%%%%%%%%%%%%%%%%%%%%%%%%%%%
%% Tangent-point parametrization

\subsection{Tangent-point parametrization}
\label{ssc:dblrootpar}

In the following it will be convenient to choose a different parametrization of the polynomial $\mathcal{J}(x^2)$. First we discuss this reparametrization for a general metric \eqref{eq:gKerrE} and later we specify it for the Lorentzian cases studied in the previous subsection.

Let us assume that for a particular value ${b_\mu=\hat{b}_\mu}$ the polynomial $X_{\mu}$ has a double root ${x_\mu=\hat{x}_\mu}$,
\begin{equation}\label{eq:doubleroot}
	X_\mu(\hat{x}_\mu)\big|_{b_{\mu}=\hat{b}_{\mu}}=0\;,
	\quad
	X'_\mu(\hat{x}_\mu)\big|_{b_{\mu}=\hat{b}_{\mu}}=0\;.
\end{equation}
If ${\hat{x}_{\mu}}$ and $\hat{b}_\mu$ are real, then the double root $\hat{x}_{\mu}$ graphically represents a tangent point where the tangent line ${2\hat{b}_\mu x}$ touches the polynomial ${\lambda\mathcal{J}(x^2)}$ and we call ${\hat{x}_{\mu}}$ tangent points. We generalize this notion also to imaginary points ${\hat{x}_{\mu}}$ with imaginary critical values $\hat{b}_\mu$, since such quantities correspond to Wick rotated tangent points and critical values of the polynomial in the Wick rotated coordinate~$r$.

Clearly, the tangent points are quantities determined by the polynomial ${\lambda\mathcal{J}}$, i.e., by the roots~${a_\mu}$, see~\eqref{eq:Jinroots}. This relation can be reversed. A~collection of $N$ tangent points ${\hat{x}_\mu}$, ${\mu=1,\,\dots,\,N}$, can be used to parametrize the polynomial ${\mathcal{J}}$. Moreover, since the polynomial $\mathcal{J}(x^2)$ is an even function of $x$, we can restrict ourselves to the parameters ${{\hat{x}_\mu}>0}$. We call this the \emph{tangent-point parametrization}.

The condition \eqref{eq:doubleroot} implies that the critical values $\hat{b}_{\mu}$ of the parameters $b_{\mu}$ are
\begin{equation}\label{eq:bhat}
	\hat{b}_\mu = \frac{\lambda\mathcal{J}(\hat{x}_\mu^2)}{2\hat{x}_\mu}\;,
\end{equation}
and tangent points ${\hat{x}_\mu}$ must satisfy
\begin{equation}\label{eq:doubrootcond}
	\lambda\mathcal{J}(\hat{x}_\mu^2)-2\lambda\hat{x}_\mu^2\mathcal{J}'(\hat{x}_\mu^2)=0
\end{equation}
for all ${\mu}$. Thus, the polynomial ${\mathcal{J}(x^2)-2x^2\mathcal{J}'(x^2)}$ of the degree~$N$ in $x^2$ has the roots $\hat{x}_\nu^2$, ${\nu=1,\,\dots,\,N}$. The coefficient of the highest order in $x^2$ is $-(2N-1)$, so the polynomial can be written as
\begin{equation}\label{eq:polyJJ}
\mathcal{J}(x^2)-2x^2\mathcal{J}'(x^2) = -(2N-1)\hat{J}(x^2)\;.
\end{equation}
Here, ${\hat{J}(x^2)}$ (and, similarly, other ``hatted'' quantities ${\hat{A}_\mu^{(k)}}$, ${\hat{U}_\mu}$, etc.) are defined in terms of parameters ${\hat{x}_\mu}$ in the same way as ${J(x^2)}$ (and ${A_\mu^{(k)}}$, ${U_\mu}$, etc., respectively) in terms of coordinates ${x_\mu}$, cf.~\eqref{eq:J} [and \eqref{eq:Amu}, \eqref{eq:Umu}, etc., respectively].

Comparing coefficients of powers of $x^2$ [see.~\eqref{eq:J}] we find
\begin{equation}\label{eq:calAA}
	(2N-2k-1)\mathcal{A}^{(k)}=(2N-1)\hat{A}^{(k)}\;,
\end{equation}
for ${k=0,\,\dots,\,N}$. These equations constitute implicit relations between the original parameters $a_{\mu}$ and the new parameters $\hat{x}_\mu$. Indeed, ${\mathcal{A}^{(k)}}$ are built from ${a_\mu}$ and ${\hat{A}^{(k)}}$ from ${\hat{x}_\mu}$, cf.~\eqref{eq:A}.

Equation \eqref{eq:calAA} can also be alternatively rewritten as an integral relation between functions $\mathcal{J}$ and~$\hat{J}$,
\begin{equation}\label{eq:calJhatJintegral}
	\mathcal{J}(x^2)=(2N-1)\,x\! \int\limits^x\!\!\frac{\hat{J}(y^2)}{y^2}\,\mathrm{d}y\;.
\end{equation}

So far we have assumed that the polynomial ${\lambda\mathcal{J}}$ has~${N}$ different tangent points. Here, we investigate this assumption in more detail. As it is difficult to solve the equations for tangent points of the polynomial ${\lambda\mathcal{J}(x^2)}$ in general, we proceed to the discussion of the Kerr-like and NUT-like choices from the previous subsection. Results are qualitatively summarized in Table \ref{tab:tangentpoints}.
\begin{table}
\centering
	{\renewcommand{\tabcolsep}{5pt} \renewcommand{\arraystretch}{1.2}
	\begin{tabular}{l|l|l}
	&\multicolumn{1}{c}{Kerr-like} & \multicolumn{1}{|c}{NUT-like} \\ \hline
	${\lambda>0}$	& $\imag\hat{r},\,\hat{x}_2,\,\dots,\,\hat{x}_{\bar{N}},\,\imag\hat{r}^{(\lambda)}$ &  $\hat{x}_1,\,\hat{x}_2,\,\dots,\,\hat{x}_{\bar{N}},\,\imag\hat{r}^{(\lambda)}$  \\
	${\lambda<0}$	& $\imag\hat{r},\,\hat{x}_2,\,\dots,\,\hat{x}_{\bar{N}},\,\hat{x}_{N}$ &  $\hat{x}_1,\,\hat{x}_2,\,\dots,\,\hat{x}_{\bar{N}},\,\hat{x}_{N}$  \\
	${\lambda= 0}$	& $\imag\hat{r},\,\hat{x}_2,\,\dots,\,\hat{x}_{\bar{N}}$ &  $\hat{x}_1,\,\hat{x}_2,\,\dots,\,\hat{x}_{\bar{N}}$  \\
	\end{tabular}
	}
	\caption{Tangent points $\hat{x}_{\mu}$ of the polynomial $\lambda\mathcal{J}(x^2)$. Imaginary tangent points are Wick rotated real tangent points of the polynomial  $-\lambda\mathcal{J}(-r^2)$.}
	\label{tab:tangentpoints}
\end{table}

Polynomial $\lambda\mathcal{J}(x^2)$ admits a full set of tangent points in all cases ($N$ points for ${\lambda\neq0}$ and $\bar{N}$ points for ${\lambda=0}$), however, some tangent points are not real but imaginary (Fig.~\ref{fig:rangex1} and Fig.~\ref{fig:rangex2}). These imaginary tangent points correspond to the Wick rotated real tangent points of the polynomial $-\lambda\mathcal{J}(-r^2)$ (Fig.~\ref{fig:ranger1} and Fig.~\ref{fig:ranger2}). In particular, the tangent point $\hat{x}_1$ is real only for the NUT-like choice, but it becomes imaginary ${\hat{x}_1=\imag\hat{r}}$ for the Kerr-like choice, where it corresponds to the Wick rotated real tangent point $\hat{r}$ of the polynomial $-\lambda\mathcal{J}(-r^2)$.

The situation also varies for different values of the cosmological constant. If ${\lambda<0}$, the polynomial $\lambda\mathcal{J}(x^2)$ has an additional real tangent point, denoted by ${\hat{x}_N}$, but for ${\lambda>0}$ it becomes imaginary ${\hat{x}_N=\imag\hat{r}^{(\lambda)}}$ and translates to the real tangent point ${\hat{r}^{(\lambda)}}$ of the polynomial  $-\lambda\mathcal{J}(-r^2)$.

Altogether, we have enough parameters to parametrize the polynomial $\mathcal{J}$ by means of the tangent points, but the gauge condition \eqref{eq:anlambda} says that these new parameters are not all independent. Since ${a_N^2}$ is a root of ${\mathcal{J}}$, it implies that
\begin{equation}
    0=\mathcal{J}\Bigl(-\frac1\lambda\Bigr)=\sum_{k=0}^{N}\mathcal{A}^{(k)}\lambda^{-N+k}\;.
\end{equation}
Employing \eqref{eq:calAA} and \eqref{eq:AAmuAmu}, we can rewrite this condition as
\begin{equation}\label{eq:eqexprxN}
\begin{split}
    0&=\sum_{k=0}^{N}\frac{2N-1}{2N{-}2k{-}1}\hat{A}^{(k)}\lambda^{-N+k}\\
     &=\sum_{k=0}^N\frac{2N-1}{2N{-}2k{-}1}
      \bigl(\hat{A}_N^{(k)}+\hat{x}_N^2 \hat{A}_N^{(k{-}1)}\bigr)\lambda^{-N+k}\;.
\end{split}
\end{equation}
Solving this equation with respect to ${\hat{x}_N^2}$, and realizing that ${\hat{A}_N^{(k)}=0}$ for ${k=-1}$ and ${k=N}$, we can express ${\hat{x}_N}$ in terms of the other tangent points ${\hat{x}_{\bar{\mu}}}$,
\begin{equation}\label{eq:constraintxhatexpl}
   \hat{x}_N^2 = -\frac{1}{\lambda}\frac
      {\displaystyle\sum_k\frac{\lambda^k \hat{\bar{A}}^{(k)}}{2N-2k-1}}
      {\displaystyle\sum_k\frac{\lambda^k \hat{\bar{A}}^{(k)}}{2N-2k-3}}
    \;,
\end{equation}
where our conventions imply ${\hat{\bar{A}}^{(k)}=\hat{A}_{N}^{(k)}}$. The expression in \eqref{eq:constraintxhatexpl} is divergent for ${\lambda\rightarrow 0}$ due to the fact that neither real nor imaginary tangent point $\hat{x}_N$ exists for~${\lambda=0}$.

%%%%%%%%%%%%%%%%%%%%%%%%%%%%%%%%%%%%%%%%%%%%%%%%%%%%%%%%%%%%%%%%%%%%%%%%%%%%%%%%%%%%%%
%% LIMITING PROCEDURE

\section{Limiting procedure}
\label{sc:limitingprocedure}

Our main goal is to investigate a situation when some of the metric functions ${X_\mu}$ have double roots. However, a well-defined metric cannot be achieved without an appropriate rescaling of coordinates ${x_\mu}$. The reason is that ${x_\mu}$ typically runs between two adjacent roots of ${X_\mu}$ and we are interested in the limit when these two roots coincide. It turns out that it is not enough to rescale just ${x_\mu}$ coordinates. We need to adjust the angular coordinates as well. In the following section we discuss both: scaling of ${x_\mu}$ and a proper transformation of Killing coordinates.

As it was discussed in Sec.~\ref{ssc:physmeancoor}, the Wick rotated coordinate $x_N$ corresponds to the radius $r$. This coordinate makes sense both above and bellow horizons, so it does not have to be restricted between roots of the metric function unlike the coordinates ${x_{\bar{\mu}}}$. Therefore, the double-root limiting procedure where the Wick rotated coordinate degenerates slightly differs from the Euclidean case. For simplicity, we restrict ourselves to the Euclidean case in this section. We will discuss the necessary changes for degeneration of the Wick rotated coordinate separately in Sec.~\ref{sc:NHLimits}.

%%%%%%%%%%%%%%%%%%%%%%%%%%%%%%%%%%%%%%%%%%%%%%%%%%%%%%%%%%%%%%%%%%%%%%%%%%%%%%%%%%%%%%
%% Scaling of coordinates x

\subsection{Scaling of coordinates ${x_\mu}$}
\label{ssc:scx}

Let us introduce a limiting procedure in which $\Ns$ of the coordinates ${x_\mu}$ are squeezed to the double root of the corresponding metric functions ${X_\mu}$. We indicate these coordinates using ``breved'' indices and complementary we denote $\Nr=N-\Ns$ coordinates, for which we do not take the limit, by ``inverse-breved'' indices:
\begin{equation}
\begin{aligned}
\mus,\nus,\,\dots &= \{\,\dots\,\text{degenerate directions}\,\dots\,\}\;,
\\
\mur,\nur,\,\dots  &= \{\,\dots\,\text{regular directions}\,\dots\,\}\;.
\end{aligned}
\end{equation}
The ranges of degenerated coordinates ${x_\mus}$ are restricted between adjacent roots ${{}^-x_\mus}$ and ${{}^+x_\mus}$ of the corresponding polynomial ${X_\mus}$,
\begin{equation}\label{eq:pmroots}
    X_\mus({}^{\pm}x_\mus)=0\;.
\end{equation}

It is straightforward to scale ${x_\mus}$ by distances between the roots, introducing rescaled coordinates~${\xi_\mus\in(-1,1)}$,
\begin{equation}\label{eq:xscl}
    x_\mus = \frac{{}^+x_\mus+{}^-x_\mus}{2} + \frac{{}^+x_\mus-{}^-x_\mus}{2}\, \xi_\mus\;.
\end{equation}

The limit in which both roots approach the real tangent points, ${{}^\pm x_\mus\to\hat{x}_\mus}$, corresponds to the parameters ${b_\mus}$ approaching its critical values, ${b_\mus\to\hat{b}_\mus}$. We can introduce a small parameter ${\eps\ll 1}$ governing expansions of ${{}^\pm x_\mus}$ and ${b_\mus}$ near their limiting values in such a way that the distances between the roots is of the first order in ${\eps>0}$,
\begin{equation}\label{eq:rootsdistance}
    {}^+x_\mus-{}^-x_\mus = 2\,\delta x_\mus\, \eps\;.
\end{equation}
The coefficients ${\delta x_\mus>0}$ control how fast the roots ${}^\pm x_\mus$ are approaching each other. We will fix them later.

\begin{figure}
    \centering
    \includegraphics[width=\columnwidth]{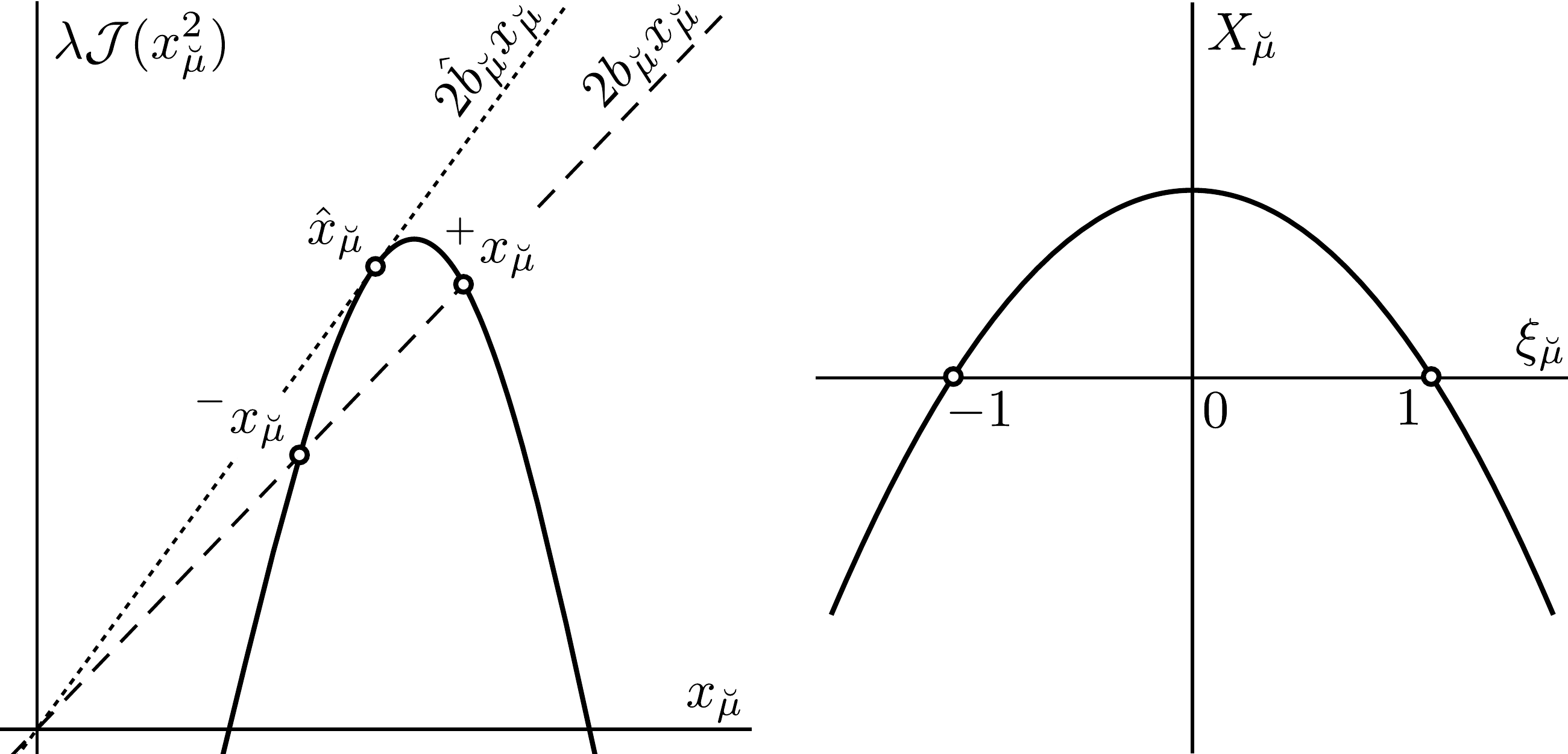}
    \caption{Approximation procedure of the polynomial $X_\mus$ is illustrated in terms of the lines passing through the origin and approaching the tangent line on the left. This is accompanied by zooming in on the region close to the tangent point $\hat{x}_\mus$, where a new coordinate $\xi_\mus$ is introduced. The resulting polynomial $X_\mus$ is approximated by a quadratic function in coordinate $\xi_\mus$ on the right.}
    \label{fig:approxX}
\end{figure}

The limiting procedure ${\varepsilon\rightarrow 0}$ thus corresponds to zooming in on a region around the tangent points $\hat{x}_\mu$ in which new coordinates $\xi_\mus$ are introduced. This zooming is accompanied by adjusting parameters $b_\mus$ so that they approach their critical values~$\hat{b}_\mus$ and the roots ${}^\pm x_\mus$ approach $\hat{x}_\mus$, see Fig.~\ref{fig:approxX}.

Functions $X_\mus\big|_{b_{\mus}=\hat{b}_{\mus}}$ can be expanded near $\hat{x}_\mus$ using Taylor expansion,
\begin{equation}\label{eq:Xhatbexp}
    X_\mus\big|_{b_{\mus}=\hat{b}_{\mus}} =
    -(2N-1)\,\lambda\,\hat{U}_\mus (x_\mus-\hat{x}_\mus)^2
    +\mathcal{O}\bigl((x_\mus-\hat{x}_\mus)^3\bigr)\;,
\end{equation}
where we used \eqref{eq:doubleroot} to eliminate the first two terms in the expansion. The factor ${\hat{U}_{\mus}}$ entered through the identity
\begin{equation}\label{eq:JJderder}
\mathcal{J}'(\hat{x}_\mus^2)+2\hat{x}_\mus^2\mathcal{J}''(\hat{x}_\mus^2) = -(2N-1)\hat{U}_\mus\;,
\end{equation}
which can be obtained by differentiation of \eqref{eq:polyJJ} and evaluation at $\hat{x}_\mus$. Employing \eqref{eq:Xhatbexp}, we can express the metric functions as
\begin{equation}\label{eq:Xexp}
 \begin{split}
    X_\mus &=
    -2(b_\mus-\hat{b}_\mus)x_\mu
    -(2N-1)\,\lambda\,\hat{U}_\mus (x_\mus-\hat{x}_\mus)^2 \\
    &\feq+ \mathcal{O}\bigl((x_\mus-\hat{x}_\mus)^3\bigr)\;.
 \end{split}
\end{equation}

Substituting general $\eps$ expansions of ${}^\pm x_\mus$ and $b_\mus$ into condition \eqref{eq:pmroots} with ${X_\mus}$ given by \eqref{eq:Xexp}, one finds that, in the first order in ${\eps}$, roots ${{}^+ x_\mus}$ and ${{}^- x_\mus}$ have the same distance from the tangent point ${\hat{x}_\mus}$,
\begin{equation}\label{eq:rootexp}
    {}^\pm x_\mus = \hat{x}_\mus\pm\delta x_\mus\, \eps+\mathcal{O}(\eps^2)\;,
\end{equation}
which implies that relation \eqref{eq:xscl} for rescaled coordinate ${\xi_\mus}$ reads\footnote{%
The middle point ${({}^+x_\mus+{}^-x_\mus)/2}$ differs from ${\hat{x}_\mus}$ in the second order in ${\eps}$, but an exact form will not be needed.}
\begin{equation}\label{eq:xsclexp}
    x_\mus = \hat{x}_\mus + \delta x_\mus\,\xi_\mus\,\eps+\mathcal{O}(\eps^2)\;.
\end{equation}
Also, we find that the parameters ${b_\mus}$ differ from their critical values only in the second order in $\eps$,
%\begin{equation}\label{eq:NUTexp}
%\begin{split}
%    %b_\mus = \hat{b}_\mus + \mathcal{O}(\eps^2)\;.
%    b_\mus
%      &=\hat{b}_\mus
%      + \lambda \bigl(\mathcal{J}'(\hat{x}_\mus^2)
%         + 2\hat{x}_\mus^2\mathcal{J}''(\hat{x}_\mus^2)\bigr)
%           \frac{\delta x_\mus}{\hat{x}_\mus}\eps^2
%      + \mathcal{O}(\eps^3)\\
%      &=\hat{b}_\mus
%      - (2N-1)\frac{\lambda}{2} \frac{\delta x_\mus}{\hat{x}_\mus}\hat{U}_\mus\eps^2
%      + \mathcal{O}(\eps^3)\;.
%\end{split}\raisetag{7ex}
%\end{equation}
\begin{equation}\label{eq:NUTexp}
    b_\mus
      =\hat{b}_\mus
      - (2N-1)\lambda \hat{U}_\mus\frac{\delta x_\mus^2}{2\hat{x}_\mus}\eps^2
      + \mathcal{O}(\eps^3)\;.
\end{equation}

In order to get the limiting metric, we need expansions of the metric functions ${X_\mus}$. Substituting expansions \eqref{eq:xsclexp} and \eqref{eq:NUTexp} into $X_\mus$ given by \eqref{eq:Xexp}, we obtain
\begin{equation}\label{eq:Xsimple}
	X_{\mus} = (2N-1)\lambda\hat{U}_{\mus}\delta x_\mus^2(1-\xi_\mus^2)\varepsilon^2
      + \mathcal{O}(\varepsilon^3)\;,
\end{equation}
The metric function $X_\mus$ is thus approximated by a quadratic function in new coordinate $\xi_\mus$ with roots ${\xi_\mus=\pm 1}$. Moreover, $X_\mus$ is of the order ${\eps^2}$ which allows us to compensate behavior ${\tb{\dd} x_\mus^{\bs{2}}\sim\eps^2\tb{\dd}\xi_\mus^{\bs{2}}}$ in metric terms ${\frac{U_\mus}{X_\mus}\tb{\dd} x_\mus^{\bs{2}}}$. However, we need to eliminate this ${\eps^2}$ behavior in metric terms proportional to ${X_\mus}$. To do so, it is necessary to rescale also the Killing coordinates.

%%%%%%%%%%%%%%%%%%%%%%%%%%%%%%%%%%%%%%%%%%%%%%%%%%%%%%%%%%%%%%%%%%%%%%%%%%%%%%%%%%%%%%
%% Scaling of Killing coordinates

\subsection{Scaling of Killing coordinates}
\label{ssc:scangles}

It turns out that all Killing coordinates $\psi_k$ have to be rescaled by $1/\eps$ to achieve a finite limit of the metric. However, among the leading terms of these $N$ coordinates only $\Ns$ of them are independent. Remaining $\Nr$ coordinates are related to the subleading order of coordinates $\psi_k$.

We start with the metric \eqref{eq:gKerrEphi} written in the angular coordinates $\phi_\mu$, introduced in \eqref{eq:psiphirel}. The coordinates $\phi_\mu$ can be associated with coordinates $x_\mu$ labeled by the same index~$\mu$. It allows us to distinguish two sets of these coordinates: degenerate coordinates $\phi_\mus$ and regular coordinates $\phi_\mur$.

Motivated by this distinction, we assume that constants $\cir{x}_\mus$ from the definition \eqref{eq:psiphirel} which are related to degenerate directions are close to $\hat{x}_\mus$. Thus they can be parametrized by rescaled constants~$\cir{\xi}_\mus$ according to \eqref{eq:xsclexp},
\begin{equation}\label{eq:xcircexp}
    \cir{x}_\mus = \hat{x}_\mus + \delta x_\mus\cir{\xi}_\mus\eps\;,\quad
    \cir{\xi}_\mus\in[-1,1]\;.
\end{equation}
Constants $\cir{\xi}_\mus$ will be parameters of the limiting metric, which may be related to the regularity of the metric at fixed points. A similar parameter is used to adjust the regularity on semiaxes in four dimensions. On the other hand, parameters $\cir{x}_\mur$ corresponding to the regular directions can be chosen arbitrarily. It turns out that the resulting metric is independent of them.

Now we rescale angular coordinates $\phi_\mus$ corresponding to degenerate directions $x_\mus$ by divergent factor $1/\eps$. Coefficients of the leading order define new finite coordinates~$\ph_\mus$,
\begin{equation}\label{eq:phdef}
    \phi_\mus = -\cir{\invbreve{J}}(\hat{x}_\mus^2)\,\ph_\mus\, \frac1\eps\;.
\end{equation}
Here, $-\cir{\invbreve{J}}(\hat{x}_\mus^2)$ are properly chosen constant factors which will simplify some expressions later. The inverse breve~$\invbreve{\phantom{x}}$ says that only regular directions are involved, but no degenerate directions. Here, it means that the polynomial~$\cir{\invbreve{J}}(x^2)$ contains only the constants $\cir{x}_\mur$,
\begin{equation}\label{eq:Jibc}
    \cir{\invbreve{J}}(x^2) = \prod_\nur (\cir{x}_\nur^2-x^2)\;.
\end{equation}
Similarly, we will introduce another breved and inverse breved quantities that are constructed using the degenerate and regular directions, respectively.

Thus, $\Ns$ coordinates $\phi_\mus$, or equivalently $\ph_\mus$, control divergent parts of the original Killing coordinates $\psi_k$. Indeed, substituting \eqref{eq:phdef} into \eqref{eq:psiphirel}, we find
\begin{equation}\label{eq:psiphlord}
    \psi_k = -\frac1\eps\sum_\nus\frac{(-\hat{x}_\nus^2)^{N-k-1}}{\hat{\breve{U}}_\nus}\ph_\nus
    +\mathcal{O}(\eps^0)\;,
\end{equation}
cf.\ also \eqref{eq:xcircexp} and \eqref{eq:Umusrexp} below.
Remaining information is encoded in $\Nr$ coordinates $\phi_\mur$.

However, to eliminate a dependence on arbitrary constants $\cir{x}_\mur$, it will be useful to combine these coordinates to another set of $\Nr$ Killing coordinates $\psir_\kr$. We employ a transformation analogous to \eqref{eq:psiphirel}, involving, however, only regular directions:
\begin{equation}\label{eq:psirphirel}
    \psir_\kr = \sum_\nur \frac{(-\cir{x}_\nur^2)^{\Nr{-}\kr{-}1}}{\cir{\invbreve{U}}_\nur}\phi_\nur\;,\quad
    \phi_\mur = \sum_\lr \cir{\invbreve{A}}^{(\lr)}_\mur\psir_\lr\;.
\end{equation}
Here, the latin indices associated with regular directions take values ${\kr,\lr,\,\ldots =0,1,\dots,\Nr{-}1}$, and similarly for indices associated with degenerate directions, ${\ks,\ls,\,\ldots =0,1,\,\dots\Ns{-}1}$. Coordinates ${\psir_\kr}$ thus encode information hidden in subleading order of the original Killing coordinates ${\psi_k}$.

Starting with the first equation in \eqref{eq:psirphirel} and substituting \eqref{eq:psiphirel} for $\phi_\nur$, we can express $\psir_\kr$ in terms of the original coordinates~$\psi_k$ (with no expansion involved) as
\begin{equation}\label{eq:psirpsirel}
    \psir_\kr = \sum_{\ls=0}^{\Ns} \cir{\breve{A}}^{(\ls)}\psi_{\ls+\kr}\;.
\end{equation}
Here, we used an identity
\begin{equation}
	\cir{A}_{\mur}^{(k)}=\sum_\lr\cir{\breve{A}}^{(k-\lr)}\cir{\invbreve{A}}_{\mur}^{(\lr)}\;,
\end{equation}
which can be proven by calculating the ${(N-k-1)}$th derivative of ${\cir{J}_{\mur}(x^2)=\cir{\invbreve{J}}_{\mur}(x^2)\cir{\breve{J}}(x^2)}$ with respect to $(-x^2)$ at ${x=0}$.

At the first sight it seems that the left-hand side of \eqref{eq:psirpsirel} is finite and the right-hand side is of order ${1/\eps}$. However, it is possible to check that the leading orders of ${\psi}$'s, given by \eqref{eq:psiphlord}, cancel in \eqref{eq:psirpsirel} and one has to take into account a contribution from the subleading order of ${\psi}$'s to obtain ${\psir_\kr}$.

We also see, that \eqref{eq:psirpsirel} employs only breved polynomials $\cir{\breve{A}}^{(\ls)}$ which contain only constants $\cir{x}_\mus$ corresponding to the degenerate sector. Any dependence on $\cir{x}_\mur$ from the regular sector has disappeared. Similarly, the relation  between $\psi_k$ and $\ph_\mus$ is also independent of $\cir{x}_\mur$ in the leading term, see  \eqref{eq:psiphlord}.

%%%%%%%%%%%%%%%%%%%%%%%%%%%%%%%%%%%%%%%%%%%%%%%%%%%%%%%%%%%%%%%%%%%%%%%%%%%%%%%%%%%%%%
%% Limit of the metric

\subsection{Limit of metric}
\label{ssc:LimitOfTheMetric}
After describing transformation from $x_\mus$ to $\xi_\mus$ and from $\psi_k$ to  $\ph_\mus,\,\psir_\mur$, we can write down the leading terms of $\eps$ expansions of various metric functions and terms of the metric. Substituting expansion \eqref{eq:xsclexp} into definitions \eqref{eq:J}, \eqref{eq:Jmu} and \eqref{eq:Umu} we obtain
\begin{equation}\label{eq:Umusrexp}
   U_\mus \approx \hat{\breve{U}}_\mus \invbreve{J}(\hat{x}_\mus^2)
     \;,\quad
   U_\mur \approx \hat{\breve{J}}(x_\mur^2) \invbreve{U}_\mur
     \;,
\end{equation}
%\begin{align}
%   U_\mus &\approx \hat{\breve{U}}_\mus \invbreve{J}(\hat{x}_\mus^2)
%     \;,&
%   U_\mur &\approx \hat{\breve{J}}(x_\mur^2) \invbreve{U}_\mur
%     \;,\label{eq:Umusrexp}\\
%   \cir{U}_\mus &\approx \hat{\breve{U}}_\mus \cir{\invbreve{J}}(\hat{x}_\mus^2)
%     \;,&
%   \cir{U}_\mur &\approx \hat{\breve{J}}(\cir{x}_\mur^2) \cir{\invbreve{U}}_\mur
%     \;,\label{eq:cUmusrexp}
%\end{align}
\begin{equation}\label{eq:Jmunuexp}
\begin{split}
   J_\mus(\cir{x}_\mus^2) &\approx \hat{\breve{U}}_\mus \invbreve{J}(\hat{x}_\mus^2)
     \;,\\
   J_\mus(\cir{x}_\nus^2) &\approx \frac{2\hat{x}_\nus\delta x_\nus}{\hat{x}_\mus^2-\hat{x}_\nus^2}
     (\xi_\nus-\cir{\xi}_\nus) \hat{\breve{U}}_\nus \invbreve{J}(\hat{x}_\nus^2)\,\eps
     \;,\;\; \mus\neq\nus\;,\mspace{-80mu}\\
   J_\mus(\cir{x}_\nur^2) &\approx \hat{\breve{J}}_\mus(\cir{x}_\nur^2) \invbreve{J}(\cir{x}_\nur^2)
     \;,\\
   J_\mur(\cir{x}_\nus^2) &\approx 2\hat{x}_\nus\delta x_\nus
     (\xi_\nus-\cir{\xi}_\nus) \hat{\breve{U}}_\nus \invbreve{J}_\mur(\hat{x}_\nus^2)\,\eps
     \;,\\
   J_\mur(\cir{x}_\nur^2) &\approx \hat{\breve{J}}(\cir{x}_\nur^2) \invbreve{J}_\mur(\cir{x}_\nur^2)
     \;,
\end{split}\raisetag{15ex}
\end{equation}
%\begin{align}
%   J_\mus(\cir{x}_\mus^2) &\approx \hat{\breve{U}}_\mus \invbreve{J}(\hat{x}_\mus^2)
%     \;,\label{eq:Jmusmusexp}\\
%   J_\mus(\cir{x}_\nus^2) &\approx \frac{2\hat{x}_\nus\delta x_\nus}{\hat{x}_\mus^2-\hat{x}_\nus^2}
%     (\xi_\nus-\cir{\xi}_\nus) \hat{\breve{U}}_\nus \invbreve{J}(\hat{x}_\nus^2)\,\eps
%     \;,\quad \mus\neq\nus\;,\label{eq:Jmusnusexp}\\
%   J_\mus(\cir{x}_\nur^2) &\approx \hat{\breve{J}}_\mus(\cir{x}_\nur^2) \invbreve{J}(\cir{x}_\nur^2)
%     \;,\label{eq:Jmusnurexp}\\
%   J_\mur(\cir{x}_\nus^2) &\approx 2\hat{x}_\nus\delta x_\nus
%     (\xi_\nus-\cir{\xi}_\nus) \hat{\breve{U}}_\nus \invbreve{J}_\mur(\hat{x}_\nus^2)\,\eps
%     \;,\label{eq:Jmurnusexp}\\
%   J_\mur(\cir{x}_\nur^2) &\approx \hat{\breve{J}}(\cir{x}_\nur^2) \invbreve{J}_\mur(\cir{x}_\nur^2)
%     \;,\label{eq:Jmurnurexp}
%\end{align}
where we denote the equality in leading order by $\approx$.

Using these formulas we can expand angular terms in the metric for degenerate and regular directions:
\begin{equation}
\begin{split}
   &\sum_k A^{(k)}_\mus \,\tb{\dd}\psi_k
     = \sum_\nu\frac{J_\mus(\cir{x}_\nu^2)}{\cir{U}_\nu}\,\tb{\dd}\phi_\nu
     \approx -\frac1\eps\,\invbreve{J}(\hat{x}_\mus^2)\,\tb{\dd}\ph_\mus
     \;,
     \\
     &\sum_k A^{(k)}_\mur \,\tb{\dd}\psi_k
     = \sum_\nu\frac{J_\mur(\cir{x}_\nu^2)}{\cir{U}_\nu}\,\tb{\dd}\phi_\nu
     \approx \sum_\kr\invbreve{A}^{(\kr)}_\mur\,\tb{\dd}\psir_\kr´	
     \\
   &\quad-\sum_\nus 2\hat{x}_\nus\delta x_\nus(\xi_\nus{-}\cir{\xi}_\nus)
       \invbreve{J}_\mur(\hat{x}_\nus^2)\,\tb{\dd}\ph_\nus
     \;,
\end{split}\raisetag{5ex}
\end{equation}
It is important that terms associated with degenerate directions are proportional to $1/\eps$ since it cancels \mbox{$\eps^2$ behavior} of metric functions $X_\mus$.

Indeed, taking into account expansions \eqref{eq:Xsimple} of $X_\mus$, \eqref{eq:Umusrexp} and just derived expressions, the terms in the metric corresponding to degenerate directions become
\begin{equation}\label{eq:xdegterms}
%    \frac{U_\mus}{X_\mus} \tb{\dd}x_\mus^{\bs 2} \approx
%    \frac1{(2N{-}1)\lambda\hat{\invbreve{J}}(\hat{x}_\mus^2)}\invbreve{J}(\hat{x}_\mus^2)
%    \,\frac1{1-\xi_\mus^2}\,\tb{\dd}\xi_\mus^{\bs 2}\;,
    \frac{U_\mus}{X_\mus} \tb{\dd}x_\mus^{\bs 2} \approx
    \frac{\invbreve{J}(\hat{x}_\mus^2)}{(2N{-}1)\lambda\hat{\invbreve{J}}(\hat{x}_\mus^2)}
    \frac{\tb{\dd}\xi_\mus^{\bs 2}}{1-\xi_\mus^2}\;,
\end{equation}
\begin{equation}\label{eq:adegterms}
\begin{split}
    &\frac{X_\mus}{U_\mus} \Bigl(\sum_k A^{(k)}_\mus \tb{\dd}\psi_k\Bigr)^{\bs 2} \approx\\
    &\quad\approx
    \bigl[(2N{-}1)\lambda\hat{\invbreve{J}}(\hat{x}_\mus^2)\delta x_\mus^2\bigr]
    \invbreve{J}(\hat{x}_\mus^2)
    \,(1-\xi_\mus^2)\,\tb{\dd}\ph_\mus^{\bs 2}\;.
\end{split}
\end{equation}

Now we can use freedom in specifying limiting parameters $\delta x_\mus$. The resulting metric simplifies if we choose
\begin{equation}\label{eq:deltaxmus}
    \delta x_\mus = \frac{1}{(2N{-}1)|\lambda\hat{\invbreve{J}}(\hat{x}_\mus^2)|}\;.
\end{equation}
With such a choice, the limiting metric reads
\begin{equation}\label{eq:limmet}
\begin{split}
    \tb{g} &= \sum_\mus \delta x_\mus |\invbreve{J}(\hat{x}_\mus^2)|\biggl(
%     \frac1{1-\xi_\mus^2}\,\tb{\dd}\xi_\mus^{\bs 2}
      \frac{\tb{\dd}\xi_\mus^{\bs 2}}{1-\xi_\mus^2}
      +(1-\xi_\mus^2)\,\tb{\dd}\ph_\mus^{\bs 2}\biggr)\\
    &\feq+\sum_\mur\biggl[
    \frac{\hat{\breve{J}}(x_\mur^2)\invbreve{U}_\mur}{X_\mur}\tb{\dd}x_\mur^{\bs 2}
    +\frac{X_\mur}{\hat{\breve{J}}(x_\mur^2)\invbreve{U}_\mur}\\
    &\feq\times\Bigl(\sum_\kr\invbreve{A}^{(\kr)}_\mur\,\tb{\dd}\psir_\kr
    -\sum_\nus 2 \hat{x}_\nus\delta x_\nus
       \invbreve{J}_\mur(\hat{x}_\nus^2)
       (\xi_\nus{-}\cir{\xi}_\nus)\,\tb{\dd}\ph_\nus
       \Bigr)^{\!\bs 2}\biggr]\;,
\end{split}\raisetag{14ex}
\end{equation}
where we used the fact that the sign of $\lambda\hat{\invbreve{J}}(\hat{x}_\mus^2)$ corresponds to the sign of $\invbreve{J}(\hat{x}_\mus^2)$ for our choice (NUT-like or Kerr-like).

We can observe that terms $(1{-}\xi_\mus^2)^{-1}\tb{\dd}\xi_\mus^{\bs 2}+(1{-}\xi_\mus^2)\tb{\dd}\ph_\mus^{\bs 2}$ describe homogeneous metrics on two-spheres. Since their prefactors $\invbreve{J}(\hat{x}_\mus^2)$ depend only on regular coordinates $x_\nur$, the limiting metric looks almost as a multiply warped geometry with the base metric given by the second sum in \eqref{eq:limmet} and the seed metrics given by two-spheres. The base metric has a similar structure as the original Kerr--NUT--(A)dS metric, however, it contains an additional nontrivial coupling to the seed sectors (terms with $\tb{\dd}\ph_\nus$) which spoils a simple warped structure.

Alternatively, we can write the metric \eqref{eq:limmet} in terms of the orthonormal frame of one-forms (cf. \eqref{eq:metricON}, \eqref{eq:ON1formsph}), which naturally remains finite in the limit. It takes the form
\begin{equation}\label{eq:ON1formslimit}
\begin{aligned}
	\tb{e}^\mus &= \bigg(\frac{1-\xi_\mus^2}{\delta x_\mus |\invbreve{J}(\hat{x}_\mus^2)|}\bigg)^{\!\!-\frac{1}{2}}\tb{\dd}\xi_\mus\;,
	\quad
	\tb{e}^\mur = \bigg(\frac{X_\mur}{\hat{\breve{J}}(x_\mur^2)\invbreve{U}_\mur}\bigg)^{\!\!-\frac{1}{2}}\tb{\dd}x_\mur\;,
	\\
	\tb{\hat{e}}^\mus &= -\frac{1}{\sigma_{\mus}}\Big(\delta x_\mus |\invbreve{J}(\hat{x}_\mus^2)|(1-\xi_\mus^2)\Big)^{\!\!\frac{1}{2}}\tb{\dd}\ph_\mus\;,
	\\
	\tb{\hat{e}}^\mur &= \bigg(\frac{X_\mur}{\hat{\breve{J}}(x_\mur^2)\invbreve{U}_\mur}\bigg)^{\!\!\frac{1}{2}}\Bigl(\sum_\kr\invbreve{A}^{(\kr)}_\mur\,\tb{\dd}\psir_\kr
	\\
    &\feq-\sum_\nus 2 \hat{x}_\nus\delta x_\nus\invbreve{J}_\mur(\hat{x}_\nus^2)(\xi_\nus{-}\cir{\xi}_\nus)\,\tb{\dd}\ph_\nus\Bigr)\;,
\end{aligned}
\end{equation}
where we denoted the sign of $\lambda\hat{\invbreve{J}}(\hat{x}_\mus^2)$ [or $\invbreve{J}(\hat{x}_\mus^2)$] by $\sigma_{\mus}$. It is easy to verify that the corresponding dual frame of vectors is
\begin{equation}\label{eq:ONvectorslimit}
\begin{aligned}
	\tb{e}_\mus &= \bigg(\frac{1-\xi_\mus^2}{\delta x_\mus |\invbreve{J}(\hat{x}_\mus^2)|}\bigg)^{\!\!\frac{1}{2}}\frac{\tb{\pp}}{\tb{\pp} \xi_\mus}\;,
	\quad
	\tb{e}_\mur = \bigg(\frac{X_\mur}{\hat{\breve{J}}(x_\mur^2)\invbreve{U}_\mur}\bigg)^{\!\!\frac{1}{2}}\frac{\tb{\pp}}{\tb{\pp} x_\mur}\;,
	\\
	\tb{\hat{e}}_\mus &= -\sigma_{\mus}\Big(\delta x_\mus |\invbreve{J}(\hat{x}_\mus^2)|(1-\xi_\mus^2)\Big)^{\!\!-\frac{1}{2}}
	\\
	&\feq\times\bigg(\frac{\tb{\pp}}{\tb{\pp} \ph_\mus}+2 \hat{x}_\mus\delta x_\mus(\xi_\mus{-}\cir{\xi}_\mus)\sum_\kr{(-\hat{x}_\mus^2)}^{\Nr-\kr-1}\frac{\tb{\pp}}{\tb{\pp} \psir_\kr}\bigg)\;,
	\\
	\tb{\hat{e}}_\mur &= \bigg(\frac{X_\mur}{\hat{\breve{J}}(x_\mur^2)\invbreve{U}_\mur}\bigg)^{\!\!-\frac{1}{2}}\sum_\kr\frac{{(-x_\mur^2)}^{\Nr-\kr-1}}{\invbreve{U}_\mur}\frac{\tb{\pp}}{\tb{\pp} \psir_\kr}\;,
\end{aligned}
\end{equation}
but it can also be calculated directly from \eqref{eq:ONvectorsph} by employing the relations analogous to \eqref{eq:Umusrexp}, \eqref{eq:Jmunuexp} with all ${x_\mu}$ and ${\cir{x}_\mu}$ swapped.

%%%%%%%%%%%%%%%%%%%%%%%%%%%%%%%%%%%%%%%%%%%%%%%%%%%%%%%%%%%%%%%%%%%%%%%%%%%%%%%%%%%%%%
%% Symmetry enhancement
\subsection{Symmetry enhancement}
\label{ssc:Symmetryenhancement}

Starting with the Killing vectors \eqref{eq:newKV}, we introduce another set of Killing vectors in directions of coordinates~$\ph_\mus$ and $\invbreve{\psi}_\kr$, which are finite in the limit ${\eps\rightarrow 0}$,
\begin{equation}\label{eq:KVlim}
\breve{\tb{r}}_{(\mus)} = -\frac{\cir{\invbreve{J}}(\hat{x}_\mus^2)}{\eps}\cir{\tb{r}}_{(\mus)}=\frac{\tb{\pp}}{\tb{\pp} \ph_\mus}\;,
\quad
\invbreve{\tb{l}}_{(\kr)}= \sum_\mur\cir{\invbreve{A}}_{\mur}^{(\kr)}\cir{\tb{r}}_{(\mur)}=\frac{\tb{\pp}}{\tb{\pp} \invbreve{\psi}_\kr}\;.
\end{equation}
Because the metric \eqref{eq:limmet} is considerably simplified in the degenerate directions, it is not surprising that the symmetry of this spacetime has been enhanced. Indeed, it can be verified with the use of the connection forms \eqref{eq:connectionformslim} that the spacetime possesses additional independent Killing vectors,

\begin{equation}
\begin{split}
\breve{\tb{r}}_{(\mus)}^{(1)} &=\sqrt{1-\xi_\mus^2}\cos{\ph_\mus} \frac{\tb{\pp}}{\tb{\pp} \xi_\mus}
+\frac{\sin{\ph_\mus}}{\sqrt{1-\xi_\mus^2}}\bigg(\xi_\mus\frac{\tb{\pp}}{\tb{\pp} \ph_\mus}
\\
&\feq+2 \hat{x}_\mus\delta x_\mus(1-\cir{\xi}_\mus\xi_\mus)\sum_\kr{(-\hat{x}_\mus^2)}^{\Nr-\kr-1}\frac{\tb{\pp}}{\tb{\pp} \psir_\kr}\bigg)\;,
\\
\breve{\tb{r}}_{(\mus)}^{(2)} &=-\sqrt{1-\xi_\mus^2}\sin{\ph_\mus} \frac{\tb{\pp}}{\tb{\pp} \xi_\mus}
+\frac{\cos{\ph_\mus}}{\sqrt{1-\xi_\mus^2}}\bigg(\xi_\mus\frac{\tb{\pp}}{\tb{\pp} \ph_\mus}
\\
&\feq+2 \hat{x}_\mus\delta x_\mus(1-\cir{\xi}_\mus\xi_\mus)\sum_\kr{(-\hat{x}_\mus^2)}^{\Nr-\kr-1}\frac{\tb{\pp}}{\tb{\pp} \psir_\kr}\bigg)\;,
\end{split}
\end{equation}
which, together with the Killing vectors $\breve{\tb{r}}_{(\mus)}^{(0)}$ (a linear combinations of $\breve{\tb{r}}_{(\mus)}$ and $\invbreve{\tb{l}}_{(\kr)}$),
\begin{equation}
\breve{\tb{r}}_{(\mus)}^{(0)}=\breve{\tb{r}}_{(\mus)} - 2 \hat{x}_\mus\delta x_\mus\cir{\xi}_\mus\sum_\kr{(-\hat{x}_\mus^2)}^{\Nr-\kr-1}\invbreve{\tb{l}}_{(\kr)}\;,
\end{equation}
generate the algebra of $\mathrm{SO}(3)$ group,
\begin{equation}
\begin{gathered}
\lieb{\tb{r}_{(\mus)}^{(1)}}{\tb{r}_{(\mus)}^{(2)}}=\tb{r}_{(\mus)}^{(0)}\;,
\quad
\lieb{\tb{r}_{(\mus)}^{(2)}}{\tb{r}_{(\mus)}^{(0)}}=\tb{r}_{(\mus)}^{(1)}\;,
\\
\lieb{\tb{r}_{(\mus)}^{(0)}}{\tb{r}_{(\mus)}^{(1)}}=\tb{r}_{(\mus)}^{(2)}\;,
\end{gathered}
\end{equation}
for each ${\mus}$, which is due to the fact that these vectors are simply the Killing vectors for the two-sphere modified by a few additional terms with $\tb{\partial}_{\invbreve{\psi}_{\invbreve{k}}}$ vectors. All the other Lie brackets of the Killing vectors vanish. In total, we have a complete set of ${3\Ns+\Nr}$ Killing vectors:
\begin{equation}\label{eq:newKVlimit}
\breve{\tb{r}}_{(\mus)}^{(0)}\;,
\;\;
\breve{\tb{r}}_{(\mus)}^{(1)}\;,
\;\;
\breve{\tb{r}}_{(\mus)}^{(2)}\;,
\;\;
\invbreve{\tb{l}}_{(\kr)}\;,
\end{equation}
which describe the explicit symmetries of the spacetime. Thus, we can conclude that the symmetry group of the spacetime is enhanced from ${\mathrm{U}(1)}$ to ${\mathrm{SO}(3)}$ for each degenerate direction.

Also, the rank-two Killing tensors \eqref{eq:newKT} of hidden symmetries corresponding to degenerate directions simplify (see \eqref{eq:Umusrexp}, \eqref{eq:Jmunuexp}, and \eqref{eq:ONvectorslimit}),
\begin{equation}
\begin{split}
 \cir{\tb{q}}_{(\mus)} &= \frac{1}{\sigma_{\mus}\delta x_\mus \cir{\invbreve{J}}(\hat{x}_\mus^2)}\biggl[(1-\xi_\mus^2)
      \frac{\tb{\pp}^{\bs 2}}{\tb{\pp} \xi_\mus}
      +\frac{1}{1-\xi_\mus^2}\bigg(\frac{\tb{\pp}}{\tb{\pp} \ph_\mus}
      \\
      &\feq+2 \hat{x}_\mus\delta x_\mus(\xi_\mus{-}\cir{\xi}_\mus)\sum_\kr{(-\hat{x}_\mus^2)}^{\Nr-\kr-1}\frac{\tb{\pp}}{\tb{\pp} \psir_\kr}\bigg)^{\!\bs 2}\biggr]\;.
\end{split}
\end{equation}
Furthermore, they can be expressed as a linear combination of the Killing vectors \eqref{eq:newKVlimit} with constant coefficients,
\begin{equation}
\begin{split}
 \cir{\tb{q}}_{(\mus)} &= \frac{1}{\sigma_{\mus}\delta x_\mus \cir{\invbreve{J}}(\hat{x}_\mus^2)}\biggl[\Big(\breve{\tb{r}}_{(\mus)}^{(0)}\Big)^{\!\bs 2}+\Big(\breve{\tb{r}}_{(\mus)}^{(1)}\Big)^{\!\bs 2}+\Big(\breve{\tb{r}}_{(\mus)}^{(2)}\Big)^{\!\bs 2}
 \\
 &\feq-4\hat{x}_\mus^2\delta x_\mus^2\bigg(\sum_\kr{(-\hat{x}_\mus^2)}^{\Nr-\kr-1}\invbreve{\tb{l}}_{(\kr)}\bigg)^{\!\bs 2}\biggr]\;,
\end{split}
\end{equation}
thus, unlike the Killing tensors of regular directions $\cir{\tb{q}}_{(\mur)}$, they are not independent.

%%%%%%%%%%%%%%%%%%%%%%%%%%%%%%%%%%%%%%%%%%%%%%%%%%%%%%%%%%%%%%%%%%%%%%%%%%%%%%%%%%%%%%
%% NUT-LIKE LIMITS

\section{NUT-like limits}
\label{sc:NUTlikeLimits}

We now turn to the particular examples of the limiting metric \eqref{eq:limmet} from the previous section. First, we consider the NUT-like choice of parameters and ranges of coordinates. The polynomial ${\mathcal{J}}$ can be parametrized by real tangent points $\hat{x}_{\bar{\mu}}$, ${\bar{\mu}=1,\,\dots,\,\bar{N}}$, and (for ${\lambda\neq0}$) either by imaginary or real tangent point $\hat{x}_{N}$, see Table~\ref{tab:tangentpoints}. This additional tangent point, however, can be expressed in terms of~$\hat{x}_{\bar{\mu}}$, cf. \eqref{eq:constraintxhatexpl}.

%%%%%%%%%%%%%%%%%%%%%%%%%%%%%%%%%%%%%%%%%%%%%%%%%%%%%%%%%%%%%%%%%%%%%%%%%%%%%%%%%%%%%%
%% NUT-like limit in one direction

\subsection{NUT-like limit in one direction}
\label{ssc:NUTlimitinonedir}

First, let us study the case where the limit is taken in one direction, ${\Ns=1}$. Then the metric \eqref{eq:limmet} reads\footnote{A combination of accents inverse breve and bar on a quantity, for instance $\invbreve{\bar{A}}^{(\invbreve{\bar{k}})}$, indicates that such a quantity is constructed out of all $x_\mu$ except $x_N$ and $x$, which is a coordinate of the degenerate direction.}
%\begin{equation}
%\begin{split}
%  \tb{g} &= -\frac{\Delta}{(r^2+\hat{x}^2)\invbreve{\bar{J}}(-r^2)}\biggl(\tb{\dd}\tau
%  +\sum_{\invbreve{\bar{k}}}\invbreve{\bar{A}}^{(\invbreve{\bar{k}})}\,\tb{\dd}\invbreve{\chi}_{\invbreve{\bar{k}}}
%  -\frac{2\hat{x}}{\delta}(\xi-\cir{\xi})\,\tb{\dd}\ph\bigg)^{\!\!\bs{2}}
%  \\
%  &\feq+\frac{(r^2+\hat{x}^2)\invbreve{\bar{J}}(-r^2)}{\Delta}\,\tb{\dd}r^{\bs{2}}
%  \\
%%  &\feq+\frac{(r^2+\hat{x}^2)\invbreve{\bar{J}}(\hat{x}^2)}{\delta}\bigg(\frac{1}{1-\xi^2}\,\tb{\dd}\xi^{\bs{2}}+(1-\xi^2)\,\tb{\dd} \ph^{\bs{2}}\biggr)
%  &\feq+\frac{(r^2+\hat{x}^2)\invbreve{\bar{J}}(\hat{x}^2)}{\delta}\bigg(\frac{\tb{\dd}\xi^{\bs{2}}}{1-\xi^2}+(1-\xi^2)\,\tb{\dd} \ph^{\bs{2}}\biggr)
%   \\
%  &\feq+\sum_{\invbreve{\bar{\mu}}}\biggl[\frac{(r^2+\hat{x}_{\invbreve{\bar{\mu}}}^2)(\hat{x}^2-x_{\invbreve{\bar{\mu}}}^2)\invbreve{\bar{U}}_{\invbreve{\bar{\mu}}}}{\mathcal{X}_{\invbreve{\bar{\mu}}}}\,\tb{\dd}x_{\invbreve{\bar{\mu}}}^{\bs 2}
%  \\
%  &\feq+\frac{\mathcal{X}_{\invbreve{\bar{\mu}}}}{(r^2+\hat{x}_{\invbreve{\bar{\mu}}}^2)(\hat{x}^2-x_{\invbreve{\bar{\mu}}}^2)\invbreve{\bar{U}}_{\invbreve{\bar{\mu}}}}\biggl(\tb{\dd}\tau
%  +\sum_{\invbreve{\bar{k}}}\big(\invbreve{\bar{A}}_{\invbreve{\bar{\mu}}}^{(\invbreve{\bar{k}}+1)}-r^2\invbreve{\bar{A}}_{\invbreve{\bar{\mu}}}^{(\invbreve{\bar{k}})}\big)\,\tb{\dd}\invbreve{\chi}_{\invbreve{\bar{k}}}
%  \\
%  &\feq+\frac{2\hat{x}}{\delta}(r^2+\hat{x}_{\invbreve{\bar{\mu}}}^2)\invbreve{\bar{J}}_{\invbreve{\bar{\mu}}}(\hat{x}^2)(\xi-\cir{\xi})\,\tb{\dd}\ph\biggr)^{\!\!\bs{2}}\biggr]\;,
%\end{split}
%\end{equation}
\begin{widetext}
\begin{equation}\label{eq:singleNUT}
\begin{split}
  \tb{g} &= -\frac{\Delta}{(r^2+\hat{x}^2)\invbreve{\bar{J}}(-r^2)}\biggl(\tb{\dd}\tau
  +\sum_{\invbreve{\bar{k}}}\invbreve{\bar{A}}^{(\invbreve{\bar{k}}+1)}\,\tb{\dd}\invbreve{\chi}_{\invbreve{\bar{k}}}
  -\frac{2\hat{x}}{\delta}\invbreve{\bar{J}}(\hat{x}^2)(\xi-\cir{\xi})\,\tb{\dd}\ph\bigg)^{\!\!\bs{2}}
+\frac{(r^2+\hat{x}^2)\invbreve{\bar{J}}(-r^2)}{\Delta}\,\tb{\dd}r^{\bs{2}}
  \\
%  &\feq+\frac{(r^2+\hat{x}^2)|\invbreve{\bar{J}}(\hat{x}^2)|}{\delta}\bigg(\frac{1}{1-\xi^2}\,\tb{\dd}\xi^{\bs{2}}+(1-\xi^2)\,\tb{\dd} \ph^{\bs{2}}\biggr)
  &\feq+\frac{(r^2+\hat{x}^2)|\invbreve{\bar{J}}(\hat{x}^2)|}{\delta}\bigg(\frac{\tb{\dd}\xi^{\bs{2}}}{1-\xi^2}+(1-\xi^2)\,\tb{\dd} \ph^{\bs{2}}\biggr)
+\sum_{\invbreve{\bar{\mu}}}\biggl[\frac{(r^2+x_{\invbreve{\bar{\mu}}}^2)(\hat{x}^2-x_{\invbreve{\bar{\mu}}}^2)\invbreve{\bar{U}}_{\invbreve{\bar{\mu}}}}{\mathcal{X}_{\invbreve{\bar{\mu}}}}\,\tb{\dd}x_{\invbreve{\bar{\mu}}}^{\bs 2}
  \\
  &\feq+\frac{\mathcal{X}_{\invbreve{\bar{\mu}}}}{(r^2+x_{\invbreve{\bar{\mu}}}^2)(\hat{x}^2-x_{\invbreve{\bar{\mu}}}^2)\invbreve{\bar{U}}_{\invbreve{\bar{\mu}}}}\biggl(\tb{\dd}\tau
  +\sum_{\invbreve{\bar{k}}}\big(\invbreve{\bar{A}}_{\invbreve{\bar{\mu}}}^{(\invbreve{\bar{k}}+1)}-r^2\invbreve{\bar{A}}_{\invbreve{\bar{\mu}}}^{(\invbreve{\bar{k}})}\big)\,\tb{\dd}\invbreve{\chi}_{\invbreve{\bar{k}}}
+\frac{2\hat{x}}{\delta}(r^2+\hat{x}^2)\invbreve{\bar{J}}_{\invbreve{\bar{\mu}}}(\hat{x}^2)(\xi-\cir{\xi})\,\tb{\dd}\ph\biggr)^{\!\!\bs{2}}\biggr]\;,
\end{split}
\end{equation}
\end{widetext}
where we dropped the index corresponding to the degenerate direction. Moreover, we renamed the Killing coordinates,
\begin{equation}
\tau=\invbreve{\psi}_0\;,
\quad
\invbreve{\chi}_{\invbreve{\bar{k}}}=\invbreve{\psi}_{\invbreve{\bar{k}}+1}\;,
\end{equation}
and introduced the symbol ${\delta= 1/\delta x}$, which can be rewritten with the help of \eqref{eq:constraintxhatexpl}, \eqref{eq:deltaxmus} as
\begin{equation}\label{eq:deltasingleNUT}
%\delta =(2N-1)|\hat{\invbreve{\bar{J}}}(\hat{x}^2)|\left(\frac{\sum_k\frac{\lambda^k \big(\hat{\invbreve{\bar{A}}}^{(k)}+\hat{x}^2\hat{\invbreve{\bar{A}}}^{(k-1)}\big)}{2N-2k-1}}{\sum_k\frac{\lambda^k \big(\hat{\invbreve{\bar{A}}}^{(k)}+\hat{x}^2\hat{\invbreve{\bar{A}}}^{(k-1)}\big)}{2N-2k-3}}+\lambda\hat{x}^2\right)\;,
\delta =(2N-1)|\hat{\invbreve{\bar{J}}}(\hat{x}^2)|\left(\frac{\displaystyle\sum_k\frac{\lambda^k \big(\hat{\invbreve{\bar{A}}}^{(k)}+\hat{x}^2\hat{\invbreve{\bar{A}}}^{(k-1)}\big)}{2N-2k-1}}{\displaystyle\sum_k\frac{\lambda^k \big(\hat{\invbreve{\bar{A}}}^{(k)}+\hat{x}^2\hat{\invbreve{\bar{A}}}^{(k-1)}\big)}{2N-2k-3}}+\lambda\hat{x}^2\right)\;.
\end{equation}
We used the fact that ${\lambda(\hat{x}^2-\hat{x}_N^2)}$ is positive for the NUT-like choice. The functions $\Delta$, $\mathcal{X}_{\invbreve{\bar{\mu}}}$ are given by \eqref{eq:DeltacalX}, with the polynomial $\mathcal{J}(x^2)$ expressed by means of the tangent-point parametrization \eqref{eq:calJhatJintegral} and $\hat{x}_N^2$ replaced by \eqref{eq:constraintxhatexpl}.

After all these substitutions, the metric \eqref{eq:singleNUT} contains the parameters $\hat{x}$, $\hat{x}_{\invbreve{\bar{\mu}}}$, $b_{\invbreve{\bar{\mu}}}$, $m$, $\lambda$, and an additional parameter ${\cir{\xi}\in[-1,1]}$ which can be set to an arbitrary value. Thus, the limiting procedure eliminated only one parameter from the original metric. In four dimensions, the spacetime reduces to the Taub--NUT--(A)dS and the choice ${\cir{\xi}=\pm 1}$ gives the metric which is regular on one of the semiaxes ${\xi=\pm 1}$.

%%%%%%%%%%%%%%%%%%%%%%%%%%%%%%%%%%%%%%%%%%%%%%%%%%%%%%%%%%%%%%%%%%%%%%%%%%%%%%%%%%%%%%
%% Taub--NUT--(A)dS

\subsection{Taub--NUT--(A)dS}
\label{ssc:TaubNUTAdS}
Another important example is when the limit is taken in all directions of the Euclidean sector, ${\Ns=N-1}$. The metric takes the form
\begin{equation}\label{eq:gTaub}
\begin{split}
  \tb{g} &= -\frac{\Delta}{\hat{\bar{J}}(-r^2)}\bigg(\tb{\dd}\tau -\sum_{\bar{\mu}}\frac{2\hat{x}_{\bar{\mu}}}{\delta_{\bar{\mu}}}(\xi_{\bar{\mu}}-\cir{\xi}_{\bar{\mu}})\,\tb{\dd}\ph_{\bar{\mu}}\bigg)^{\!\!\bs{2}}
+ \frac{\hat{\bar{J}}(-r^2)}{\Delta}\,\tb{\dd}r^{\bs{2}}
  \\
%  &\feq + \sum_{\bar{\mu}}\frac{r^2+\hat{x}_{\bar{\mu}}^2}{\delta_{\bar{\mu}}}\bigg(\frac{1}{1-\xi_{\bar{\mu}}^2}\,\tb{\dd}\xi_{\bar{\mu}}^{\bs{2}}+(1-\xi_{\bar{\mu}}^2)\,\tb{\dd} \ph_{\bar{\mu}}^{\bs{2}}\bigg)\;,
  &\feq + \sum_{\bar{\mu}}\frac{r^2+\hat{x}_{\bar{\mu}}^2}{\delta_{\bar{\mu}}}\bigg(\frac{\tb{\dd}\xi_{\bar{\mu}}^{\bs{2}}}{1-\xi_{\bar{\mu}}^2}+(1-\xi_{\bar{\mu}}^2)\,\tb{\dd} \ph_{\bar{\mu}}^{\bs{2}}\bigg)\;,
\end{split}
\end{equation}
where we denoted the temporal coordinate $\invbreve{\psi}_0$ by $\tau$ and introduced ${\delta_{\bar{\mu}}= 1/\delta x_{\bar{\mu}}}$, which reads [cf. \eqref{eq:constraintxhatexpl}, \eqref{eq:deltaxmus}]
\begin{equation}\label{eq:deltabarmu}
%\delta_{\bar{\mu}} =(2N-1)\Bigg(\frac{\sum_k\frac{1}{2N-2k-1}\lambda^k \hat{\bar{A}}^{(k)}}{\sum_k\frac{1}{2N-2k-3}\lambda^k \hat{\bar{A}}^{(k)}}+\lambda\hat{x}_{\bar{\mu}}^2\Bigg)
\delta_{\bar{\mu}} =(2N-1)\left(\frac{\displaystyle\sum_k\frac{\lambda^k \hat{\bar{A}}^{(k)}}{2N-2k-1}}{\displaystyle\sum_k\frac{\lambda^k \hat{\bar{A}}^{(k)}}{2N-2k-3}}+\lambda\hat{x}_{\bar{\mu}}^2\right)\;.
\end{equation}
Again, we employed ${\lambda(\hat{x}_{\bar{\mu}}^2-\hat{x}_N^2)>0}$, which is a consequence of the NUT-like choice. The function~$\Delta$ is given by \eqref{eq:DeltacalX}, \eqref{eq:calJhatJintegral}, where $\hat{x}_N^2$ can be replaced by the expression \eqref{eq:constraintxhatexpl}. Alternatively, it can be rewritten (after some algebraic manipulation) as
\begin{equation}
\Delta =r\!\!\int\limits^{r}\!\big(\delta_1-(2N-1)\lambda(\hat{x}_{1}^2+s^2)\big)
\frac{\hat{\bar{J}}(-s^2)}{s^2}\,\mathrm{d}s-2mr\;.
\end{equation}
The metric \eqref{eq:gTaub} corresponds to the higher-dimensional analogy of the Taub--NUT--(A)dS metric. It agrees with the geometry of ``multiply nutty'' spacetimes which was found in \cite{MannStelea:2006}. The only difference is that the conditions for constants $\delta_{\bar{\mu}}$ from \cite{MannStelea:2006} are solved by the explicit expression \eqref{eq:deltabarmu}. The metric contains the parameters $\hat{x}_{\bar{\mu}}$, $m$, $\lambda$, and the constants ${\cir{\xi}_{\bar{\mu}}\in[-1,1]}$, which can be set arbitrarily.

We can observe that by switching off the parameters~$\hat{x}_{\bar{\mu}}$, the metric takes the form
\begin{equation}\label{eq:gswoffnut}
	\tb{g}=-f\tb{\dd}\tau^{\bs{2}}
+ \frac{1}{f}\,\tb{\dd}r^{\bs{2}}
+ \sum_{\bar{\mu}}\frac{r^2}{2N-3}\bigg(\frac{\tb{\dd}\xi_{\bar{\mu}}^{\bs{2}}}{1-\xi_{\bar{\mu}}^2}+(1-\xi_{\bar{\mu}}^2)\,\tb{\dd} \ph_{\bar{\mu}}^{\bs{2}}\bigg)\;,
\end{equation}
with the metric function
\begin{equation}
	f=1-\lambda r^2-\frac{2m}{r^{2N-3}}\;.
\end{equation}
Although \eqref{eq:gswoffnut} has the same radial dependence in the Lorentzian sector as the Schwarzschild--Tangherlini solution, it differs in the Euclidean sector. In the case of Schwarzschild--Tangherlini, the Euclidean part is given by the homogeneous metric on $2(N-1)$-sphere instead of a sum of two-spheres, but they both reduce to the Schwarzschild metric in four dimensions. Also, \eqref{eq:gswoffnut} for ${m=0}$ is not a maximally symmetric spacetime in higher than four dimensions.

%%%%%%%%%%%%%%%%%%%%%%%%%%%%%%%%%%%%%%%%%%%%%%%%%%%%%%%%%%%%%%%%%%%%%%%%%%%%%%%%%%%%%%
%% NEAR-HORIZON LIMITS

\section{Near-horizon limits}
\label{sc:NHLimits}
The double-root limiting procedure from Sec.~\ref{sc:limitingprocedure} can be generalized to the case where the degenerated direction is in the Lorentzian sector. It means that the radial coordinate $r$ is squeezed to the real double root of the polynomial $\Delta$, which is either $\hat{r}$ or $\hat{r}^{(\lambda)}$. The former exists only for the Kerr-like choice and corresponds to the position where the inner and outer horizons merge, while the latter represents the coinciding outer and cosmological horizons, which is present only if ${\lambda>0}$. We restrict ourselves to $\hat{r}$, which is more physically interesting, however, we will discuss the case $\hat{r}^{(\lambda)}$ in connection with the completely degenerate case in Sec.~\ref{sc:CompleteDegeneration}. For this reason, we focus on the Kerr-like choice first.

%%%%%%%%%%%%%%%%%%%%%%%%%%%%%%%%%%%%%%%%%%%%%%%%%%%%%%%%%%%%%%%%%%%%%%%%%%%%%%%%%%%%%%
%% Limiting procedure in Lorentzian sector

\subsection{Limiting procedure in Lorentzian sector}
\label{ssc:limlor}

In order to study limits in Lorentzian sector, we have to take into account a few very important distinctions that occur here. As was discussed above, the radius $r$ does not have to be restricted between roots of the metric function. Actually, one can set the value of mass to its critical value ${\hat{m}}$, which corresponds to the double root ${\hat{r}}$, without scaling the coordinate $r$ at all, obtaining thus the extreme black hole.

However, it is also interesting to zoom to the neighborhood of the extreme horizon and to study so-called a near-horizon limit of the black hole solution~\cite{BardeenHorowitz:1999}. Indeed, one can rescale the radial coordinate in such a way that the degenerating horizon remains in the zoomed region. Since the range of $r$ is unrestricted, there is a greater freedom to squeeze the radial coordinate and introduce rescaled coordinate near the extreme horizon.

\begin{figure}
    \centering
    \includegraphics[width=\columnwidth]{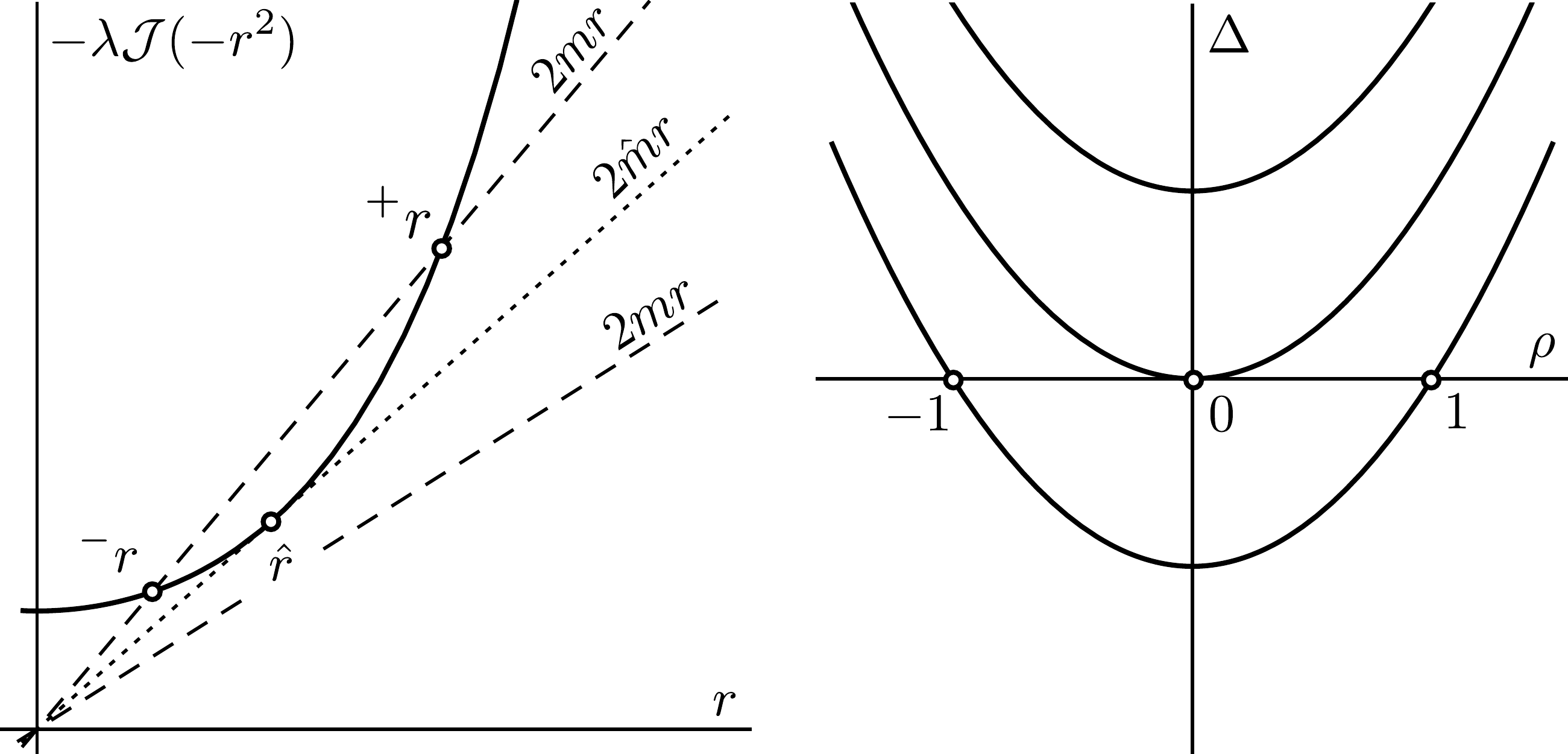}
    \caption{Three distinct approximation procedures of the polynomial $\Delta$ are illustrated in terms of lines passing through the origin on the left. The line approaching the tangent line has either two intersections (two real roots), no intersection (two complex conjugate roots), or is tangent from the very beginning (a double root). This is accompanied by zooming in on the region close to the tangent point $\hat{r}$, where a new coordinate $\rho$ is introduced. The resulting polynomial $\Delta$ is approximated by one of the three quadratic functions in coordinate $\rho$ on the right.}
    \label{fig:approxDelta}
\end{figure}

We study three possible limits leading to the extreme case which differ by a correlation between the coordinate scaling and how we approach the critical value ${\hat{m}}$, see Fig.~\ref{fig:approxDelta}.
In so called \emph{extreme limit}, one simply sets ${m=\hat{m}}$ and scales the radial coordinate independently \cite{BardeenHorowitz:1999}. In \emph{subextreme limit}, one approaches ${m\to\hat{m}}$ in such a way that ${\Delta}$ has two real roots near the critical value ${\hat{r}}$ \cite{Zaslavskii:1998}. Thus, there are two horizons which coincide in the limit, and the coordinate scaling is adjusted to these horizons. Finally, in \emph{superextreme limit}, one approaches ${m\to\hat{m}}$ with no horizons before the limit, see e.g. \cite{Hejda:dipl}. Thus, ${\Delta}$ does not have real roots approaching ${\hat{r}}$. Instead, it has two complex conjugated roots which coincide in real extreme limit and the coordinate scaling is adjusted to them.

Another important difference arises due to the structure of tangent points for the Kerr-like choice, see Table~\ref{tab:tangentpoints}. In this case, the tangent point ${\hat{x}_1=\imag \hat{r}}$ of ${\mathcal{J}(x^2)}$ is imaginary and its imaginary part $\hat{r}$ represents the real tangent point of ${\mathcal{J}(-r^2)}$. Thus the limit ${r\approx\hat{r}}$ is just the Wick rotated limit ${x_N\approx\hat{x}_1}$. It is, however, a limit of a different type than the limits ${x_\mu\approx\hat{x}_\mu}$ which we studied in Sec.~\ref{sc:limitingprocedure}. Fortunately, this distinction requires only small modifications in our formulas.

%%%%%%%%%%%%%%%%%%%%%%%%%%%%%%%%%%%%%%%%%%%%%%%%%%%%%%%%%%%%%%%%%%%%%%%%%%%%%%%%%%%%%%
%% Extreme near-horizon limit

\subsection{Extreme near-horizon limit}
\label{ssc:extremenearhorizon}
A common near-horizon limiting procedure usually begins with the extreme case where mass is set to the critical mass,
\begin{equation}
	m=\hat{m}\;,
\end{equation}
and the horizons ${{}^+r}$, ${{}^-r}$ simply coincide ${{{}^+r}={{}^-r}=\hat{r}}$. Then we zoom in on this extreme horizon by introducing the rescaled coordinate ${\rho\in\mathbb{R}}$,
\begin{equation}
	r=\hat{r} + \delta r\rho\eps+\mathcal{O}(\eps^2)\;,
\end{equation}
where ${\eps\ll1}$ is a small positive parameter and  ${\delta r>0}$ is a factor which will be specified below. The polynomial $\Delta$ is then approximated by a quadratic function in the order $\eps^2$,
\begin{equation}\label{eq:deltaextreme}
	\Delta=-(2N-1)\lambda\hat{J}_1(-\hat{r}^2)\delta r^2 \rho^2 \eps^2 +\mathcal{O}(\eps^3)\;,
\end{equation}
where $\hat{x}_N^2$ is expressible in terms of other constants via \eqref{eq:constraintxhatexpl}.

As before, the transformation of the coordinate $r$ must be accompanied by a proper transformation of the Killing coordinate $\phi_N$,\footnote{The coordinate ${\phi_N}$ can also be expressed in terms of the coordinates \eqref{eq:psicoordlorentz} as ${\phi_N=\mathcal{T}+\sum_{\bar{k}}\cir{\bar{A}}^{(\bar{k}+1)}\chi_{\bar{k}}}$.}
\begin{equation}\label{eq:angtransextr}
\phi_N=\cir{\bar{J}}(-\hat{r}^2)\,t\,\frac{1}{\eps}
\end{equation}
with $t$ being a new temporal coordinate.

We do not assume any particular scaling of the parameter $\cir{x}_N$. We simply set
\begin{equation}
    \cir{r}=\hat{r}\;,
\end{equation}
where $\cir{r}$ denotes the Wick rotated parameter ${\cir{x}_N=i\cir{r}}$.

Employing \eqref{eq:deltaextreme}, the formulas analogous to \eqref{eq:Umusrexp}, \eqref{eq:Jmunuexp}, taking the limit ${\eps\rightarrow0}$, and fixing
\begin{equation}\label{eq:fixdeltaextr}
	\delta r=\frac{1}{-(2N-1)\lambda\hat{J}_1(-\hat{r}^2)}\;,
\end{equation}
we obtain the metric
\begin{equation}\label{eq:gNH}
\begin{split}
	\tb{g} &= \frac{\bar{J}(-\hat{r}^2)}{\delta}\biggl(-\rho^2\,\tb{\dd}t^{\bs{2}}+\frac{\tb{\dd}\rho^{\bs{2}}}{\rho^2}\biggr)
+\sum_{\bar{\mu}}\Bigg[\frac{(\hat{r}^2+x_{\bar{\mu}}^2)\bar{U}_{\bar{\mu}}}{\mathcal{X}_{\bar{\mu}}}\,\tb{\dd}x_{\bar{\mu}}^{\bs{2}}
\\
&\feq+\frac{\mathcal{X}_{\bar{\mu}}}{(\hat{r}^2+x_{\bar{\mu}}^2)\bar{U}_{\bar{\mu}}}\bigg(\sum_{\bar{k}}\bar{A}^{(\bar{k})}_{\bar{\mu}}\,\tb{\dd}\bar{\psi}_{\bar{k}}-\frac{2\hat{r}}{\delta}\bar{J}_{\bar{\mu}}(-\hat{r}^2)\rho\,\tb{\dd}t\bigg)^{\!\!\bs{2}}\Bigg]\;,
\end{split}
\end{equation}
where we introduced the symbol ${\delta=1/\delta r}$ which can be expressed as [cf. \eqref{eq:fixdeltaextr}, \eqref{eq:constraintxhatexpl}]
\begin{equation}\label{eq:deltaextr}
%	\delta=(2N-1)\hat{\bar{J}}_1(-\hat{r}^2)\left(\frac{\sum_k\frac{\lambda^k \big(\hat{\bar{A}}_1^{(k)}-\hat{r}^2\hat{\bar{A}}_1^{(k-1)}\big)}{2N-2k-1}}{\sum_k\frac{\lambda^k \big(\hat{\bar{A}}_1^{(k)}-\hat{r}^2\hat{\bar{A}}_1^{(k-1)}\big)}{2N-2k-3}}-\lambda\hat{r}^2\right)\;.
	\delta=(2N-1)\hat{\bar{J}}_1(-\hat{r}^2)\left(\frac{\displaystyle\sum_k\frac{\lambda^k \big(\hat{\bar{A}}_1^{(k)}-\hat{r}^2\hat{\bar{A}}_1^{(k-1)}\big)}{2N-2k-1}}{\displaystyle\sum_k\frac{\lambda^k \big(\hat{\bar{A}}_1^{(k)}-\hat{r}^2\hat{\bar{A}}_1^{(k-1)}\big)}{2N-2k-3}}-\lambda\hat{r}^2\right)\;.
\end{equation}
The positiveness of $\delta r$ (and $\delta$) follows from the Kerr-like choice in which ${-\lambda(\hat{r}^2+\hat{x}_N^2)}$ is positive.

Functions $\mathcal{X}_{\bar{\mu}}$ are given by \eqref{eq:DeltacalX}, \eqref{eq:calJhatJintegral}, where $\hat{x}_N^2$ is replaced by the expression \eqref{eq:constraintxhatexpl}. The metric \eqref{eq:gNH} represents the near-horizon limit of the extreme Kerr--NUT--(A)dS spacetimes. This result was derived earlier in \cite{LuMeiPope:2009}, where authors used a standard parametrization in terms of $a_\mu$ instead of tangent points $\hat{x}_\mu$. We see that the terms ${-\rho^2\tb{\dd}t^{\bs{2}}+\rho^{-2}\tb{\dd}\rho^{\bs{2}}}$ in degenerated sector describe the metric of the two-dimensional anti-de Sitter spacetime.

The frame of one-forms and the dual frame of vectors [cf. \eqref{eq:ON1formsph} and \eqref{eq:ONvectorsph}] are
\begin{equation}\label{eq:ON1formslimitNH}
\begin{aligned}
	\tb{e}^{\bar\mu} &=  \bigg(\frac{\mathcal{X}_{\bar{\mu}}}{(\hat{r}^2+x_{\bar{\mu}}^2)\bar{U}_{\bar{\mu}}}\bigg)^{\!\!-\frac{1}{2}}\tb{\dd}x_{\bar\mu}\;,
	\;\;
	\tb{e}^N = \imag\bigg(\frac{-\delta}{\bar{J}(-\hat{r}^2)}\rho^2\bigg)^{\!\!-\frac{1}{2}}\tb{\dd}\rho\;,
	\\
	\tb{\hat{e}}^{\bar\mu} &= \bigg(\frac{\mathcal{X}_{\bar{\mu}}}{(\hat{r}^2+x_{\bar{\mu}}^2)\bar{U}_{\bar{\mu}}}\bigg)^{\!\!\frac{1}{2}}\bigg(\sum_{\bar{k}}\bar{A}^{(\bar{k})}_{\bar{\mu}}\,\tb{\dd}\bar{\psi}_{\bar{k}}-\frac{2\hat{r}}{\delta}\bar{J}_{\bar{\mu}}(-\hat{r}^2)\rho\,\tb{\dd}t\bigg)\;,
	\\
	\tb{\hat{e}}^N &=\bigg(\frac{\bar{J}(-\hat{r}^2)}{-\delta}\rho^2\bigg)^{\!\!\frac{1}{2}} \tb{\dd}t\;,
\end{aligned}
\end{equation}
and
\begin{equation}\label{eq:ONvectorslimitNH}
\begin{aligned}
	\tb{e}_{\bar\mu} &= \bigg(\frac{\mathcal{X}_{\bar{\mu}}}{(\hat{r}^2+x_{\bar{\mu}}^2)\bar{U}_{\bar{\mu}}}\bigg)^{\!\!\frac{1}{2}}\frac{\tb{\pp}}{\tb{\pp} x_{\bar\mu}}\;,
	\;\;
	\tb{e}_N = -\imag\bigg(\frac{-\delta}{\bar{J}(-\hat{r}^2)}\rho^2\bigg)^{\!\!\frac{1}{2}}\frac{\tb{\pp}}{\tb{\pp} \rho}\;,
	\\
	\tb{\hat{e}}_{\bar\mu} &= \bigg(\frac{\mathcal{X}_{\bar{\mu}}}{(\hat{r}^2+x_{\bar{\mu}}^2)\bar{U}_{\bar{\mu}}}\bigg)^{\!\!-\frac{1}{2}}\sum_{\bar{k}}\frac{{(-x_{\bar\mu}^2)}^{\bar{N}-\bar{k}-1}}{\bar{U}_{\bar\mu}}\frac{\tb{\pp}}{\tb{\pp} \bar{\psi}_{\bar{k}}}\;,
	\\
	\tb{\hat{e}}_N &= \bigg(\frac{\bar{J}(-\hat{r}^2)}{-\delta}\rho^2\bigg)^{\!\!-\frac{1}{2}}\bigg(\frac{\tb{\pp}}{\tb{\pp}t}+\frac{2}{\delta}\rho\sum_{\bar{k}}\hat{r}^{2(\bar{N}-\bar{k})-1}\frac{\tb{\pp}}{\tb{\pp}\bar{\psi}_{\bar{k}}}\bigg)  \;,
\end{aligned}
\end{equation}
respectively.

%%%%%%%%%%%%%%%%%%%%%%%%%%%%%%%%%%%%%%%%%%%%%%%%%%%%%%%%%%%%%%%%%%%%%%%%%%%%%%%%%%%%%%
%% Subextreme and superextreme limits

\subsection{Subextreme and superextreme limits}
\label{ssc:subsuperextreme}

We can also start the limiting procedure with a general subextreme mass ${m>\hat{m}}$ for which the polynomial $\Delta$ has two real roots ${}^+r$ and ${}^-r$ corresponding to outer and inner horizons respectively. Still, we describe a limit in which the horizons approach the extreme radius ${}^+r,{}^-r\rightarrow\hat{r}$. We scale the radial coordinate $r$ by introducing a new rescaled coordinate ${\rho_{+}\in\mathbb{R}}$,
\begin{equation}\label{eq:rcoorsub}
r=\frac{{}^+r+{}^-r}{2}+\frac{{}^+r-{}^-r}{2}\rho_{+}
\end{equation}
Also, we introduce a small positive parameter ${\eps\ll1}$ which controls how fast the roots are approaching each other (cf. \eqref{eq:rootsdistance}),
\begin{equation}\label{eq:rdifsub}
	{}^+r-{}^-r=2\delta{r}\eps\;.
\end{equation}
Along with the squeezing of the radial coordinate, we assume that the mass decreases to its critical value ${m\rightarrow\hat{m}}$. As in \eqref{eq:xsclexp} and \eqref{eq:NUTexp}, we can derive the expansions of $r$ and $m$ in $\eps$,
\begin{equation}
	r=\hat{r}+\delta r\rho_{+}\eps+\mathcal{O}(\eps^2)\;,
\end{equation}
\begin{equation}
	m=\hat{m}-(2N-1)\lambda\hat{J}_1(-\hat{r}^2)\frac{\delta r^2}{2\hat{r}}\eps^2+\mathcal{O}(\eps^3)\;.
\end{equation}
We find that $\Delta$ is approximated by a quadratic function
\begin{equation}\label{eq:deltasub}
	\Delta=-(2N-1)\lambda\hat{J}_1(-\hat{r}^2)\delta r^2 (\rho_{+}^2-1) \eps^2 +\mathcal{O}(\eps^3)\;,
\end{equation}
which differs from the extreme case \eqref{eq:deltaextreme} only by the term ${(\rho_{+}^2-1)}$ instead of $\rho^2$.

Alternatively, we can start with the superextreme mass ${m<\hat{m}}$, without inner or outer horizons. However, we assume that the roots ${}^+r$ and ${}^-r$ are complex conjugate, but they are still approaching the real extreme value $\hat{r}$. Since the difference of two complex conjugate numbers is imaginary, we appropriately modify the relations \eqref{eq:rdifsub}, \eqref{eq:rcoorsub}, introducing a rescaled coordinate ${\rho_{-}\in\mathbb{R}}$:
\begin{equation}
	{}^+r-{}^-r=2\imag\delta{r}\eps\;,
\end{equation}
\begin{equation}\label{eq:rrhominussup}
	r=\frac{{}^+r+{}^-r}{2}-\imag\frac{{}^+r-{}^-r}{2}\rho_{-}\;.
\end{equation}
The factor ${-\imag}$ in \eqref{eq:rrhominussup} corresponds to the Wick rotation around the point ${({}^+r+{}^-r)/2}$ which projects the roots to the real axis where we introduce the coordinate $\rho_{-}$.

Following the steps before, we find that the coordinate $r$ is expanded in the same way,
\begin{equation}
	r=\hat{r}+\delta r\rho_{-}\eps+\mathcal{O}(\eps^2)\;
\end{equation}
and the expansion of $m$ differs in sign,
\begin{equation}
	m=\hat{m}+(2N-1)\lambda\hat{J}_1(-\hat{r}^2)\frac{\delta r^2}{2\hat{r}}\eps^2+\mathcal{O}(\eps^3)\;.
\end{equation}
Finally, we obtain the approximation of the polynomial~$\Delta$,
\begin{equation}\label{eq:deltasuper}
	\Delta=-(2N-1)\lambda\hat{J}_1(-\hat{r}^2)\delta r^2 (\rho_{-}^2+1) \eps^2 +\mathcal{O}(\eps^3)\;,
\end{equation}
which differs by the term ${(\rho_{-}^2+1)}$.

Applying the same procedure as before [$\delta r$ is chosen according to \eqref{eq:fixdeltaextr}], we end up with the metrics
\begin{equation}\label{eq:gNHpm}
\begin{split}
	\tb{g} &= \frac{\bar{J}(-\hat{r}^2)}{\delta}\biggl(-(\rho_\pm^2\mp 1)\,\tb{\dd}t_\pm^{\bs{2}}+\frac{\tb{\dd}\rho_\pm^{\bs{2}}}{\rho_\pm^2\mp 1}\biggr)
\\
&\feq+\sum_{\bar{\mu}}\Bigg[\frac{(\hat{r}^2+x_{\bar{\mu}}^2)\bar{U}_{\bar{\mu}}}{\mathcal{X}_{\bar{\mu}}}\,\tb{\dd}x_{\bar{\mu}}^{\bs{2}}
\\
&\feq+\frac{\mathcal{X}_{\bar{\mu}}}{(\hat{r}^2+x_{\bar{\mu}}^2)\bar{U}_{\bar{\mu}}}\bigg(\sum_{\bar{k}}\bar{A}^{(\bar{k})}_{\bar{\mu}}\,\tb{\dd}\bar{\psi}_{\pm\bar{k}}-\frac{2\hat{r}}{\delta}\bar{J}_{\bar{\mu}}(-\hat{r}^2)\rho_\pm\,\tb{\dd}t_\pm\bigg)^{\!\!\bs{2}}\Bigg]\;,
\end{split}
\end{equation}
where we distinguished the two cases by adding the subscripts $\pm$ to some coordinates.  The factor $\delta$ is still given by \eqref{eq:deltaextr}. In what follows, we show that both metrics \eqref{eq:gNHpm} and the original one \eqref{eq:gNH} actually represent the same spacetime, but in different coordinate systems. The three systems of coordinates are similar to the ones which are often used in four-dimensional Robinson--Bertotti spacetime, see e.g. \cite{Lapedes:1978}.

Let us start by expressing the $\mathrm{AdS}_2$ in Poincar\'{e} and two other types of static coordinates
\begin{equation}
\tb{g}_{\mathrm{AdS}_2} = -\rho^2\,\tb{\dd}t^{\bs{2}}+\frac{\tb{\dd}\rho^{\bs{2}}}{\rho^2}=-(\rho_\pm^2\mp 1)\,\tb{\dd}t_\pm^{\bs{2}}+\frac{\tb{\dd}\rho_\pm^{\bs{2}}}{\rho_\pm^2\mp 1}\;,
\end{equation}
where the transformations are given by the relations
\begin{equation}\label{eq:trplusminus}
\begin{split}
t_+ &=-\frac{1}{2}\log\bigg|t^2-\frac{1}{\rho^2}\bigg|\;,
\\
\cot{t_-}&=-\frac{1}{2t}\bigg(t^2-\frac{1}{\rho^2}-1\bigg)\;,
\end{split}
\quad
\begin{split}
\rho_+ &=-\rho t\vphantom{\bigg|}\;,
\\
\rho_- &=\frac{\rho}{2}\bigg(t^2-\frac{1}{\rho^2}+1\bigg)\;.
\end{split}
\end{equation}

These are the correct coordinate transformations of the degenerate sector, however, they produce wrong terms in the other parts of the full metric. Fortunately, they can be compensated by an appropriate transformation of the coordinates $\bar{\psi}_{\pm\bar{k}}$,\footnote{Alternatively, \eqref{eq:trphires} can be written in terms of the logarithmic function with the use of the identity: ${\Re\artanh{x}=\frac{1}{2}\log\Big|\frac{x+1}{x-1}\Big|}$.}
\begin{equation}\label{eq:trphires}
\begin{split}
\bar{\psi}_{+\bar{k}}&=\bar{\psi}_{\bar{k}}-\frac{2}{\delta}\hat{r}^{2(\bar{N}-\bar{k})-1}\Re\artanh\rho t\;,
\\
\bar{\psi}_{-\bar{k}}&=\bar{\psi}_{\bar{k}}-\frac{2}{\delta}\hat{r}^{2(\bar{N}-\bar{k})-1}\Re\artanh \frac{2\rho t}{\rho^2 t^2+\rho^2+1}\;.
\end{split}
\end{equation}
Thus, we confirmed that both limits, lead to the extreme near-horizon geometry \eqref{eq:gNH}. Transformations \eqref{eq:trplusminus}, \eqref{eq:trphires} are generalizations of their four-dimensional analogues, which can be found, e.g., in \cite{Hejda:dipl}.

The supplementary transformations \eqref{eq:trphires} can be derived as follows: We start with a general transformation of the form (no changes of $x_{\bar{\mu}}$ are needed)
\begin{equation}\label{eq:trphi}
\bar{\psi}_{\pm\bar{k}}=\bar{\psi}_{\pm\bar{k}}(t,\rho,\bar{\psi}_{\bar{l}})\;.
\end{equation}
By expressing the metric \eqref{eq:gNHpm} in the original coordinates $t$, $\rho$, $\bar{\psi}_{\bar{l}}$ [by means of \eqref{eq:trplusminus}, \eqref{eq:trphi}], and comparing with \eqref{eq:gNH} (the nontrivial part is only the one inside the round bracket in third line), we get the equations
\begin{equation}\label{eq:difeqphi}
\begin{split}
\sum_{\bar{k}}\bar{A}^{(\bar{k})}_{\bar{\mu}}\frac{\pp\bar{\psi}_{\pm\bar{k}}}{\pp t} &=\frac{2\hat{r}}{\delta}\bar{J}_{\bar{\mu}}(-\hat{r}^2) T_{\pm}\;,
\\
\sum_{\bar{k}}\bar{A}^{(\bar{k})}_{\bar{\mu}}\frac{\pp\bar{\psi}_{\pm\bar{k}}}{\pp\rho} &=\frac{2\hat{r}}{\delta}\bar{J}_{\bar{\mu}}(-\hat{r}^2) R_{\pm}\;,
\\
\sum_{\bar{k}}\bar{A}^{(\bar{k})}_{\bar{\mu}}\frac{\pp\bar{\psi}_{\pm\bar{k}}}{\pp\bar{\psi}_{\bar{l}}} &=\bar{A}^{(\bar{l})}_{\bar{\mu}}\;.
\end{split}
\end{equation}
Here, we introduced the shorthands
\begin{equation}
\begin{split}
T_{+} &=\frac{\rho}{\rho^2 t^2-1}\;,
\\
T_{-} &=\frac{\frac{\rho}{2}\big(t^2-\frac{1}{\rho^2}-1\big)}{\frac{\rho^2}{4}\big(t^2-\frac{1}{\rho^2}+1\big)^{\!\!2}+1}\;,
\end{split}
\;
\begin{split}
R_{+} &=\frac{t}{\rho^2 t^2-1}\;,
\\
R_{-} &=\frac{\frac{t}{2}\big(t^2-\frac{1}{\rho^2}+1\big)}{\frac{\rho^2}{4}\big(t^2-\frac{1}{\rho^2}+1\big)^{\!\!2}+1}\;.
\end{split}
\end{equation}
Equations \eqref{eq:difeqphi} can be inverted with the help of \eqref{eq:AmuU1} and integrated out, so we obtain \eqref{eq:trphires}.

The inverse transformations to $\pm$ coordinates are
\begin{equation}\label{eq:invtrp}
\begin{gathered}
	t =\pm\frac{\rho_{+}}{\sqrt{|\rho_{+}^2{-}1|}}\,\mathrm{e}^{-t_{+}}\;,\;\;
	\rho =\mp\sqrt{|\rho_{+}^2{-}1|}\,\mathrm{e}^{t_{+}}\;,
	\\
	 \bar{\psi}_{\bar{k}}=\bar{\psi}_{+\bar{k}}-\frac{2}{\delta}\hat{r}^{2(\bar{N}-\bar{k})-1}\Re\artanh{\rho_{+}}\;,
\end{gathered}
\end{equation}
and
\begin{equation}\label{eq:invtrm}
\begin{gathered}
	t =\frac{\sqrt{\rho_{-}^2{+}1}\sin{t_{-}}}{\rho_{-}+\sqrt{\rho_{-}^2{+}1}\cos{t_{-}}}\;,\;\;
	\rho =\rho_{-}+\sqrt{\rho_{-}^2{+}1}\cos{t_{-}}\;,
	\\
	 \bar{\psi}_{\bar{k}}=\bar{\psi}_{-\bar{k}}+\frac{2}{\delta}\hat{r}^{2(\bar{N}-\bar{k})-1}\Re\artanh\frac{\sin{t_{-}}}{\sqrt{\rho_{-}^2{+}1}+\rho_{-}\cos{t_{-}}}\;.
\end{gathered}
\end{equation}

%%%%%%%%%%%%%%%%%%%%%%%%%%%%%%%%%%%%%%%%%%%%%%%%%%%%%%%%%%%%%%%%%%%%%%%%%%%%%%%%%%%%%%
%% Symmetry enhancement

\subsection{Symmetry enhancement}
\label{ssc:SymmetryenhancementNH}

Since the form of the metric \eqref{eq:gNH} is simplified in the degenerated Lorentzian sector, we can expect the enhanced symmetry as with the Euclidean case in Sec.~\ref{ssc:Symmetryenhancement}. It is actually a well-known fact that the near-horizon limit of the Kerr--NUT--(A)dS spacetime has enhanced symmetry in four dimensions as well as in higher dimensions \cite{Chernyavsky:2014,XuYue:2015}.

From the extreme limiting procedure, we find the Killing vectors associated with the coordinates $t$ and $\bar{\psi}_{\bar{k}}$,
\begin{equation}\label{eq:KVfromexlim}
\frac{\tb{\pp}}{\tb{\pp} t} =\frac{\cir{\bar{J}}(-\hat{r}^2)}{\eps}\cir{\tb{r}}_{(N)}\;,
\quad
\frac{\tb{\pp}}{\tb{\pp} \bar{\psi}_{\bar{k}}}= \sum_{\bar{\mu}}\cir{\bar{A}}_{\bar{\mu}}^{(\bar{k})}\cir{\tb{r}}_{(\bar{\mu})}\;.
\end{equation}
Similarly, the subextreme and superextreme limiting procedures lead to the coordinate systems with the temporal coordinates $t_{+}$ and $t_{-}$. The corresponding Killing vectors together with \eqref{eq:KVfromexlim} form an independent set. Thus, we define
\begin{equation}
    \tb{t}_{(\circ)}=\frac{1}{\sqrt{2}}\frac{\tb{\pp}}{\tb{\pp} t}\;,
    \;\;
    \tb{t}_{(+)}=\frac{\tb{\pp}}{\tb{\pp} t_{+}}\;,
    \;\;
    \tb{t}_{(-)}=\frac{\tb{\pp}}{\tb{\pp} t_{-}}\;,
    \;\;
    \tb{s}_{(\bar{k})}=\frac{\tb{\pp}}{\tb{\pp} \bar{\psi}_{\bar{k}}}\;.
\end{equation}
With the help of the inverse transformations \eqref{eq:invtrp} and \eqref{eq:invtrm}, the vectors $\tb{t}_{(+)}$ and $\tb{t}_{(-)}$ can also be expressed in terms of the coordinates $t$, $\rho$, $\bar{\psi}_{\bar{k}}$, see \eqref{eq:KVtpmx} below.

Although these Killing vectors are independent, their algebra has an unusual form,
\begin{equation}\label{eq:algebralim}
\begin{gathered}
    \lieb{\tb{t}_{(+)}}{\tb{t}_{(-)}}=\sqrt{2}\tb{t}_{(\circ)}-\tb{t}_{(-)}\;,
    \quad
    \lieb{\tb{t}_{(+)}}{\tb{t}_{(\circ)}}=\tb{t}_{(\circ)}\;,
    \\
    \lieb{\tb{t}_{(-)}}{\tb{t}_{(\circ)}}=\frac{1}{\sqrt{2}}\tb{t}_{(+)}\;,
\end{gathered}
\end{equation}
where we omitted the vectors $\tb{s}_{(\bar{k})}$, which Lie-commute with all Killing vectors. Obviously, we can improve \eqref{eq:algebralim} by choosing a different set of Killing vectors, in particular, a different combination of vectors $\tb{t}_{(A)}$.

A form of the Lie brackets is given by the structure constants $C_{AB}{}^C$,
\begin{equation}
    \lieb{\tb{t}_{(A)}}{\tb{t}_{(B)}} = -\sum_{C}C_{AB}{}^{C}\,\tb{t}_{(C)}\;.
\end{equation}
Structure constants naturally define the Killing form
\begin{equation}
    K_{AB}=-\frac{1}{2}\sum_{C,D}C_{AC}{}^D C_{BD}{}^C\;,
\end{equation}
which can be regarded as a metric tensor on the Lie algebra. Thus, the inverse Killing metric is given by the inverse matrix $K^{-1}{}^{AB}$ and it can be lifted to the spacetime manifold,
\begin{equation}
    \tb{K}^{-1}=\sum_{A,B} K^{-1}{}^{AB}\,\tb{t}_{(A)}\tb{t}_{(B)}\;.
\end{equation}

The Killing metric with respect to the frame $\tb{t}_{(+)}$, $\tb{t}_{(-)}$, $\tb{t}_{(\circ)}$ reads\footnote{The symbol $\vee$ denotes the symmetric product of two vectors. For two vectors $\tb{A}$ and $\tb{B}$, it is given by ${\tb{A}\vee\tb{B}=\tb{A}\tb{B}+\tb{B}\tb{A}}$.}
\begin{equation}
\tb{K}^{-1}=-\tb{t}_{(+)}^{\bs 2}+\sqrt{2}\,\tb{t}_{(-)}\vee\tb{t}_{(\circ)}-2\,\tb{t}_{(\circ)}^{\bs 2}\;.
\end{equation}
%\begin{equation}
%K_{\text{lim}} = \left[ \begin{array}{ccc}
%-1 & 0 & 0 \\
%0 & 1 & 1/\sqrt{2} \\
%0 & 1/\sqrt{2} & 0 \end{array} \right]\;.
%\end{equation}
From this we see that the set of vectors is not chosen well and it can be adjusted twofold; to the orthonormal frame or to the null frame with respect to the Killing form.

Let us introduce the Killing vectors
\begin{equation}
    \tb{t}_{(\times)}=\tb{t}_{(-)}-\sqrt{2}\tb{t}_{(\circ)}\;,
    \qquad
    \tb{t}_{(\bullet)}=\sqrt{2}\tb{t}_{(-)}-\tb{t}_{(\circ)}\;.
\end{equation}
Then, the Killing-orthonormal frame corresponds to the choice $\tb{t}_{(+)}$, $\tb{t}_{(\times)}$, $\tb{t}_{(-)}$, which gives the algebra
\begin{equation}\label{eq:algortho}
\begin{gathered}
    \lieb{\tb{t}_{(\times)}}{\tb{t}_{(-)}}=\tb{t}_{(+)}\;,
    \quad
    \lieb{\tb{t}_{(-)}}{\tb{t}_{(+)}}=\tb{t}_{(\times)}\;,
    \\
    \lieb{\tb{t}_{(\times)}}{\tb{t}_{(+)}}=\tb{t}_{(-)}\;.
\end{gathered}
\end{equation}
It can be easily seen that the corresponding Killing form is diagonal in this frame,
\begin{equation}\label{eq:KMortho}
\tb{K}^{-1}=-\tb{t}_{(+)}^{\bs 2}-\tb{t}_{(\times)}^{\bs 2}+\tb{t}_{(-)}^{\bs 2}\;.
\end{equation}
%\begin{equation}
%K_{\text{orth}} = \left[ \begin{array}{ccc}
%-1 & 0 & 0 \\
%0 & -1 & 0 \\
%0 & 0 & +1 \end{array} \right]\;.
%\end{equation}
Another common choice is the Killing-null frame $\tb{t}_{(+)}$, $\tb{t}_{(\bullet)}$, $\tb{t}_{(\circ)}$, which leads to the algebra
\begin{equation}\label{eq:algnull}
\begin{gathered}
    \lieb{\tb{t}_{(\bullet)}}{\tb{t}_{(\circ)}}=\tb{t}_{(+)}\;,
    \quad
    \lieb{\tb{t}_{(\bullet)}}{\tb{t}_{(+)}}=\tb{t}_{(\bullet)}\;,
    \\
    \lieb{\tb{t}_{(+)}}{\tb{t}_{(\circ)}}=\tb{t}_{(\circ)}\;.
\end{gathered}
\end{equation}
With respect to this frame, the Killing form is
\begin{equation}\label{eq:KMnull}
\tb{K}^{-1}=-\tb{t}_{(+)}^{\bs 2}-\tb{t}_{(\bullet)}\vee\tb{t}_{(\circ)}\;.
\end{equation}
%\begin{equation}
%K_{\text{null}} = \left[ \begin{array}{ccc}
%-1 & 0 & 0 \\
%0 & 0 & +1 \\
%0 & +1 & 0 \end{array} \right]\;.
%\end{equation}

\begin{figure*}
    \centering
    \includegraphics[width=\textwidth/\real{5.15}]{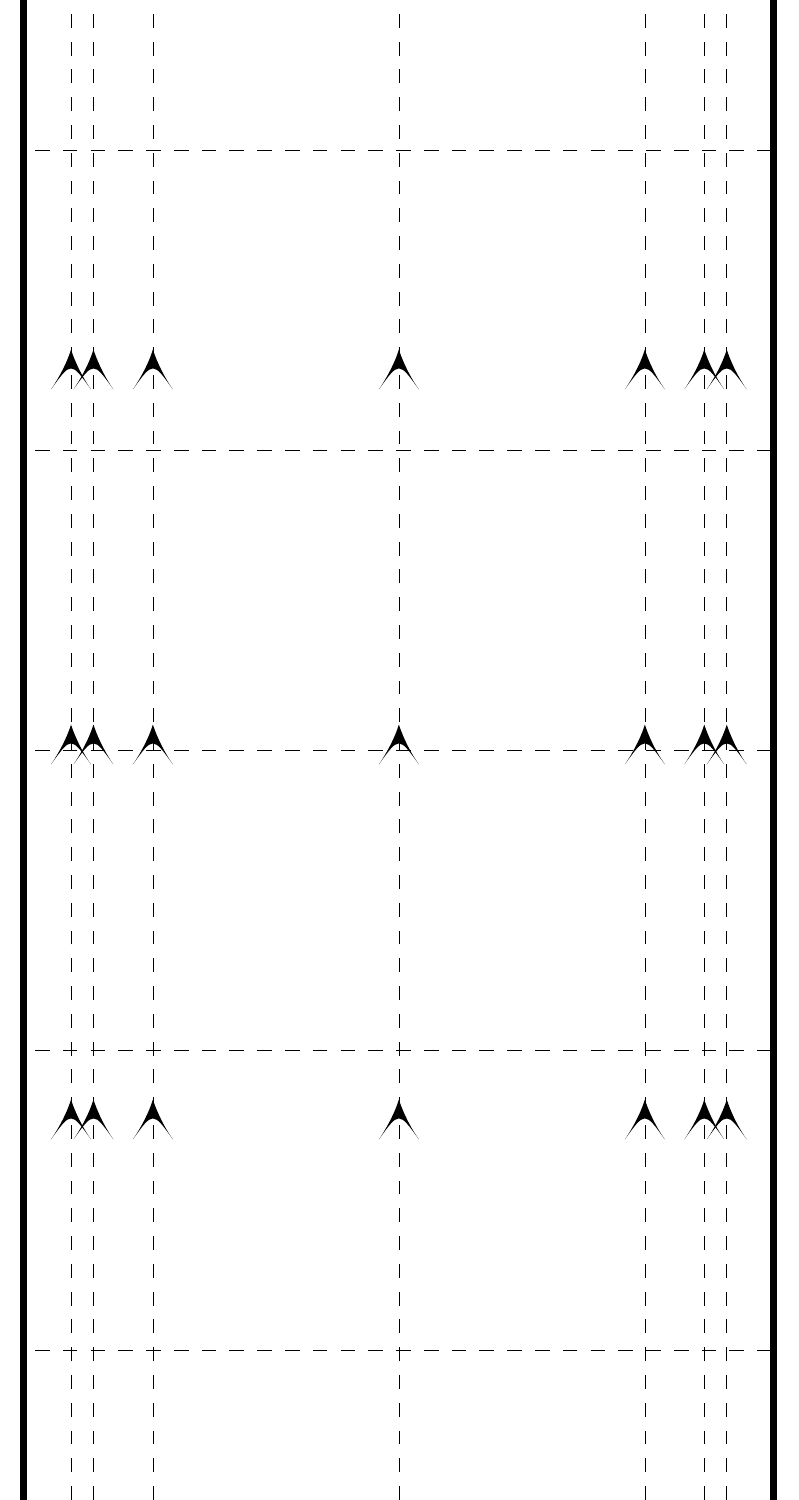}
    \includegraphics[width=\textwidth/\real{5.15}]{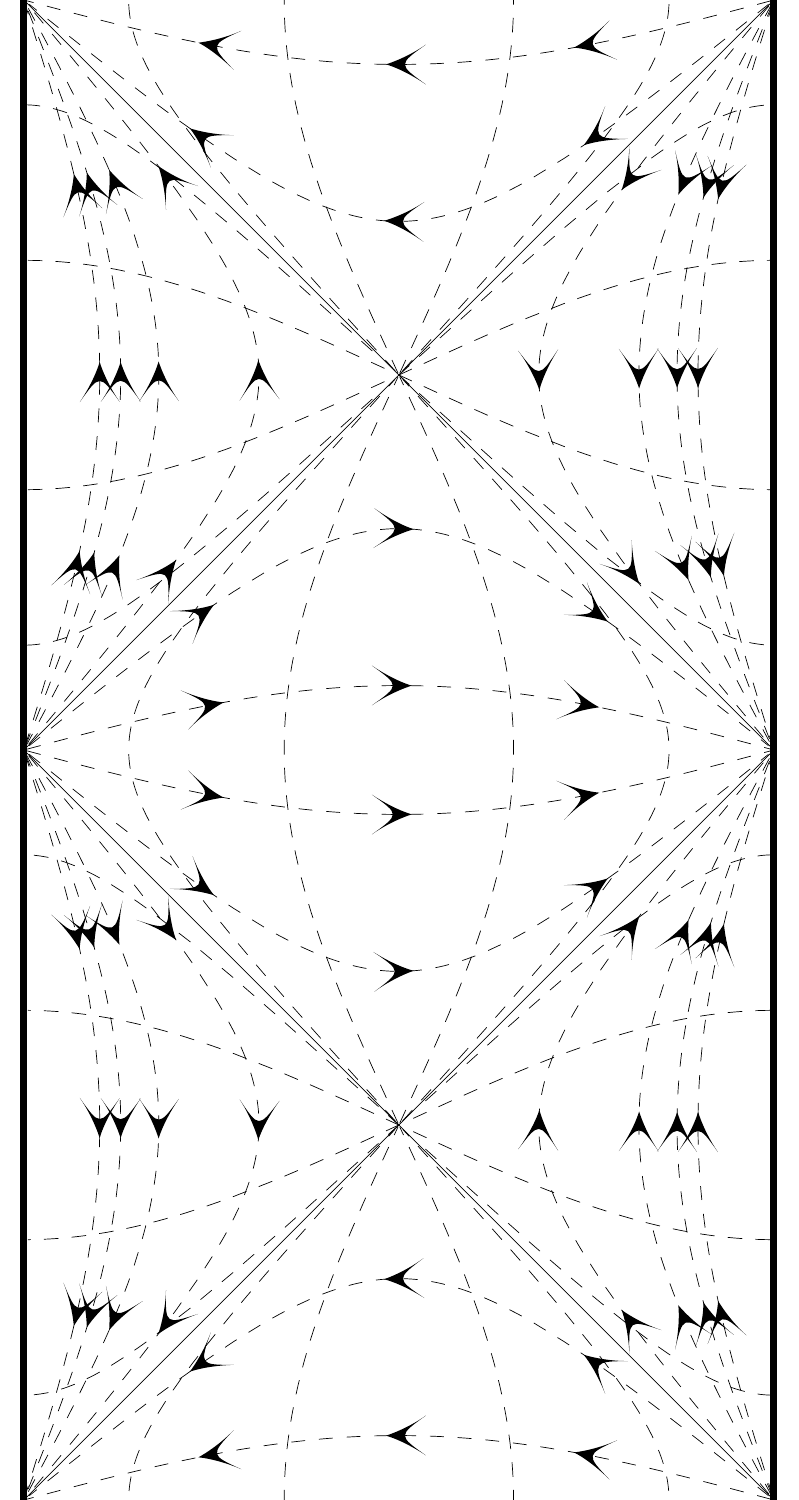}
    \includegraphics[width=\textwidth/\real{5.15}]{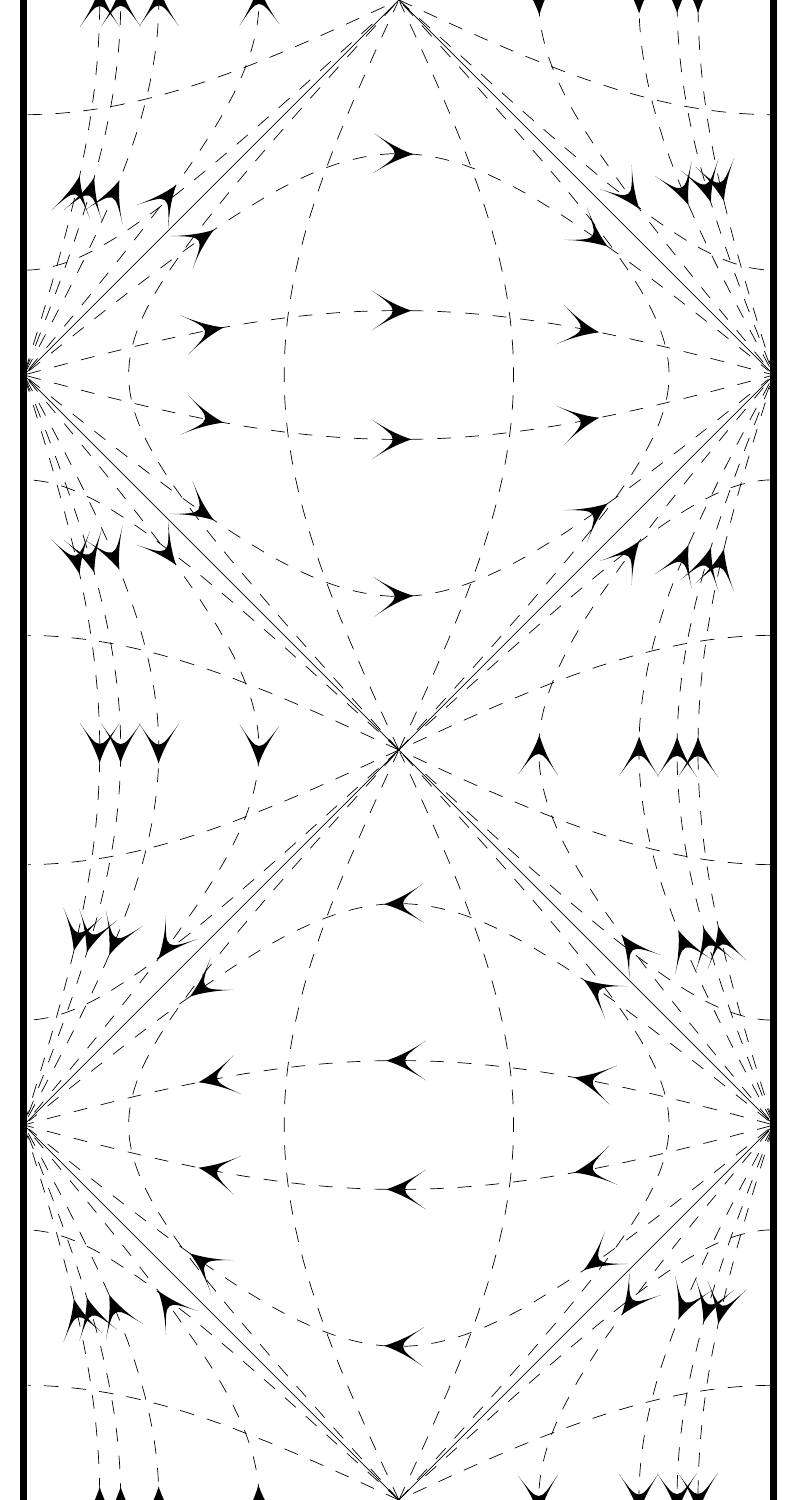}
    \includegraphics[width=\textwidth/\real{5.15}]{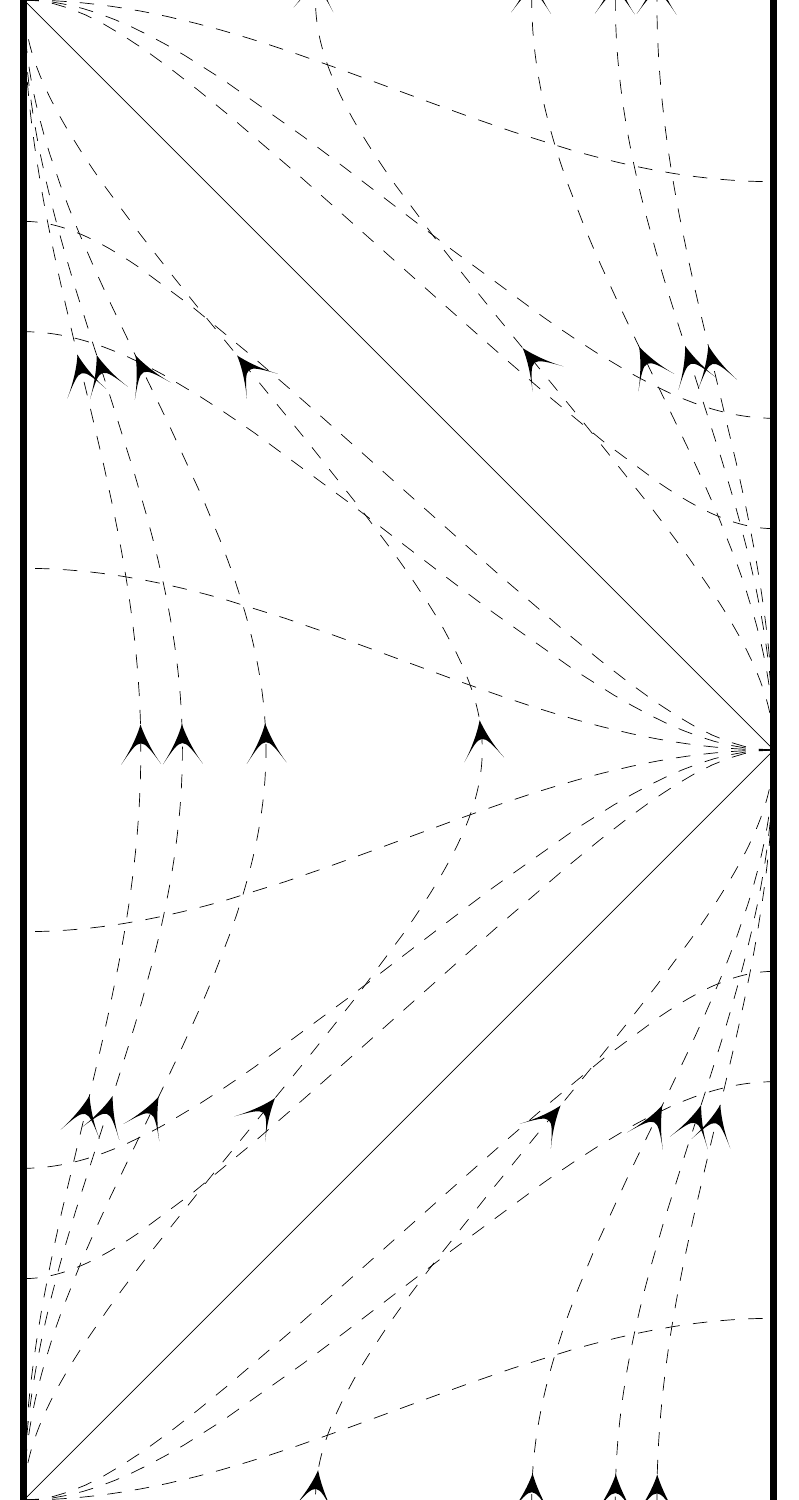}
    \includegraphics[width=\textwidth/\real{5.15}]{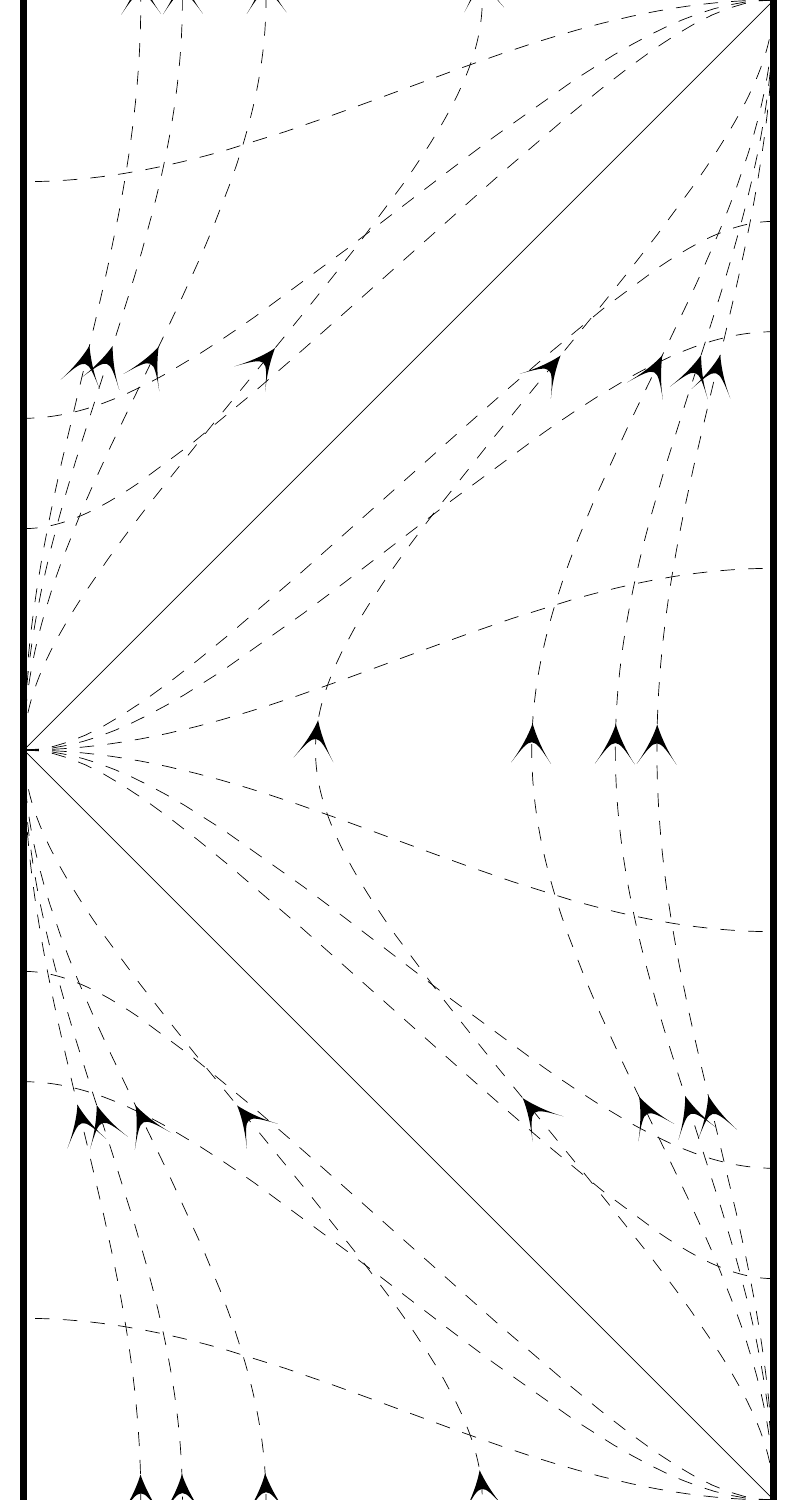}
    \caption{Conformal diagrams of the $\mathrm{AdS}_2$ spacetime with various static coordinate charts. Arrows denotes the Killing vectors $\tb{t}_{(+)}$, $\tb{t}_{(-)}$, $\tb{t}_{(\times)}$, $\tb{t}_{(\circ)}$, $\tb{t}_{(\bullet)}$ (from left to right) projected to the $\mathrm{AdS}_2$ directions. Left and right vertical lines are the infinities of $\mathrm{AdS}_2$, while the diagonal lines denote the Killing horizons of the corresponding $\mathrm{AdS}_2$ Killing vectors. The diagram should continue in vertical directions, since only the spatial (horizontal) directions are compactified.}
    \label{fig:conf}
\end{figure*}

In \eqref{eq:algortho} (or \eqref{eq:algnull}) we recognize the algebra of ${\mathrm{SL(2,\mathbb{R})}\sim\mathrm{SO(1,2)}}$ group. In fact, these Killing vectors are actually the Killing vectors of the two-dimensional anti-de Sitter spacetime with a few additional terms involving $\tb{\partial}_{\bar{\psi}_{\bar{k}}}$ directions. For example, the Killing-orthonormal frame $\tb{t}_{(+)}$, $\tb{t}_{(\times)}$, $\tb{t}_{(-)}$ in the Poincar\'{e} coordinates reads
\begin{equation}\label{eq:KVtpmx}
\begin{aligned}
\tb{t}_{(+)} &=\rho\frac{\tb{\pp}}{\tb{\pp}\rho}-t\frac{\tb{\pp}}{\tb{\pp}t}\;,
\\
%\tb{t}_{(\times)} &=\frac{1}{2}\bigg(\frac{1}{\rho^2}+t^2-1\bigg)\frac{\tb{\pp}}{\tb{\pp}t}-\rho t\frac{\tb{\pp}}{\tb{\pp}\rho}
%\\
%&\feq+\frac{2\hat{r}}{\delta}\frac{1}{\rho}\sum_{\bar{k}}\hat{r}^{2(\bar{N}-\bar{k}-1)}\frac{\tb{\pp}}{\tb{\pp}\bar{\psi}_{\bar{k}}}\;,
%\\
%\tb{t}_{(-)} &=\frac{1}{2}\bigg(\frac{1}{\rho^2}+t^2+1\bigg)\frac{\tb{\pp}}{\tb{\pp}t}-\rho t\frac{\tb{\pp}}{\tb{\pp}\rho}
%\\
%&\feq+\frac{2\hat{r}}{\delta}\frac{1}{\rho}\sum_{\bar{k}}\hat{r}^{2(\bar{N}-\bar{k}-1)}\frac{\tb{\pp}}{\tb{\pp}\bar{\psi}_{\bar{k}}}\;,
%\\
\tb{t}_{\binom{\times}{-}} &=\frac{1}{2}\bigg(\frac{1}{\rho^2}+t^2\mp1\bigg)\frac{\tb{\pp}}{\tb{\pp}t}-\rho t\frac{\tb{\pp}}{\tb{\pp}\rho}
\\
&\feq+\frac{2}{\delta}\frac{1}{\rho}\sum_{\bar{k}}\hat{r}^{2(\bar{N}-\bar{k})-1}\frac{\tb{\pp}}{\tb{\pp}\bar{\psi}_{\bar{k}}}\;.
\end{aligned}
\end{equation}
The projections of the Killing vectors to the ${\mathrm{AdS}_2}$ directions are shown in conformal diagrams in Fig.~\ref{fig:conf}.

Altogether the explicit symmetries of the near-horizon geometry are fully described by a set of ${3+\bar{N}}$ Killing vectors, e.g.,
\begin{equation}\label{eq:KVnhlimito}
\tb{t}_{(+)}\;,
\;\;
\tb{t}_{(\times)}\;,
\;\;
\tb{t}_{(-)}\;,
\;\;
\tb{s}_{(\bar{k})}\;,
\end{equation}
or alternatively
\begin{equation}\label{eq:KVnhlimitn}
\tb{t}_{(+)}\;,
\;\;
\tb{t}_{(\bullet)}\;,
\;\;
\tb{t}_{(\circ)}\;,
\;\;
\tb{s}_{(\bar{k})}\;.
\end{equation}
This indicates that the symmetry is enhanced from ${\mathbb{R}\times\mathrm{U}(1)^{\bar{N}}}$ to ${SL(2,\mathbb{R})\times\mathrm{U}(1)^{\bar{N}}}$.

Moreover, the rank-two Killing tensor \eqref{eq:newKT} in the degenerate direction is also simplified [see \eqref{eq:Umusrexp}, \eqref{eq:Jmunuexp}, and \eqref{eq:ONvectorslimitNH}]
\begin{equation}\label{eq:nhQexpl}
\cir{\tb{q}}_{(N)}=\frac{\delta}{\cir{\bar{J}}(-\hat{r}^2)}\biggl[\rho^2\frac{\tb{\pp}}{\tb{\pp}\rho}^{\!\!\!\bs{2}}-\frac{1}{\rho^2}\bigg(\frac{\tb{\pp}}{\tb{\pp}t}+\frac{2}{\delta}\rho\sum_{\bar{k}}\hat{r}^{2(\bar{N}-\bar{k})-1}\frac{\tb{\pp}}{\tb{\pp}\bar{\psi}_{\bar{k}}}\bigg)^{\!\!\bs{2}}\biggr]\;.
\end{equation}
It can be easily verified that such a tensor can be actually written as a combination of the Killing vectors \eqref{eq:KVnhlimito},
\begin{equation}
\begin{aligned}
\tb{q}_{(N)} &=\frac{\delta}{\cir{\bar{J}}(-\hat{r}^2)}\biggl[\tb{t}_{(+)}^{\bs 2}+\tb{t}_{(\times)}^{\bs 2}-\tb{t}_{(-)}^{\bs 2}\\
&\feq-\Big(\frac{2}{\delta}\sum_{\bar{k}}\hat{r}^{2(\bar{N}-\bar{k})-1}\tb{s}_{(\bar{k})}\Big)^{\!\bs{2}}\biggr]\;,
\end{aligned}
\end{equation}
or in terms of the Killing vectors \eqref{eq:KVnhlimitn} as
\begin{equation}
\begin{aligned}
\tb{q}_{(N)} &=\frac{\delta}{\cir{\bar{J}}(-\hat{r}^2)}\biggl[\tb{t}_{(+)}^{\bs 2}-\tb{t}_{(\circ)}\vee\tb{t}_{(\bullet)}\\
&\feq-\Big(\frac{2}{\delta}\sum_{\bar{k}}\hat{r}^{2(\bar{N}-\bar{k})-1}\tb{s}_{(\bar{k})}\Big)^{\!\bs{2}}\biggr]\;,
\end{aligned}
\end{equation}
which agrees with the result obtained in \cite{XuYue:2015}. The part of the Killing tensor which does not contain the directions $\tb{s}_{(\bar{k})}$ resembles the form of the Killing metric, cf. \eqref{eq:KMortho}, \eqref{eq:KMnull}.

%%%%%%%%%%%%%%%%%%%%%%%%%%%%%%%%%%%%%%%%%%%%%%%%%%%%%%%%%%%%%%%%%%%%%%%%%%%%%%%%%%%%%%
%% COMPLETE DEGENERATION

\section{Complete degeneration}
\label{sc:CompleteDegeneration}

Finally, we would like to discuss the limit when all directions degenerate, ${\Ns=N}$. The complete degeneration is possible only in the NUT-like choice for ${\lambda>0}$, because only then all coordinates ${x_1,\,\dots,\,x_{\bar{N}},\,r}$ possess the corresponding real tangent points ${\hat{x}_1,\,\dots,\,\hat{x}_{\bar{N}},\,\hat{r}^{(\lambda)}}$, see Table~\ref{tab:tangentpoints}, so we can expand these coordinates around such double roots.

As with the NUT-like limits, we introduce a rescaled coordinate $\xi_{\bar{\mu}}$,
\begin{equation}
    x_{\bar{\mu}} = \frac{{}^+x_{\bar{\mu}}+{}^-x_{\bar{\mu}}}{2} + \frac{{}^+x_{\bar{\mu}}-{}^-x_{\bar{\mu}}}{2}\, \xi_{\bar{\mu}}\;.
\end{equation}
and parametrize the convergence of the roots ${}^+x_{\bar{\mu}}$, ${}^-x_{\bar{\mu}}$, by the positive parameter ${\eps\ll 1}$,
\begin{equation}
    {}^+x_{\bar{\mu}}-{}^-x_{\bar{\mu}} = 2\,\delta x_{\bar{\mu}}\, \eps\;,
\end{equation}
where ${\delta x_{\bar{\mu}}>0}$. As in \eqref{eq:xsclexp} and \eqref{eq:NUTexp}, the expansions of $x_{\bar{\mu}}$, $b_{\bar{\mu}}$ read
\begin{equation}
	x_{\bar{\mu}} = \hat{x}_{\bar{\mu}} + \delta x_{\bar{\mu}}\xi_{\bar{\mu}}\eps+\mathcal{O}(\eps^2)\;.
\end{equation}
\begin{equation}
    b_{\bar{\mu}}
      =\hat{b}_{\bar{\mu}}
      - (2N-1)\lambda \hat{U}_{\bar{\mu}}\frac{\delta x_{\bar{\mu}}^2}{2\hat{x}_{\bar{\mu}}}\eps^2
      + \mathcal{O}(\eps^3)\;,
\end{equation}
which leads to the approximations of the polynomials $\mathcal{X}_{\bar{\mu}}$ [cf. \eqref{eq:Xsimple}],
\begin{equation}
	\mathcal{X}_{\bar{\mu}} = -(2N-1)\lambda\hat{U}_{\bar{\mu}}\delta x_{\bar{\mu}}^2(1-\xi_{\bar{\mu}}^2)\,\varepsilon^2
      + \mathcal{O}(\varepsilon^3)\;.
\end{equation}

The radial coordinate can be expanded by any of three procedures described in Sec.~\ref{sc:NHLimits}, but, for simplicity, we use the procedure which starts with the extreme case, see Sec.~\ref{ssc:extremenearhorizon}. This time, however, we begin with a different critical value,
\begin{equation}
	m=\hat{m}^{(\lambda)}\;
\end{equation}
which corresponds to the extreme horizon $\hat{r}^{(\lambda)}$ and introduce the rescaled coordinate $\rho$ in the vicinity of $\hat{r}^{(\lambda)}$,
\begin{equation}\label{eq:trrr}
	r=\hat{r}^{(\lambda)} + \delta r\rho\eps+\mathcal{O}(\eps^2)\;,
\end{equation}
where $\delta r>0$.
The polynomial $\Delta$ is then approximated~as
\begin{equation}
	\Delta=-(2N-1)\lambda\hat{J}_N(-{\hat{r}^{(\lambda)}}^2)\delta r^2 \rho^2 \eps^2 +\mathcal{O}(\eps^3)\;.
\end{equation}
In order to obtain a finite metric, the transformation \eqref{eq:trrr} must be supplemented with an appropriate transformation of the Killing coordinates,
\begin{equation}
\phi_{\bar{\mu}}=\ph_{\bar{\mu}}\,\frac{1}{\eps}\;,
\quad
\phi_N=t\,\frac{1}{\eps}\;.
\end{equation}
We simply set ${\cir{x}_\mu=\hat{x}_\mu}$, but other scaling of these parameters would lead to the same resulting metric. Taking the limit ${\eps\rightarrow0}$ and fixing
\begin{equation}
	\delta x_{\bar{\mu}}=\delta r=\frac{1}{\delta}\;,
	\quad
	\delta=(2N-1)\lambda\;,
\end{equation}
we obtain the resulting metric
\begin{equation}
  \tb{g} =  \frac{1}{\delta}\biggl[-\frac{\tb{\dd}\rho^{\bs{2}}}{\rho^2}+\rho^2\,\tb{\dd} t^{\bs{2}}+\sum_{\bar{\mu}}\bigg(\frac{\tb{\dd}\xi_{\bar{\mu}}^{\bs{2}}}{1-\xi_{\bar{\mu}}^2}+(1-\xi_{\bar{\mu}}^2)\,\tb{\dd} \ph_{\bar{\mu}}^{\bs{2}}\bigg)\biggr]\;.
\end{equation}
It is a direct product spacetime of two-dimensional de Sitter spacetime and homogeneous metrics on two-spheres, i.e. a generalization of a four-dimensional Nariai spacetime. Also, it is not surprising that the rescaled radial coordinate $\rho$ plays a role of time, since in the extreme spacetime of coinciding outer horizon ${{}^+r}$ and cosmological horizon ${{}^{+}r^{(\lambda)}}$, the coordinate $r$ is timelike.

%%%%%%%%%%%%%%%%%%%%%%%%%%%%%%%%%%%%%%%%%%%%%%%%%%%%%%%%%%%%%%%%%%%%%%%%%%%%%%%%%%%%%%
%% CONCLUSIONS

\section{Conclusions}
\label{sc:Conc}
In this paper, several limits of the Kerr--NUT--(A)dS spacetimes were investigated. 

Although the Kerr--NUT--(A)dS spacetimes have been widely studied in the last decade, their full interpretation is not yet achieved. There is still missing understanding of the geometry in the presence of NUT parameters. In Sec.~\ref{sc:Kerr} we have given a rather extensive discussion of the parameter choices leading to physically interesting subcases of the Kerr--NUT--(A)dS geometry. We have identified the ``Kerr-like'' and ``NUT-like'' choices which generalize analogical dichotomy of Kerr and Taub--NUT solution in four dimensions. We hope that this discussion makes at least a small step in further understanding the Kerr--NUT--(A)dS metric.

However, the main goal of the paper was investigating nontrivial limits of the Kerr--NUT--(A)dS solution, namely the limits of the double roots of the metric functions ${X_\mu}$. Such limits generalize two interesting cases known from the four dimensions: the Taub--NUT limit and the near-horizon limit of the extreme Kerr black hole. For the purpose of the limiting procedure we have reparametrized the solution using so-called tangent-point parametrization. In this language it was possible to control the limiting process when two roots of the metric function coincide at the critical value. The limiting procedure had to be accompanied by a proper rescaling of coordinates which effectively zooms in on the neighborhood of the critical points of the metric functions.

We have presented limits along arbitrary number of planes of rotations, which give rather complicated geometries. Next, we have discussed particular cases, recovering, for example, the ``multiply nutty'' spacetimes of \cite{MannStelea:2006}.

When the ``double-root'' limit is taken for the metric function governing the position of the horizons, it leads to the near-horizon limit \cite{BardeenHorowitz:1999}. In our discussion we have identified three different ways how one can zoom in on the near-horizon region. It naturally leads to the near-horizon geometry written in different coordinates, each of them adjusted to a different static Killing vector. 

We have identified enhanced symmetry of the limiting spacetimes and showed that the hidden symmetries of the original spacetime encoded by Killing tensors become reducible to a richer structure of the explicit symmetries given by Killing vectors.

Finally, we have also studied the ``double-root'' limit taken in all coordinates ${x_\mu}$ and obtained a rather trivial direct product spacetime generalizing four-dimensional Nariai spacetime.

%%%%%%%%%%%%%%%%%%%%%%%%%%%%%%%%%%%%%%%%%%%%%%%%%%%%%%%%%%%%%%%%%%%%%%%%%%%%%%%%%%%%%%
%% ACKNOWLEDGMENTS

\section*{Acknowledgments}
This work was supported by the project of excellence of the Czech Science Foundation No.~\mbox{14-37086G}. I.K. was also supported by the Charles University Grants No.~\mbox{GAUK-196516} and No.~\mbox{SVV-260320}. We acknowledge the hospitality of the Theoretical Physics Institute of the University of Alberta where this work began. We also appreciate the financial support of the Natural Sciences and Engineering Research Council of Canada and the Killam Trust. Authors would like to thank Valeri Frolov, David Kubiz\v{n}\'{a}k, and Filip Hejda for valuable discussions.

%%%%%%%%%%%%%%%%%%%%%%%%%%%%%%%%%%%%%%%%%%%%%%%%%%%%%%%%%%%%%%%%%%%%%%%%%%%%%%%%%%%%%%
%% APPENDIX

\appendix

%%%%%%%%%%%%%%%%%%%%%%%%%%%%%%%%%%%%%%%%%%%%%%%%%%%%%%%%%%%%%%%%%%%%%%%%%%%%%%%%%%%%%%
%% Functions $J$, $A$, $U$

\section{Functions $J$, $A$, $U$}
\label{ap:ui}
Throughout the paper we use many polynomial functions, such as $J(x^2)$, $A^{(k)}$, $U_\mu$, and their generalizations. Although these are just polynomial functions, it turns out that they satisfy important identities which appear in almost any calculations regarding the Kerr--NUT--(A)dS spacetimes. We use here the notation which was established mainly in \cite{ChenLuPope:2006,KrtousEtal:2016} and is widely used in many related papers. All definitions come in two variants, as polynomials of variables $x_\mu$ and $a_\mu$. The most important functions are functions $J$, which give rise to both~$A$ and~$U$. Their simplest variants are defined by
\begin{equation}\label{eq:J}
\begin{gathered}
J(x^2) =\prod_\mu(x_\mu^2-x^2)=\sum_{k=0}^N A^{(k)}{(-x^2)}^{N-k}\;,
\\
\mathcal{J}(a^2) =\prod_\mu(a_\mu^2-a^2)=\sum_{k=0}^N \mathcal{A}^{(k)}{(-a^2)}^{N-k}\;,
\end{gathered}
\end{equation}
where the coefficients are
\begin{equation}\label{eq:A}
A^{(k)}=\sum_{\mathclap{\substack{\mu_1,\,\dots,\,\mu_k\\ \mu_1<\dots<\mu_k}}}x_{\mu_1}^2\dots x_{\mu_k}^2\;,
\quad
\mathcal{A}^{(k)}=\sum_{\mathclap{\substack{\mu_1,\,\dots,\,\mu_k\\ \mu_1<\dots<\mu_k}}}a_{\mu_1}^2\dots a_{\mu_k}^2\;.
\end{equation}
We can generalize \eqref{eq:J} by omitting an index~$\mu$,
\begin{equation}\label{eq:Jmu}
\begin{gathered}
J_\mu(x^2) =\prod_{\substack{\nu\\ \nu\neq\mu}}(x_\nu^2-x^2)=\sum_k A_{\mu}^{(k)}{(-x^2)}^{N-k-1}\;,
\\
\mathcal{J}_\mu(a^2) =\prod_{\substack{\nu\\ \nu\neq\mu}}(a_\nu^2-a^2)=\sum_k \mathcal{A}_{\mu}^{(k)}{(-a^2)}^{N-k-1}\;,
\end{gathered}
\end{equation}
which generate the functions
\begin{equation}\label{eq:Amu}
A_\mu^{(k)}=\sum_{\mathclap{\substack{\nu_1,\,\dots,\,\nu_k\\ \nu_1<\dots<\nu_k\\ \nu_j\neq\mu}}}x_{\nu_1}^2\dots x_{\nu_k}^2\;,
\quad
\mathcal{A}_\mu^{(k)}=\sum_{\mathclap{\substack{\nu_1,\,\dots,\,\nu_k\\ \nu_1<\dots<\nu_k\\ \nu_j\neq\mu}}}a_{\nu_1}^2\dots a_{\nu_k}^2\;.
\end{equation}
Similarly, we could also define the functions
$J_{\mu\nu}(x^2)$, $\mathcal{J}_{\mu\nu}(x^2)$, $A_{\mu\nu}^{(k)}$, $\mathcal{A}_{\mu\nu}^{(k)}$ by skipping the indices $\mu$, $\nu$. Besides this we set
\begin{equation}
\begin{gathered}
	A^{(0)}=A_\mu^{(0)}=A_{\mu\nu}^{(0)}=\dots=1\;,
	\\
	\mathcal{A}^{(0)}=\mathcal{A}_\mu^{(0)}=\mathcal{A}_{\mu\nu}^{(0)}=\dots=1\;,
\end{gathered}
\end{equation}
and
\begin{equation}
\begin{gathered}
	J_\mu(x^2)\big|_{N=1}=J_{\mu\nu}(x^2)\big|_{N=2}=\dots=1\;,
	\\
	\mathcal{J}_\mu(a^2)\big|_{N=1}=\mathcal{J}_{\mu\nu}(a^2)\big|_{N=2}=\dots=1\;.
\end{gathered}
\end{equation}
We also assume that the functions $J$ and $A$ are zero if the indices ${\mu,\nu}$ overflow. Finally, the special case of \eqref{eq:Jmu} are the functions
\begin{equation}\label{eq:Umu}
\begin{gathered}
	U_\mu=J_\mu(x_\mu^2)=\prod_{\substack{\nu\\ \nu\neq\mu}}(x_\nu^2-x_\mu^2)\;,
	\\
	\mathcal{U}_\mu=\mathcal{J}_\mu(a_\mu^2)=\prod_{\substack{\nu\\ \nu\neq\mu}}(a_\nu^2-a_\mu^2)\;.
\end{gathered}
\end{equation}

These functions satisfy the following identities:
\begin{equation}\label{eq:AAmuAmu}
	A^{(k)} =A_{\mu}^{(k)}+x_\mu^2 A_{\mu}^{(k-1)}\;,
\end{equation}
\begin{equation}\label{eq:AmuAmuAmunu}
	A_\mu^{(k)} =A_{\mu\nu}^{(k)}+x_\nu^2 A_{\mu\nu}^{(k-1)}\;,
\end{equation}
\begin{equation}\label{eq:AmuU1}
	\sum_\mu A_\mu^{(l)}\frac{{(-x_\mu^2)}^{N-k-1}}{U_\mu}=\delta_k^l\;,
\end{equation}	
\begin{equation}\label{eq:AmuU2}
	\sum_k A_\mu^{(k)}\frac{{(-x_\nu^2)}^{N-k-1}}{U_\nu}=\delta_\mu^\nu\;,
\end{equation}
\begin{equation}\label{eq:JJ}
\sum_\kappa\frac{J_\nu(a_{\kappa}^2)}{\mathcal{U}_{\kappa}}\frac{\mathcal{J}_\kappa(x_{\mu}^2)}{U_{\mu}}=\delta^\mu_\nu\;,
\end{equation}
\begin{equation}\label{eq:Ader}
 	A_{\mu,\nu}^{(k)}=\frac{2x_\nu}{x_\nu^2-x_\mu^2}\big(A_{\mu}^{(k)}-A_{\nu}^{(k)}\big)\;,
\end{equation}
\begin{equation}\label{eq:Uder}
 	U_{\mu,\nu} =\delta_{\mu\nu}\sum_{\substack{\rho\\ \rho\neq\mu}}\frac{2x_\mu}{x_\mu^2{-}x_\rho^2}U_{\mu}+(1{-}\delta_{\mu\nu})\frac{2x_\nu}{x_\nu^2{-}x_\mu^2}U_{\mu}\;.
\end{equation}

\section{Connection forms}
\label{ap:cononeforms}
From the first structure equations,
\begin{equation}\label{eq:streqI}
	\tb{\dd}\tb{e}^a +\sum_{b}\tb{\omega}^a{}_b\wedge\tb{e}^b=0\;,
	\quad
	\tb{\omega}_{ab} + \tb{\omega}_{ba}=0\;,
\end{equation}
and by employing \eqref{eq:Uder} and \eqref{eq:Ader}, we can obtain the connection forms \cite{HamamotoEtal:2007}
\begin{equation}\label{eq:connectionforms}
\begin{split}
	\tb{\omega}_{\mu\nu} &= (1{-}\delta_{\mu\nu})\frac{1}{x_\nu^2{-}x_\mu^2}\bigg[x_\nu \bigg(\frac{X_\nu}{U_\nu}\bigg)^{\!\!\frac{1}{2}}\tb{e}^{\mu}+x_\mu\bigg(\frac{X_\mu}{U_\mu}\bigg)^{\!\!\frac{1}{2}}\tb{e}^{\nu}\bigg]\;,
	\\
	\tb{\omega}_{\ubars\mu\ubars\nu} &= (1{-}\delta_{\mu\nu})\frac{1}{x_\nu^2{-}x_\mu^2}\bigg[x_\mu\bigg(\frac{X_\nu}{U_\nu}\bigg)^{\!\!\frac{1}{2}}\tb{e}^{\mu}+x_\nu\bigg(\frac{X_\mu}{U_\mu}\bigg)^{\!\!\frac{1}{2}}\tb{e}^{\nu}\bigg]\;,
	\\
	\tb{\omega}_{\mu\ubars\nu} &= \delta_{\mu\nu}\bigg[\!-\bigg(\bigg(\frac{X_\mu}{U_\mu}\bigg)^{\!\!\frac{1}{2}}\bigg)_{\!\!,\mu}\tb{\hat{e}}^\mu+\sum_{\substack{\rho\\ \rho\neq\mu}}\frac{x_\mu}{x_\mu^2{-}x_\rho^2}\bigg(\frac{X_\rho}{U_\rho}\bigg)^{\!\!\frac{1}{2}}\tb{\hat{e}}^{\rho}\bigg]
	\\
	 &\feq+(1{-}\delta_{\mu\nu})\frac{x_\mu}{x_\mu^2{-}x_\nu^2}\bigg[\bigg(\frac{X_\nu}{U_\nu}\bigg)^{\!\!\frac{1}{2}}\tb{\hat{e}}^{\mu}-\bigg(\frac{X_\mu}{U_\mu}\bigg)^{\!\!\frac{1}{2}}\tb{\hat{e}}^\nu\bigg]\;,	 
\end{split}
\end{equation}
where the underscored indices of the connection forms correspond to $\tb{\hat{e}}$'s while the ordinary ones correspond to~$\tb{e}$'s.

By taking the limit that is described in Sec.~\ref{sc:limitingprocedure}, we find the connection forms of the limiting metric \eqref{eq:limmet} with respect to the orthonormal frame \eqref{eq:ON1formslimit}. The nonvanishing connection forms are
\begin{widetext}
\begin{equation}\label{eq:connectionformslim}
\begin{gathered}
\begin{aligned}
	\tb{\omega}_{\mur\nur} &= \frac{1}{x_\nur^2{-}x_\mur^2}\bigg[x_\nur\bigg(\frac{X_\nur}{\hat{\breve{J}}(x_{\nur}^2)\invbreve{U}_{\nur}}\bigg)^{\!\!\frac{1}{2}}\tb{e}^{\mur}
	 +x_\mur\bigg(\frac{X_\mur}{\hat{\breve{J}}(x_{\smash{\mur}}^2)\invbreve{U}_{\mur}}\bigg)^{\!\!\frac{1}{2}}\tb{e}^{\nur}\bigg]\;,
	\\
	\tb{\omega}_{\ubars\mur\ubars\nur} &= \frac{1}{x_\nur^2{-}x_\mur^2}\bigg[x_\mur\bigg(\frac{X_\nur}{\hat{\breve{J}}(x_{\nur}^2)\invbreve{U}_{\nur}}\bigg)^{\!\!\frac{1}{2}}\tb{e}^{\mur}
	 +x_\nur\bigg(\frac{X_\mur}{\hat{\breve{J}}(x_{\smash{\mur}}^2)\invbreve{U}_{\mur}}\bigg)^{\!\!\frac{1}{2}}\tb{e}^{\nur}\bigg]\;,
	\\
	\tb{\omega}_{\mur\ubars\nur} &=\frac{x_\mur}{x_\mur^2{-}x_{\nur}^2}\bigg[\bigg(\frac{X_\nur}{\hat{\breve{J}}(x_{\nur}^2)\invbreve{U}_{\nur}}\bigg)^{\!\!\frac{1}{2}}\tb{\hat{e}}^{\mur}-\bigg(\frac{X_\mur}{\hat{\breve{J}}(x_{\mur}^2)\invbreve{U}_{\mur}}\bigg)^{\!\!\frac{1}{2}}\tb{\hat{e}}^{\nur}\bigg]\;,
	\\
	\tb{\omega}_{\mur\ubars\mur} &= -\bigg(\bigg(\frac{X_\mur}{\hat{\breve{J}}(x_{\mur}^2)\invbreve{U}_{\mur}}\bigg)^{\!\!\frac{1}{2}}\bigg)_{\!\!,\mur}\tb{\hat{e}}^{\mur}+\sum_{\substack{{\invbreve\rho}\\ {\invbreve\rho}\neq\mur}}\frac{x_\mur}{x_\mur^2{-}x_{\invbreve\rho}^2}\bigg(\frac{X_{\invbreve\rho}}{\hat{\breve{J}}(x_{\invbreve\rho}^2)\invbreve{U}_{\invbreve\rho}}\bigg)^{\!\!\frac{1}{2}}\tb{\hat{e}}^{\invbreve\rho}\;,
\end{aligned}
\qquad\quad
\begin{aligned}
	\tb{\omega}_{\mus\nur} &= \frac{x_\nur}{x_\nur^2{-}\hat{x}_\mus^2}\bigg(\frac{X_\nur}{\hat{\breve{J}}(x_{\nur}^2)\invbreve{U}_{\nur}}\bigg)^{\!\!\frac{1}{2}}\tb{e}^{\mus}\;,
	\\
	\tb{\omega}_{\ubars\mus\ubars\nur} &= \frac{\hat{x}_\mus}{x_\nur^2{-}\hat{x}_\mus^2}\bigg(\frac{X_\nur}{\hat{\breve{J}}(x_{\nur}^2)\invbreve{U}_{\nur}}\bigg)^{\!\!\frac{1}{2}}\tb{e}^{\mus}\;,
	\\
	\tb{\omega}_{\mus\ubars\nur} &= \frac{\hat{x}_\mus}{\hat{x}_\mus^2{-}x_\nur^2}\bigg(\frac{X_\nur}{\hat{\breve{J}}(x_{\nur}^2)\invbreve{U}_{\nur}}\bigg)^{\!\!\frac{1}{2}}\tb{\hat{e}}^{\mus}\;,
	\\
	\tb{\omega}_{\mur\ubars\nus} &= \frac{x_\mur}{\hat{x}_\nus^2{-}x_\mur^2}\bigg(\frac{X_\mur}{\hat{\breve{J}}(x_{\mur}^2)\invbreve{U}_{\mur}}\bigg)^{\!\!\frac{1}{2}}\tb{\hat{e}}^{\nus}\;,
	\\
\end{aligned}
\\
\tb{\omega}_{\mus\ubars\mus} = \xi_\mus\Big(\delta x_\mus |\invbreve{J}(\hat{x}_\mus^2)|(1{-}\xi_\mus^2)\Big)^{\!\!-\frac{1}{2}}\tb{\hat{e}}^{\mus}
	 +\sum_{\invbreve\rho}\frac{\hat{x}_\mus}{\hat{x}_\mus^2{-}x_{\invbreve\rho}^2}\bigg(\frac{X_{\invbreve\rho}}{\hat{\breve{J}}(x_{\invbreve\rho}^2)\invbreve{U}_{\invbreve\rho}}\bigg)^{\!\!\frac{1}{2}}\tb{\hat{e}}^{\invbreve\rho}\;, \quad\mur\neq\nur \;.
\end{gathered}
\end{equation}
\end{widetext}

%%%%%%%%%%%%%%%%%%%%%%%%%%%%%%%%%%%%%%%%%%%%%%%%%%%%%%%%%%%%%%%%%%%%%%%%%%%%%%%%%%%%%%
%% REFERENCES

%\bibliographystyle{apsrev4-1} %leave this commented out when using longbibliography option
%\bibliography{references.bib}


\end{document}